\documentclass[a4paper,12pt,twoside,openright]{UGthesis}
\usepackage{fancyhdr}
\usepackage[dvips]{graphicx}
\usepackage{cite}
\usepackage{amssymb}
\usepackage[mathscr]{eucal}
\usepackage[errorshow]{tracefnt}
\usepackage{amsmath}
\usepackage{amssymb}
\usepackage{amsthm}
\usepackage{eucal}
\usepackage{eufrak}

\usepackage{epsf}
\usepackage{amsmath}
\usepackage{graphicx}
\usepackage{color}
\usepackage{psfrag}
\usepackage{amssymb}

\usepackage{ifpdf}

\newcommand{\beqs}{\begin{eqnarray}}
\newcommand{\eeqs}{\end{eqnarray}}

\newcommand{\cN}{{\cal N}}

\newcommand{\beq}{\begin{equation}}
\newcommand{\beqn}{\begin{equation*}}
\newcommand{\eeq}{\end{equation}}
\newcommand{\eeqn}{\end{equation*}}
\newcommand{\beqa}{\begin{eqnarray}}
\newcommand{\beqan}{\begin{eqnarray*}}
\newcommand{\eeqa}{\end{eqnarray}}
\newcommand{\eeqan}{\end{eqnarray*}}
\newcommand{\bdm}{\begin{displaymath}}
\newcommand{\edm}{\end{displaymath}}

\newcommand{\ba}{\begin{array}}
\newcommand{\ea}{\end{array}}

\newcommand\ffam{\sffamily}
\newcommand\fser{\bfseries}
\newcommand\fsh{\upshape}

\newcommand\rctr{\renewcommand{\theenumi}{\roman{enumi}}}

\newcommand\nn{\nonumber}

\newcommand\benu{\begin{enumerate}}
\newcommand\eenu{\end{enumerate}}
\newcommand\bit{\begin{itemize}}
\newcommand\eit{\end{itemize}}

\newcommand{\Th}{T_{\rm h}}
\newcommand{\Uh}{U_{\rm h}}
\newcommand{\qh}{q_{\rm h}}
\newcommand{\Pf}{\mathrm{Pf}\,}

\newcommand{\tr}{{\rm tr}}
\newcommand{\Tr}{{\rm Tr}}
\newcommand{\im}{{\rm Im}}

\newcommand{\IZ}{{\mathbb Z }}

\newcommand{\ov}{\overline}
\newcommand{\Ncal}{{\cal N }}

\newcommand{\ABFLP}{{\sffamily\bfseries Paper~I}}
\newcommand{\Petersson}{{\sffamily\bfseries Paper~II}}
\newcommand{\AFP}{{\sffamily\bfseries Paper~III}}
\newcommand{\FP}{{\sffamily\bfseries Paper~IV}}
\newcommand{\PeterssonN}{{\sffamily\bfseries Paper~V}}
\newcommand{\PSU}{{\sffamily\bfseries Paper~VI}}
\newcommand{\AFPold}{{\sffamily\bfseries Paper~VII}}
\newcommand{\GNP}{{\sffamily\bfseries Paper~VIII}}

\numberwithin{equation}{section}

\RequirePackage{hyperref}

\begin{document}

%
%




%
%

\thispagestyle{empty}
\begin{center}
\LARGE{\bf   D-instantons and Dualities\\
in String Theory  \\[8mm]}
 \large{C}\normalsize{HRISTOFFER} \large{P}\normalsize{ETERSSON}\\[8mm]
\small{\emph{Physique Th\'eorique et Math\'ematique,\\
Universit\'e Libre de Bruxelles, C.P. 231, 1050 Bruxelles, Belgium}}\\[2mm] 
and\\[2mm] 
\small{\emph{International Solvay Institutes, Bruxelles, Belgium.}}\\[8mm] 
\end{center}

\centerline{\bf Abstract}

\medskip
\normalsize
\noindent
This introductory part of the author's PhD compilation thesis discusses non-perturbative aspects of string theory, with focus on D-instantons which play key roles in the context of string phenomenology and dualities. By translating ideas from ordinary instanton calculus in field theory, D-instanton calculus is formulated and applied to supersymmetric gauge theories realized as world volume theories of spacefilling D-branes in non-trivial backgrounds. We show that, for configurations in which certain fermionic D-instanton zero modes are either lifted or projected out, new couplings, which may be forbidden at the perturbative level, are generated in the effective action, even though their origin does not admit an obvious interpretation in terms of ordinary field theory. Such couplings are of great relevance for semi-realistic MSSM/GUT models since they can correspond to Yukawa couplings, Majorana mass terms for right-handed neutrinos or Polonyi terms, relevant for supersymmetry breaking. In the last part, we make use of various string dualities in order to compute multi-instanton corrections which are otherwise technically difficult to obtain from explicit D-instanton calculations. We compute one-loop diagrams with BPS particles in type IIA string theory configurations, involving spacefilling D-branes and orientifolds, and obtain exact quantum corrections in the dual type IIB picture, including infinite series of D-instanton contributions. By lifting the one-loop calculations to the M-theory picture, we obtain a geometric understanding of the non-perturbative sector of a wide range of gauge theories and elucidate the underlying symmetries of the effective action.

\newpage

\begin{center}
  {\fsh\ffam\fser Thesis for the degree of Doctor of Philosophy}
\end{center}


\begin{center}
{\upshape\sffamily\bfseries\huge 
\noindent
D-instantons and Dualities \\[2mm] 
in String Theory}
 \\[4mm]
\end{center}

\vspace*{2mm}
\begin{center}
        \rule{110mm}{2pt}
\end{center}

\vspace*{4mm}
\begin{center}
  {\fsh\ffam\fser\Large Christoffer Petersson}\\
\end{center}
\vspace{10mm}

\begin{center}
\includegraphics[width=6cm]{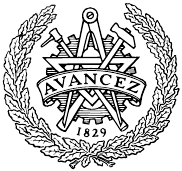}\\
\end{center}

\vspace{10mm}

\begin{center}
        {\ffam\fsh 
Fundamental Physics\\*[1mm]
        Chalmers University of Technology\\*[1mm]
        G\"oteborg, Sweden 2010}
\end{center}

\newpage

\noindent This PhD compilation thesis consists of an introductory text and eight appended published papers.
The introductory text is based on the following six papers, referred to as {\sffamily\bfseries Paper I--VI}:

\vspace{0.2cm}

\begin{enumerate}

\item[{\sffamily\bfseries I}]
	R.~Argurio, M.~Bertolini, G.~Ferretti, A.~Lerda and C.~Petersson,\\
	{\it Stringy instantons at orbifold singularities,} \\
	{JHEP} {\bf 0706:067} (2007)  \lbrack{\tt arXiv:0704.0262 [\tt hep-th]}].
	
\item[{\sffamily\bfseries II}]
	C.~Petersson,\\
	{\it Superpotentials from stringy instantons without orientifolds,}\\
          {JHEP} {\bf 0805:078} (2008)  \lbrack{\tt arXiv:0711.1837 [\tt hep-th]}].	

\item[{\sffamily\bfseries III}]
	R.~Argurio, G.~Ferretti and C.~Petersson,\\
	{\it Instantons and toric quiver gauge theories,}\\
         {JHEP} {\bf 0807:123} (2008)   \lbrack{\tt arXiv:0803.2041 [\tt hep-th]}].	
           
\item[{\sffamily\bfseries IV}]
	G.~Ferretti and C.~Petersson,\\ 
	{\it Non-perturbative effects on a fractional D3-brane,}\\
         {JHEP} {\bf 0903:040} (2009) \lbrack{\tt arXiv:0901.1182 [\tt hep-th]}].
     
\item[{\sffamily\bfseries V}]
	C.~Petersson,\\ 
	{\it Holomorphic corrections from wrapped euclidean branes,}\\
         {Nucl.\,Phys.\,B} {\bf 192-193:169-171} (2009).    
         
\item[{\sffamily\bfseries VI}]
	C.~Petersson, P.~Soler and A.~Uranga,\\ 
	{\it D-instanton and polyinstanton effects from type I' D0-brane loops,}\\
         {JHEP} {\bf 1006:089} (2010)  \lbrack{\tt arXiv:1001.3390 [\tt hep-th]}].         

\end{enumerate}

\vspace{0.2cm}

\noindent In addition, during the course of my PhD I have also published the following two papers, referred to as {\sffamily\bfseries Paper VII--VIII}, whose topics lie outside the main line of focus of the introductory text: 

\vspace{0.2cm}

\begin{enumerate} 

\item[{\sffamily\bfseries VII}]
	R.~Argurio, G.~Ferretti and C.~Petersson,\\
	{\it Massless fermionic bound states and the gauge/gravity correspondence,}\\
	{JHEP} {\bf 0603:043} (2006)  [{\tt hep-th/0601180}].

\item[{\sffamily\bfseries VIII}]
	U.~Gran, B.~E.~W.~Nilsson and C.~Petersson,\\ 
	{\it On relating multiple M2 and D2-branes,}\\
         {JHEP} {\bf 0810:067} (2008)  \lbrack{\tt arXiv:0804.1784 [\tt hep-th]}].	

\end{enumerate}

\newpage

%
%
\vspace*{0.5cm}

\centerline{\ffam\fser\Large Acknowledgments}
\medskip
\smallskip

\normalsize
\noindent
Let me begin by thanking Gabriele Ferretti for being the best supervisor a PhD student could possibly have. Our numerous discussions, collaborations and your constant support have been  invaluable to me. I am also truly grateful to Bengt E.\,W.\,Nilsson for vital contributions to my scientific education, collaboration and continuous encouragement. I would also like to thank my other extraordinary collaborators: Riccardo Argurio, Matteo Bertolini, Ulf Gran, Alberto Lerda, Pablo Soler and Angel Uranga. Working with you has been immensely rewarding and fun. 

A big thanks goes to my sidekick Daniel Persson for friendship, countless conversations and valuable comments on the manuscript. Thanks also to Jakob Palmkvist for being the Tintin scholar you are. I would also like to express my sincere gratitude to Per Arvidsson, Ling Bao, Victor Bengtsson, Lars Brink, Martin Cederwall, Ludde Edgren, Erik Flink, Rainer Heise, M{\aa}ns Henningson, Kate Larsson, Robert Marnelius, Fredrik Ohlsson, Per Salomonsson, Niclas Wyllard and all the other present and past members of the Department of Fundamental Physics for making my PhD years memorable and very enjoyable.   

I also wish to thank the CERN theory group for providing an exciting and inspiring working atmosphere during my year as a Marie Curie Fellow. In particular, I am very grateful to Luis Alvarez-Gaume, Pablo Camara, Bruce Campbell, Gia Dvali, Steve Giddings, Fernando Marchesano, Sara Pasquetti, Fernando Quevedo, Panteleimon Tziveloglou, Giovanni Villadoro, and Edward Witten for many interesting and illuminating discussions. Moreover, throughout the course of my PhD I have  benefitted greatly from conversations with Marcus Berg, Ralph Blumenhagen, Joe Conlon, Jose F.\,Morales, Eric Plauschinn, Maximilian Schmidt-Sommerfeld, Paolo di Vecchia and Timo Weigand.   

\emph{My research is currently supported by IISN-Belgium (conventions \\4.4511.06, 4.4505.86 and 4.4514.08), by the ``Communaut\'e Fran\c{c}aise de Belgique" through the ARC program and by a ``Mandat d'Impulsion Scientifique" of the F.R.S.-FNRS.}

\tableofcontents

\cleardoublepage
\pagestyle{fancy}
\renewcommand{\chaptermark}[1]{\markboth{Chapter \thechapter\ \ \ #1}{#1}}
\renewcommand{\sectionmark}[1]{\markright{\thesection\ \ #1}}
\lhead[\fancyplain{}{\sffamily\thepage}]%
  {\fancyplain{}{\sffamily\rightmark}}
\rhead[\fancyplain{}{\sffamily\leftmark}]%
  {\fancyplain{}{\sffamily\thepage}}
\cfoot{}
\setlength\headheight{14pt}

\rctr

%
%

\chapter{Introduction and Motivation}

\section{Why Expect New Physics at High Energies?}

The Standard Model of particle physics has so far been successfully tested with enormous precision. At the same time, a vast number of experiments concerning gravitational effects are very accurately accounted for by Einstein's general theory of relativity. However, even though we have two such successful theories, there are theoretical and experimental reasons to believe that these theories are only low energy effective theories, valid up to some energy scale at which they should be replaced with a more fundamental theory. For example, the Standard Model does not describe gravitational interactions and Einstein's description of gravity breaks down at high energies. In this section we will discuss some of the main features of these two theories as well as a few key reasons to believe that there exists a more fundamental theory that is capable of describing all physical processes at all energy scales in a unified way.

 The Standard Model is an $SU(3)\times SU(2)\times U(1)$ gauge theory, describing how three generations of quarks and leptons interact via the strong, weak and electromagnetic interactions  \cite{Glashow:1961tr,Weinberg:1967tq,Salam}.  
This theory is a particular example of a quantum field theory (QFT), which is a framework based on the principles of quantum mechanics and special relativity that describes matter and forces as quantum fields whose excitations correspond to the elementary  particles \cite{Weinberg:1995mt,Weinberg:1996kr,Peskin:1995ev}. 

Gauge invariance forbids explicit mass terms for the gauge bosons and since only the left handed fermions transform non-trivially under the $SU(2)$ factor, also fermion mass terms are forbidden. Instead, in order to account for the observed spectrum of particle masses, the gauge bosons of the weak interactions, as well as the fermions, are rendered massive by the process of spontaneous electroweak symmetry breaking of the $SU(2)\times U(1)$ group down to the electromagnetic abelian group. The energy scale at which this process takes place defines  the electroweak scale, which  is known from experiments to be around 246 GeV \cite{Amsler:2008zzb}. The microscopic origin of the mechanism triggering this breaking is not yet known, but the Standard Model provides a simple candidate by including an elementary scalar $SU(2)$ doublet Higgs field that acquires a non-vanishing vacuum expectation value at the electroweak scale and spontaneously breaks the electroweak symmetry \cite{Englert:1964et,Higgs:1964pj,Guralnik:1964eu}. It is through interactions with the Higgs field that the Standard Model particles acquire their masses. Moreover, this mechanism predicts the existence of a physical massive scalar particle, the Higgs boson, and an experimental detection of it would complete the Standard Model.
It is one of the main tasks for the Large Hadron Collider (LHC) at CERN in Switzerland to verify, or exclude, the existence of the Higgs boson.  

Perhaps the most exciting task for the LHC is however to search for new physics beyond the Standard Model, and there are reasons to expect that such new physics will be discovered by the LHC. In contrast to the mass terms for the gauge bosons of the weak interactions and the fermions, which are protected by gauge and chiral symmetries above the electroweak scale, there is no symmetry in the Standard Model that protects the Higgs mass term. This implies that the natural scale for the quantum corrections to the Higgs mass is given by the highest scale at which the theory is still valid, possibly  the Planck scale ($10^{19}$ GeV).  Unless there exists another scale, slightly above the electroweak scale, where new physics appears, an unnatural fine-tuning between these quantum corrections and the tree-level Higgs mass parameter is required in order for the Higgs mass to be of the order of the electroweak scale. 

The leading candidate for such new physics is based on the principle of supersymmetry, which is a spacetime symmetry that relates bosonic and fermionic degrees of freedom (see e.g. \cite{Wess:1992cp,Martin:1997ns,Weinberg:2000cr} for introductions). In its minimal version, called the minimal supersymmetric Standard Model (MSSM),  
supersymmetry predicts that every Standard Model particle has a supersymmetric partner with the same mass and charges, but which differ with half a unit of spin.\footnote{In order to avoid gauge anomalies caused by the fermionic superpartner of the Higgs scalar, the MSSM also includes a second Higgs doublet. Also, due to the holomorphy of   couplings in the superpotential, a second Higgs is required in order to generate masses for both down-type and up-type quarks.} This implies that supersymmetry, in its unbroken phase, relates  any scalar mass to the mass of its fermionic superpartner, which is protected by chiral symmetry. 
However, since superpartners have not yet been detected, they must have masses greater than their corresponding Standard Model particles. Because of this lack of mass degeneracy, if supersymmetry is a feature of nature, it must be in a broken phase. Moreover, in order to not be in conflict with experimental data, supersymmetry must be broken spontaneously by some other ``hidden'' sector of fields and then mediated to the visible MSSM sector, where soft terms  are generated. Soft terms are explicit supersymmetry breaking operators in the MSSM action which do not spoil the nice ultraviolet behavior of the theory. Therefore, in order to understand the microscopic origin of the soft terms, we need to understand the mechanism that triggers and mediates supersymmetry breaking.

 An independent reason for why the MSSM is appealing concerns dark matter. 
 Astrophysical and cosmological observations suggests that around 21\% of the total energy density of the observable universe consists of an unknown form of matter, dark matter, while ordinary baryonic matter only accounts for around 4\% \cite{Komatsu:2008hk}.\footnote{The remaining 75\% is attributed to dark energy, which is the unknown form of energy that causes the expansion of our universe to accelerate. Dark energy may be interpreted as the vacuum energy of empty space, corresponding to a small positive cosmological constant in Einstein's equations. A microscopic explanation of the value of the cosmological constant is currently lacking.} In contrast to the Standard Model, the MSSM provides candidate particles for dark matter.

Moreover, in the MSSM, the three gauge couplings run, according to the renormalization group equations, towards a unified value at high energies. 
Hence, in contrast to the Standard Model where the gauge couplings do not unify, the MSSM is compatible with the idea of a grand unified theory (GUT), where the three gauge groups are unified into a single one at high energies.

The GUT idea addresses the question concerning the arbitrariness that seems to be involved in the Standard Model. For example, the Standard Model provides no fundamental explanation for some of its characteristic features such as the particular choice of gauge group and matter representations. Furthermore, after choosing the gauge group and matter content there are still more than twenty free parameters, e.g. the values of the gauge coupling constants and the Yukawa couplings, which are determined by experiments. It would of course be more satisfying if these parameters were calculable in a more fundamental theory that contains fewer parameters; ideally none at all. 
   
 Even though the MSSM (possibly with a GUT extension) is a good candidate for new physics\footnote{For examples of different supersymmetric extensions of the Standard Model, see my more recent work \cite{Petersson:2011in,Petersson:2012dp,Bellazzini:2012mh,Petersson:2012nv,Dudas:2012fa,Berg:2012cg,Dudas:2013mia,D'Hondt:2013fql,Ferretti:2013wya,Calibbi:2014pza,Petersson:2014faa,Petersson:2014oga,Ferretti:2015dea,Ferretti:2015ala,Petersson:2015rza,Petersson:2015mkr}.},  it is a candidate within the framework of QFT and it does not address the fundamental problem of incorporating gravitational interactions. This is perhaps the most obvious reason to believe that the Standard Model is not a fundamental theory. It should instead be viewed as an effective theory, valid only at energy scales where gravitational effects can be neglected. Of course, at the energies accessible at current accelerators, gravitational effects can be neglected to a very good approximation.
   
  At macroscopic scales, Einstein's general theory of relativity has been experimentally verified to high accuracy. It is a classical theory that describes gravity as a geometric property of spacetime by determining how matter and energy curve spacetime via Einstein's equations \cite{WeinbergCos,Misner:1974qy,Wald:1984rg}. One immediate concern that arises when looking at these equations is that one side corresponds to matter, which is known to be governed by the laws of quantum mechanics at microscopic scales, while the other side corresponds to spacetime geometry, which is instead treated in a strictly classical way. 

Another problem with Einstein's equations is that they admit solutions which contain singularities, e.g.~inside black holes and at the big bang. Since such singularities correspond to regions in the universe where the curvature is strong and the relevant energy scale is very high, both gravitational and quantum effects are relevant. Motivated by the success of QFT in the context of the Standard Model, one might be tempted to apply the QFT framework to general relativity. However, in contrast to the Standard Model, general relativity is not renormalizable and it ceases to make sense at high energies since infinities encountered in loop amplitudes can not be cured.
 Hence, it should be viewed as an effective theory valid only up to the scale where quantum effects become important (i.e. around the Planck scale) at which we should replace Einstein's theory with a more fundamental theory which is consistent with quantum mechanics.

In summary, in this section we have  provided a variety of reasons of why to expect the existence of a fundamental framework treating all interactions, including gravitational ones, in a unified way, consistent with quantum mechanics. Such a framework should be valid at all energy scales and moreover, in order to be in agreement with experiments, it should reduce to the Standard Model or general relativity in certain low energy limits. In the remainder of this thesis we will discuss superstring theory (referred to as string theory in the following), which is a  framework beyond QFT that  provides a quantum description of gravity and where supersymmetry plays a key role.

\section{String Theory}

String theory starts from the assumption that the most fundamental constituents of nature are small vibrating one-dimensional strings, instead of zero-dimensional point particles, as is the case for QFT  (see \cite{Green:1987sp,Green:1987mn,Polchinski:1998rq,Polchinski:1998rr} for standard introductions to string theory). The extended nature of the strings provide a natural geometric cutoff scale, traditionally denoted by $\alpha'$, which is the basic reason for the absence of ultraviolet divergencies and singularities in string theory. This dimensionful parameter $\alpha'$ is the only parameter that enters the theory; all dimensionless parameters arise as vacuum expectation values of dynamical fields. For example, the string coupling constant $g_s$ arises from the vacuum expectation value of a certain scalar field called the dilaton. 

When $g_s$ is small, string theory admits a perturbative description in which we can perform an expansion in powers of $g_s$. In this description strings are moving in spacetime, tracing out two-dimensional worldsheets, in contrast to point particles which trace out one-dimensional worldlines. The strings interact by splitting and joining worldsheets. A string diagram where three strings interact corresponds to a single worldsheet that splits into two.  A perturbative expansion in $g_s$ corresponds to summing over all worldsheets which interpolate between the initial and final string configuration. The sum is organized according to the genus of each worldsheet and the power to which $g_s$ is raised is determined by the genus. For example, tree level string interactions correspond to worldsheets of genus zero, while one-loop corrections arise from genus one worldsheets.

Upon quantization of the string, the different vibrational modes are interpreted as different particles in spacetime.  The quanta of gravity, the graviton, is found among the massless modes of closed strings and gauge particles are found among the massless modes of open strings. Hence, string theory provides a unified description for gravitational\footnote{Let us mention that even though string theory provides a valid description of gravitational interactions up to arbitrary energies, in contrast to general relativity, it treats the metric (at least in its current formulation) only perturbatively, as fluctuations around a fixed background, instead of as a dynamical object.}  and gauge interactions. 
Moreover, in order to also have fermionic degrees of freedom, we impose the principle of supersymmetry and obtain a supersymmetric spectrum. At length scales well above the cutoff scale $\alpha'$, the strings appear to be pointlike and the low energy physics is well approximated by only the massless excitations. Also, at such scales, string diagrams reduce to (sums of) ordinary Feynmann diagrams for particles.  The effective theory provided by the massless modes correspond to supersymmetric gravity and gauge theories (i.e.~supergravities and super Yang-Mills (SYM) theories).      

In this perturbative picture, one finds five consistent string theories, all of which are formulated in ten-dimensional spacetime; Type IIA and IIB, with thirtytwo supercharges, and type I, heterotic $SO(32)$ and heterotic $E_8\times E_8$, with sixteen supercharges. However, there is overwhelming evidence  that there exists a unique underlying quantum theory, called M-theory \cite{Witten:1995ex} (see also \cite{Hull:1994ys}). Not much is known about the microscopic structure of M-theory but it is believed to admit an eleven dimensional description which reduces, at low energies, to eleven dimensional supergravity with thirtytwo supercharges. Moreover, the five string theories can be obtained from M-theory in different limits. Since many of the features of eleven dimensional supergravity that we will discuss are also believed to be features of the yet-unknown eleven dimensional quantum theory, we will also oftentimes refer to it as M-theory. One of the reasons why M-theory is difficult to analyze is because it does not admit a perturbative description, since it does not contain any (small) parameters or scalar fields. On the other hand, this is very appealing since we argued in the previous section that a fundamental theory of nature should contain as few parameters as possible. 

We can relate M-theory to the five different string theories by means of compactification, namely by formulating the theory on a product space where one part consists of the external spacetime and the other of a compact internal space. For example, we can consider M-theory compactified on a circle down to ten dimensions. Since this compactification does not break any supersymmetry it can only be related to one of the two type II string theories. It turns out that this compactification is equivalent to the ten-dimensional type IIA string theory with the string coupling constant related to the radius of the M-theory circle \cite{Witten:1995ex}. In particular, the strong coupling limit of the IIA theory corresponds to the large radius limit, implying that this extra dimension is not seen in perturbation theory where we assume the string coupling to be small. 

Moreover, there is a perturbative symmetry (T-duality) between the type II theories, stating that they are equivalent upon circle compactification and inversion of the two radii. This nine-dimensional equivalence suggests that the type IIB must be obtainable directly from M-theory. It turns out that M-theory compactified on a two-dimensional torus is equivalent to IIB string theory compactified on a circle, with the (complexified) string coupling constant identified with the complex structure of the M-theory torus. This torus has an $SL(2,\mathbb{Z})$ invariance group which in the type IIB picture becomes a symmetry that involves the string coupling constant. This symmetry is called S-duality and it states that the strong coupling limit of the type IIB theory is related to the weak coupling limit of the same theory via an $SL(2,\mathbb{Z})$ transformation. 

Consider instead M-theory compactified on a circle modded out by a reflection under a diameter, i.e. an $S^1/\mathbb{Z}_2$-interval. This compactification breaks half of the supercharges and thus can only be related to one of the string theories with sixteen supercharges. It turns out that, at the two endpoints of the $S^1/\mathbb{Z}_2$-interval, we find two ten-dimensional hyperplanes with an $E_8$ vector multiplet associated to each one \cite{Horava:1995qa}. This suggests that this compactification is equivalent to the heterotic $E_8\times E_8$ theory with the string coupling constant related to the size of the interval and the strong coupling limit corresponding to the large interval limit. Finally, the heterotic $E_8\times E_8$ theory is T-dual to the heterotic $SO(32)$ theory, and the latter admits a strong coupling description in terms of the type I theory at weak coupling and vice versa \cite{Witten:1995ex,Polchinski:1995df}. 

By using this web of dualities, which becomes more complex in lower dimensions, we can use perturbative string theory to gain knowledge about M-theory. However, in order to learn about the most fundamental properties of M-theory it is plausible that a more complete formulation of string theory, beyond the perturbative one, is required. In an attempt to take a step in this direction, in this thesis we will study how certain non-pertubative effects arise in string theory. Since such effects are by definition not captured by perturbation theory we need other methods to treat them, some of which concern the dualities described above.

\section{D-branes}

From the discussion in the previous section it is clear that non-perturbative states are important in the context of dualities. For example, mapping a weakly coupled type IIB string theory to a dual strongly coupled type IIB string theory  implies a mapping of the perturbative states, relevant for the former, to some non-perturbative states, relevant for the latter. This suggests that there exist states in the theory which are heavy at weak coupling but become light at strong coupling.  

A key role in the analysis of non-perturbative states in string theory is played by D$p$-branes \cite{Polchinski:1995mt}. On one hand, D$p$-branes can be viewed as solitonic solutions with $p$ extended spatial dimensions in string theories at low energies, and on the other hand, they can be viewed as hypersurfaces upon which the endpoints of open strings are attached. The interactions of the massless open string modes and the couplings to the massless closed string modes can be summarized in the following (bosonic part of the) effective world volume action for a D$p$-brane, at leading order in the string coupling,
\begin{eqnarray}
\label{Dp}
S_{Dp} & = & \frac{1}{(2\pi)^p (\alpha')^{\frac{p+1}{2}} }\int_{\mathcal{M}_{p+1}}
\left[ e^{-\phi} \sqrt{\det (g+B+2\pi\alpha' F)} - i\, \mathcal{C} \, e^{B+2\pi\alpha' F} \right] \nn \\
\end{eqnarray}
where $\phi$ is the dilaton, $\mathcal{M}_{p+1}$ denotes the $(p+1)$-dimensional  world volume, $g$ is the metric induced on the D$p$-brane, $B$ is the Neveu-Schwarz Neveu-Schwarz (NS-NS) 2-form potential pulled back to the world volume, $F$ is the world volume gauge field strength and $\mathcal{C} =C_{p+1}+C_{p-1}+C_{p-3}+\cdots$ denotes a formal sum of Ramond-Ramond (R-R) form potentials. For later convenience, we have written (\ref{Dp}) in euclidean signature. Also, we have also suppressed the A-roof genus since it will not be relevant for our purposes. 

The first part of (\ref{Dp}) is the so-called Dirac-Born-Infeld (DBI) action which, at low energies (i.e.~when $\alpha'$-corrections can be neglected), reduces to the ordinary kinetic terms for a $(p+1)$-dimensional abelian gauge theory. The second part of (\ref{Dp}) is the topological (i.e.~metric-independent) Wess-Zumino (WZ) action. The string coupling constant is given by $g_s=\left<e^\phi\right>$ and hence, the tension of the D$p$-brane is of order $1/g_s$, implying that these states are very heavy at weak coupling, as is characteristic for solitons.

By considering a stack of $N$ coincident D$p$-branes in flat space, and taking the fermionic degrees of freedom into account, one finds that the low energy limit of the D$p$-brane world volume action is given by a $(p+1)$-dimensional maximally supersymmetric $U(N$) Yang-Mills theory. The maximal supersymmetry implies the existence of sixteen unbroken supercharges and reflects the fact that D-branes are BPS objects that break one half of the 32 bulk supercharges of the type II theories. In this way, D-branes provide a geometric understanding of non-abelian gauge theories. 

In configurations where two stacks of D-branes are intersecting, chiral matter can arise at the intersection point, from the open strings that stretch between the two stacks. Furthermore, since such stacks can intersect multiple times, the chiral matter can have multiple copies, hence providing a geometric understanding of how  multiple generations of chiral matter can arise \cite{Berkooz:1996km,Blumenhagen:2000wh,Aldazabal:2000dg}. 

In (\ref{Dp}) it is seen that a D$p$-brane couples minimally to the R-R $C_{p+1}$-potential. If the dimensions transverse to the D$p$-branes are compact (or if there are none, as in the case for $p=9$), Gauss law requires the R-R charge to be cancelled. Such a cancellation can be achieved by introducing orientifold planes (O$p$-planes) in the background \cite{Sagnotti:1987tw,Gimon:1996rq}. Orientifolds  are non-dynamical hypersurfaces that flip the orientation of the strings and may carry negative R-R charge. 

By engineering configurations with intersecting D-branes and orientifolds in an appropriate way, it is possible to find models that  share many of the features of the MSSM, see e.g. \cite{Cvetic:2001nr}.  One problem that arises in several semi-realistic D-brane models is that certain desirable couplings, such as mass terms and Yukawa couplings, are not generated in the perturbative analysis. As will be discussed in this thesis, such couplings can however be generated by non-perturbative effects.

We have discussed the fact that D-branes admit descriptions both in terms of gauge degrees of freedom, arising from the open string sector, and in terms of gravitational degrees of freedom, arising from the closed string sector. By studying these two descriptions in the case of D3-branes, in a decoupling limit, Maldacena \cite{Maldacena:1997re} was led to the conjecture that  $\mathcal{N}=4$ conformal SYM theory in four dimensions is dual to type IIB string theory compactified on $AdS_5 \times S_5$.  In a similar spirit as discussed above, where a strongly coupled string theory  admitted a dual weakly coupled description, this conjecture  suggests that a (otherwise elusive) strongly coupled gauge theory is equivalent to a tractable weakly coupled higher dimensional gravitational theory.  

As a step towards the ambitious goal of finding a gravitational theory dual to the theory of strong interactions, quantum chromodynamics (QCD), much work has been done to find gauge/gravity pairs with less supersymmetry and without conformal symmetry. As will be discussed in this thesis, one way to reduce the amount if supersymmetry is to choose a singular background geometry instead of flat space and place a stack of $N$ D3-branes at the tip of the singularity. For instance, when the background has a conifold singularity, it is shown in \cite{Klebanov:1998hh} that the corresponding IIB supergravity solution is dual to an $\mathcal{N}=1$ $U(N)\times U(N)$ conformal SYM theory with bi-fundamental matter. As will also be discussed in this thesis, one way to break the conformal symmetry is to also place fractional D3-branes at the singularity, corresponding to wrapping D5-branes on a two-cycle with vanishing geometric volume in the singular limit. By placing $M$ fractional D3-branes at the conifold singularity the gauge group becomes $U(N+M)\times U(N)$, the two gauge couplings start to run and the dual gravity solution is modified \cite{Klebanov:2000nc}. The study of this renormalization group flow towards low energies from the dual gravitational perspective provides a geometric understanding of phenomena such as chiral symmetry breaking and confinement \cite{Klebanov:2000hb}. Similar backgrounds have also been used to study  dynamical supersymmetry breaking \cite{Berenstein:2005xa,Franco:2005zu,Bertolini:2005di}.\footnote{\AFPold~was devoted to constructing a method to search for a normalizable fermionic zero mode in similar backgrounds which has the interpretation of a goldstino mode in the dual gauge theory where supersymmetry has been spontaneously broken.}   

Another gauge/gravity pair, which has attracted interest recently, concerns three-dimensional $\mathcal{N}=8$ superconformal field theories which have been conjectured to live on the world volume of a stack of M2-branes. M2-branes are stable extended objects in M-theory 
which, upon circle compactification, correspond to either fundamental strings or D2-branes depending on whether the M2-branes are wrapped or not on the circle. Since M-theory is the strong coupling limit of type IIA, the world volume theory of unwrapped M2-branes is believed to correspond to the infrared fixed point to which the non-conformal $\mathcal{N}=8$ SYM theory living on D2-branes flows at strong coupling. In \cite{Bagger:2007jr,Gustavsson:2007vu}, an $\mathcal{N}=8$ superconformal Chern-Simons theory coupled to matter with gauge group $SO(4)$ was constructed.\footnote{In order to generalize this theory, with the motivation to describe gauge groups of arbitrary rank and hence an arbitrary number of M2-branes, it was proposed in \GNP~(see also  \cite{Gomis:2008uv,Benvenuti:2008bt}) to relax the assumption of the existence of a positive definite metric on the gauge algebra.} By relaxing the assumption of maximal supersymmetry, the authors of \cite{Aharony:2008ug} constructed a $U(N)\times U(N)$ superconformal Chern-Simons matter theory with Chern-Simons levels $k$ and $-k$, believed to describe a stack of $N$ M2-branes at an $\mathbb{C}^4/\mathbb{Z}_k$ orbifold singularity. This theory was conjectured to be dual to M-theory on $AdS_4\times S^7/\mathbb{Z}_k$ and moreover, that the $\mathcal{N}=6$ supersymmetry preserved for $k>2$ is enhanced to $\mathcal{N}=8$ for $k=1,2$.

\section{D-instantons}

The generic way to compute scattering amplitudes in string theory is to assume weak coupling and perform a perturbative expansion in powers of $g_s$, as was discussed above. At each order in the $g_s$-expansion there is an additional perturbative expansion in terms of powers of $\alpha'$. Note that since $\alpha'$ is dimensionful, it only make sense to talk about an expansion where $\alpha'$ is accompanied by appropriate powers of momenta which, from the effective spacetime action point of view, corresponds to an expansion in derivatives. In \cite{Shenker:1990uf} it was suggested that at a fixed order in $\alpha'$ the closed string perturbation series at large genus diverges and that the remedy is to complete the series by adding terms of the order $e^{-1/g_s}$. By Taylor expanding $e^{-1/g_s}$ in terms of a small $g_s$, we see that the expansion coefficients vanish. Hence, terms of the order $e^{-1/g_s}$ can not arise perturbatively, they are non-perturbative.  
Note that in certain  backgrounds, one also expects effects of the order $e^{-1/\alpha'}$, corresponding to worldsheet instantons, arising from euclidean fundamental strings wrapping two-cycles \cite{Dine:1986zy,Dine:1987bq}.     

In \cite{Polchinski:1994fq}, it was realized that scattering amplitudes in the presence of open strings with Dirichlet boundary conditions in all directions can be of the order $e^{-1/g_s}$. The factor $-1/g_s$ arises as the amplitude for a vacuum disk, corresponding to a tree level open string worldsheet  satisfying Dirichlet boundary conditions in all spacetime directions. Hence, the point-like object to which the open string is attached is completely localized in spacetime. The scattering amplitude can be multiplied by any number $n$ of such disks, accompanied by a symmetry factor $1/n!$, and the exponentiation arises from the sum of all such contributions. 

After the discovery of D-branes as solitons charged under R-R fields \cite{Polchinski:1995mt}, the localized object responsible for the generation of such non-perturbative terms was identified with a D($-1$)-instanton, which is a solutions to euclidean type IIB supergravity \cite{Gibbons:1995vg}, and has an action given by (\ref{Dp}) for $p=-1$,
 \begin{eqnarray}
\label{D-1}
S_{D(-1)} & = & -2\pi i \tau 
\end{eqnarray}
where
\begin{equation}
\label{ }
\tau=C_0 +i \, e^{-\phi}\,.
\end{equation}

D($-1$)-instantons have been found to play a crucial role in connection with dualities. One example was provided in \cite{Green:1997tv} where the quantum corrections to the quartic Riemann curvature  term ($R^4$) in ten-dimensional type IIB supergravity was studied. This $R^4$ term is an example of a higher derivative correction (i.e.~$\alpha'$-correction) to the low energy supergravity action and it is known to receive a perturbative one loop correction (in $g_s$) \cite{Green:1981ya}  to the tree level term \cite{Gross:1986iv}. In \cite{Green:1997tv}, a non-perturbative correction was found to arise from a single D($-1$)-instanton effect. Moreover, in order for the complete quantum corrected $R^4$ term to respect the type IIB S-duality group $SL(2,\mathbb{Z})$, it was conjectured that there should exist an infinite sum of non-perturbative corrections arising from D($-1$)-instantons.  

In the case where the background geometry admits non-trivial cycles, D-instanton effects can arise from any euclidean D-brane whose entire world volume wraps the non-trivial cycles (ED-brane) \cite{Becker:1995kb,Harvey:1996ir}.  Such objects are nevertheless completely localized in the external spacetime. In this thesis, since we will  mainly be interested in D-instanton corrections to four dimensional  gauge theories, let us realize a gauge sector by considering a spacefilling D$p$-brane wrapped on a non-trivial $(p-3)$-cycle $\Sigma_{(p-3)}$, i.e. consider (\ref{Dp}) with world volume $\mathcal{M}_{p+1}=\mathbb{R}^4\times \Sigma_{(p-3)}$. Such an action would contain a term from the second WZ-part of (\ref{Dp}) with the structure,
\begin{equation}
\label{FwF}
\int_{\mathbb{R}^4\times \Sigma_{(p-3)} } C_{p-3}\wedge F \wedge F = \int_{ \Sigma_{(p-3)} } C_{p-3} ~ \int_{\mathbb{R}^4} F \wedge F\, .
\end{equation}
 In the non-abelian case, arising from multiple branes, we recognize  (\ref{FwF}) as the $\theta$-parameter times the instanton number. In YM theory, an instanton is a topologically non-trivial solution to the euclidean equations of motion. The instanton action is finite and characterized by an integer, the instanton number, that is (up to a coefficient) given by the four-dimensional part of (\ref{FwF}) \cite{Belavin:1975fg}. The $\theta$-parameter in  (\ref{FwF}) is given by the R-R potential $C_{p-3}$ integrated over the cycle $\Sigma_{(p-3)}$. Such an R-R potential couples minimally to an ED($p-4$)-brane wrapped on $\Sigma_{(p-3)}$. Thus, (\ref{FwF}) implies that if the gauge configuration on the D$p$-branes has non-vanishing instanton number, (\ref{FwF})  provides a source term for a non-vanishing number of ED($p-4$)-branes wrapped on the $\Sigma_{(p-3)}$-cycle  \cite{Witten:1995gx,Douglas:1995bn}. 

In particular, by also expanding the first part of (\ref{Dp}) to quadratic order in $F$ we obtain the usual (euclidean) gauge field action. By combining this action with (\ref{FwF}) it is possible to complexify the four dimensional gauge coupling such that it has the following structure \cite{Billo':2008pg},
\begin{eqnarray}
\label{tauDp}
\tau & = & \frac{1}{(2\pi)^{p-3} (\alpha')^{\frac{p-3}{2}} }\int_{ \Sigma_{(p-3)} }
\left[   i\, e^{-\phi} \sqrt{\det (g)} +C_{p-3} \right] \, 
\end{eqnarray}
where we have set $B=0$. On the other hand, the action for an ED($p-4$)-brane wrapping the $\Sigma_{(p-3)}$-cycle can be obtained directly from (\ref{Dp}) by replacing $p$ for $(p-4)$, 
 \begin{eqnarray}
\label{EDp}
S_{ED(p-4)} & = & \frac{1}{(2\pi)^{p-4} (\alpha')^{\frac{p-3}{2}} }\int_{ \Sigma_{(p-3)} }
\left[ e^{-\phi} \sqrt{\det (g)} - i\, C_{p-3} \right] 
\end{eqnarray}
 where we have set $F=B=0$. By comparing (\ref{EDp}) to (\ref{tauDp}) we see that the action for the wrapped ED($p-4$)-brane is given in terms of the complexified gauge coupling of the 
$(p+1)$-dimensional gauge theory living on the D$p$-brane world volume, $S_{ED(p-4)}=-2\pi i \tau$. This is not surprising since both the gauge coupling on the D$p$-brane and the ED($p-4$)-brane action in this case depend on the volume of the same cycle. Also, this is completely analogous to a YM instanton which has an action given by,
\begin{equation}
\label{ }
S_{\mathrm{YM\,inst}}=\frac{8\pi^2}{g_{\mathrm{YM}}^2}-i\theta_{\mathrm{YM}}=-2\pi i \tau_{\mathrm{YM}}~
\end{equation}
where $\tau_{\mathrm{YM}}$ is the complexified YM coupling constant,
\begin{equation}
\label{ }
\tau_{\mathrm{YM}}=\frac{4\pi i}{g_{\mathrm{YM}}^2} + \frac{\theta_{\mathrm{YM}}}{2\pi}~
\end{equation}

As discussed above, introducing D-branes in flat space implies that the background geometry becomes curved. In other words, D-branes act as sources for closed string fields such as the graviton. From the worldsheet perspective this implies that the graviton has a tadpole on a disk with boundary along the D-branes.  In fact, the large distance expansion of the classical supergravity D-brane solution can be obtained by multiplying this massless tadpole with a free graviton propagator and taking the Fourier transform \cite{DiVecchia:1997pr}. Hence, due to the presence of D-branes the graviton acquires a non-trivial profile. For example, the D($-1$)-instanton generated non-perturbative correction to the $R^4$ term was obtained in \cite{Green:1997tv}  by considering graviton tadpoles on disks with boundaries along the D($-1$)-instanton.  

We will now consider the open string realization of these ideas.  A finite instanton solution in ordinary YM theories corresponds to having a gauge field with a non-trivial profile that approaches pure gauge at infinity \cite{Belavin:1975fg}. For a stack of D$3$-branes we have, to lowest order in $g_s$, only open string worldsheets corresponding to disks with boundaries along the D$3$-branes. On these disks, the gauge field does not have a tadpole and hence no non-trivial profile. The gauge theory, in the low energy limit, can therefore be viewed as being in the trivial vacuum (with instanton number zero). However, by introducing D($-1$)-instantons in the world volume of the D$3$-branes, new disks with boundaries either completely or partially along the D($-1$)-instantons arise. They correspond to open strings with either both endpoints on the D($-1$)-branes or with one endpoint on the D($-1$)-instantons and one on the D$3$-branes. In \cite{Billo:2002hm}, it was shown that the gauge field in the world volume theory of the D$3$-branes has a tadpole on the latter type of disk, with mixed boundary conditions. Also, it was shown that the large distance expansion of the ordinary YM instanton profile \cite{Belavin:1975fg} is obtained by multiplying this tadpole with a massless gauge field propagator and taking the Fourier transform. Thus, when D($-1$)-instantons are present in the world volume of D3-branes, the four-dimensional gauge theory can be viewed as being in a non-trivial vacuum.

The massless states of the open strings that have at least one endpoint on one of the D($-1$)-instantons carry no momentum since at least one of their endpoints do not have any longitudinal Neumann direction. Instead, these modes correspond to non-dynamical moduli fields on which the D($-1$)-instanton background depends. The interactions among the massless moduli modes of open strings in a configuration of $k$ D($-1)$-instantons in the world volume of $N$ D3-branes reproduce the (supersymmetric version of the) moduli space for the $k$-instanton sector in $\mathcal{N}=4$ $U(N)$ SYM constructed by Atiyah, Hitchin, Drinfeld and Manin (ADHM) \cite{Atiyah:1978ri}. Hence, string theory  provides a physical and geometric understanding of the ADHM-construction \cite{Witten:1995gx,Douglas:1995bn}. In this thesis, in addition to discussing how this construction arises in string theory in the $\mathcal{N}=4$ case, we will study D-instanton effects in $\mathcal{N}=1$ gauge theories by placing the D3/D($-1$) system at a singular point in the background geometry.

Even though the analogy between wrapped ED-branes and ordinary YM instantons is striking, there are subtle differences. While a YM instanton requires a non-trivial gauge theory whose fields provide the necessary background, an ED-brane is a geometrical object that exists independently of the spacefilling D-branes giving rise to the gauge theory. This fact opens up the possibility of considering an ED-brane wrapped on a cycle that is not populated by any spacefilling D-brane \cite{Witten:1996bn}. This corresponds to the case when the cycle $\Sigma_{(p-3)}$ in (\ref{EDp}) is not the same cycle as the one in (\ref{tauDp}). Hence, in this case the instanton action for the ED-brane is not tied to the gauge coupling of the gauge theory and moreover, such an ED-brane can no longer be interpreted as a YM instanton from the point of view of the gauge theory on the spacefilling D-branes. In this thesis we will discuss under which circumstances such an ``exotic" or ``stringy'' instanton configuration gives rise to new terms in the effective action. Such terms can be of great relevance for MSSM/GUT phenomenology, moduli stabilization and supersymmetry breaking  \cite{Blumenhagen:2006xt}~--\cite{Cvetic:2007sj}. 
It turns out that several of these couplings are perturbatively  forbidden. 
Thus,  even though instanton effects are in general highly supressed at weak coupling, they will in this case be of leading order.

Another setting in which instantons can provide leading order effects concerns string compactifications. 
One strategy to make contact with the four dimensional universe we observe is to compactify six of the spatial dimensions on a compact manifold. This implies that the four dimensional physics will  depend on the choice of manifold.  One problem that immediately arises in this process is that the possible deformations (which cost no energy) of the shapes and sizes of the cycles in the compact manifold appear in the effective four dimensional action as massless scalar moduli fields, which are in conflict with experimental observations. 

One way to lift this vacuum degeneracy is to turn on vacuum expectation values for the R-R and NS-NS  field strengths along the cycles. Since these cycles can no longer be deformed without a cost in energy, a four dimensional potential for the moduli fields is induced, rendering the scalars massive \cite{Gukov:1999ya}. In the most well-studied example, where the type IIB theory is compactified on a Calabi-Yau manifold with orientifold planes, the presence of three-form fluxes induces a potential for the dilaton and the complex structure moduli but not for the K$\ddot{\mathrm{a}}$hler moduli \cite{Dasgupta:1999ss,Giddings:2001yu}. In order to lift these remaining flat directions, non-perturbative effects are required   \cite{Kachru:2003aw}, e.g. D-instantons. 

In fact, D-instantons are generically present in string compactifications and D-brane models and in certain configurations they improve the model while in others they might spoil it. In either case, they are there and one should take them into account in order to have control over the four dimensional effective theory.  

For many amplitudes in string theory, one expects D-instanton contributions from all instanton numbers.  Therefore, in order to obtain exact quantum corrected results  and understand the underlying symmetries, one is required to consider infinite series of instanton corrections, as in the case of the $R^4$ term discussed above. 
A generic problem with explicit multi-instanton computations is however that they become  technically very challenging for high instanton numbers  and one is oftentimes forced to use other methods in order to  compute them. One such method involves making use of string dualities. 

In QFT it is well known that a soliton in $d+1$ dimensions with finite energy corresponds to an instanton in $d$ dimensions with finite action \cite{Polyakov:1976fu}. A string theory analogue of this is provided by starting from one of the type II  theories compactified down to $(d+1)$ spacetime dimensions. 
Consider a solitonic D$p$-brane which has all its spatial directions wrapped on a non-trivial $p$-cycle $\Sigma_p$ of the background geometry, i.e. (\ref{Dp}) with world volume $\mathcal{M}_{p+1}=\mathbb{R}\times \Sigma_p$ where $\mathbb{R}$ denotes the time direction. Such a zero-dimensional D-particle corresponds to a black hole 
in the $(d+1)$-dimensional spacetime. If we further compactify time on a circle down to $d$
dimensions, it is possible for the one-dimensional (euclidean) worldline of the D-particle to wrap around the circle. By performing a T-duality along the circle direction one obtains an ED$(p-1)$-brane wrapped on $\Sigma_p$, corresponding to a D-instanton in the dual $d$-dimensional theory.

This suggests that we can start from one side and perform a one-loop calculation with circulating D-particles with arbitrary R-R charge and Kaluza-Klein (KK) momenta along the circle. By performing a T-duality in the circle direction the contribution from the D-particles turns into contributions of D-instantons for all instanton numbers \cite{Ooguri:1996me}. For example, one way to obtain the infinite sum of D($-1$)-instanton corrections to the $R^4$ term in type IIB mentioned above is to start from type IIA compactified on a circle and perform a one-loop calculation with circulating D0-particles. Upon T-duality along the circle, the contribution from the D0-particles turns into the infinite series of D($-1$)-instanton corrections conjectured in \cite{Green:1997tv}, in agreement with $SL(2,\mathbb{Z})$ invariance.

Alternatively, we can make use of the fact that a D0-particle of arbitrary R-R charge in type IIA corresponds to an eleven dimensional graviton with arbitrary KK momenta along the M-theory circle \cite{Witten:1995ex}. Hence, one may compactify eleven dimensional supergravity on a two-torus and compute a one-loop amplitude with circulating gravitons with arbitrary KK momenta along the torus directions \cite{Green:1997as}.  By shrinking the volume of the two-torus while keeping the complex structure fixed, the amplitude  reproduces the infinite series of D($-1$)-instantons in type IIB. In this thesis we will apply these ideas to  backgrounds that involve spacefilling D-branes and orientifolds as well.

\newpage
\section{Outline}

In chapter 2 we discuss the general treatment of D-instantons in gauge theories realized in string theory. We focus on D$(-1)$-instantons located in the world volume of D3-branes in flat space. By analyzing the instanton zero mode structure arising from the massless modes of open strings we recover the construction of gauge instantons in QFT. By translating ideas from instanton calculus in QFT, we develop methods to compute D-instanton effects in string theory.  

Chapter 3 is devoted to D-instanton effects in four-dimensional $\mathcal{N}=1$ gauge theories. Such theories are realized by placing the D3/D$(-1)$ system at an orbifold singularity. We define the moduli space integral from which non-perturbative superpotentials can be calculated. We discuss those instantons which admit a direct interpretation in terms of ordinary gauge instantons and those that  do not. The latter type, called stringy instantons, generically carries unwanted fermionic zero modes which obstruct their contribution to the effective superpotential. This provides the setting for \ABFLP, \Petersson,  \FP~and \PeterssonN~where we  analyze the  circumstances under which these fermionic zero modes are either projected out by an orientifold or lifted by introducing an additional D3-brane or background fluxes.  
 
 Chapter 4 generalizes the orbifold setup and discusses  D-instanton configurations in gauge theories arising from D-branes probing generic toric singularities. We provide the general recipe for constructing multi-instanton moduli spaces in any toric gauge theories. In \AFP~we apply this recipe to a wide class of gauge theories and compute novel stringy instanton effects.

In chapter 5 we discuss methods based on string theory dualities that allow us to compute infinite sums of D-instanton corrections. In particular, we discuss how the perturbative and non-perturbative corrections to type IIB gauge couplings in eight dimensions can be obtained by computing one-loop amplitudes with BPS particles in the dual type IIA  theory as well as in eleven-dimensional supergravity. We also discuss how these ideas suggest a resolution to a puzzle concerning so-called polyinstantons. This chapter sets the stage for \PSU, in which we apply these methods and obtain complete quantum corrections to gauge and gravitational couplings in eight- and four-dimensional gauge theories.

\chapter{D-instanton Calculus}

In  this chapter we start by analyzing the massless open string spectrum for a type IIB system with $N$ D3-branes and $k$ D($-1$)instantons in flat ten-dimensional euclidean spacetime \cite{Green:2000ke,Dorey:2002ik,Billo:2002hm}. The interactions of these massless modes and their relation to the ADHM construction is described. In the last section we discuss the structure of the moduli space integral which is the main tool to compute D-instanton corrections to the effective four dimensional action. 

\section{The D3/D($\mathbf{-1}$) System}

In a D3/D($-1$) system there are three different types of open strings. In this section we will review the massless spectrum for these sectors. The massless modes of the open strings with at least one endpoint on a D($-1$)-instanton do not carry any momentum and are therefore  instanton moduli rather than dynamical spacetime fields.  

\subsubsection{The Gauge Sector}

The gauge sector consists of the massless modes of the open strings with both endpoints attached to  the D3-branes. The presence of the D3 branes breaks the Wick rotated Lorentz group $SO(10)$ to $SO(4)\times SO(6)$.

In the bosonic sector, we obtain a gauge field, $A^{\mu}$ ($\mu=0,1,2,3$), with indices along the  world volume of the D3-branes, and six real scalars $X^a$ ($a=4,\cdots,9$) with indices along the transverse directions. For later convenience we will complexify these scalar and instead consider 
$\Phi^i$ ($i=1,2,3$), now with indices along the three complex transverse directions. This can be done by writing the 6 scalars $X^a$ first as an $SU(4)\cong SO(6)$ antisymmetric matrix $X_{AB}=(\Sigma^a)_{AB}X_a$, where $(\Sigma^a)_{AB}$ is the chiral off-diagonal block in the six-dimensional gamma-matrices, and then identifying $\Phi^i=\frac{1}{2}\epsilon^{ijk} X_{jk}$ and $\Phi^\dagger_i=X_{i4}$.

In the fermionic sector we get the gauginos, $\Lambda^{\alpha A}$ and $\bar\Lambda_{\dot\alpha A}$ (the $\alpha$ and $\dot\alpha$ indices take two values while $A=1,2,3,4$), where $\alpha$/$\dot\alpha$ denote $SO(4)$ Weyl spinor indices of positive/negative chirality transforming in the fundamental representation under the respective factor of $SU(2)_L\times SU(2)_R \cong SO(4)$ of the Lorentz group. The index $A$ upstairs/downstairs denote $SO(6)$ Weyl spinor indices of negative/positive chirality which transform in the fundamental/anti-fundamental representation of the transverse rotation group $SU(4)\cong SO(6)$. We have chosen the ten dimensional chirality of the fermionic fields to be negative. 

Since these open strings have the possibility to  begin and end on one of the $N$ D3-branes, the Chan-Paton factors of the gauge sector open string states are $N\times N$ matrices corresponding to the adjoint representation of $U(N$). The massless modes in the gauge sector form a four-dimensional $U(N$) $\mathcal{N}$=4 SYM multiplet \cite{Witten:1995im}. 
In $\mathcal{N}$=1 language, this corresponds to a vector multiplet, consisting of the gauge field and one gaugino (e.g. the $A=4$ component) and three chiral superfields, consisting of the three complex scalars and the remaining three gauginos, i.e.~the $A=1,2,3$ components.

\subsubsection{The Neutral Sector}

The fields in the neutral sector correspond to the massless modes of the open strings with both ends attached to the $k$ D(-1)-branes. These fields are neutral in the sense that they do not transform under the gauge group of the D3-branes. They do however transform in the adjoint representation of the auxiliary instanton gauge group $U(k$). In the same way as the four dimensional $\mathcal{N}$=4 SYM theory can be obtained from a dimensional reduction of the $\mathcal{N}$=1 SYM theory in ten dimensions \cite{Brink:1976bc}, 
the neutral sector can be obtained by continuing the reduction down to zero dimensions. 

The bosonic fields in this sector are denoted by $a_\mu$ and $\chi^a$, corresponding to the positions of the D$(-1)$-instantons the directions longitudinal and transverse, respectively, to the
D3-branes. We will oftentimes use the complexified version of $\chi^a$, denoted by $s^i$. The fermionic zero modes are $M^{\alpha A}$ and
$\lambda_{\dot\alpha A}$. For later convenience we  also introduce a real auxiliary field
 $D^c$, $c=1,2,3$ transforming in the adjoint representation of $SU(2)_R$. 
 All these
fields are $k \times k$ matrices and can be found in Table 2.1. 

\begin{table} [ht]
\begin{center}
\begin{tabular}{c||c|c|c|clclcl}
&$SU(2)_L$&$SU(2)_R$&$SU(4)$&$U(k)$&$U(N)$  \\ 
\hline   \hline  $a_\mu$  & ${\bf 2}$&${\bf 2}$&${\bf 1}$&${\bf adj}$&$~~{\bf 1}$ \\  
\hline $\chi^a$ & ${\bf 1}$&${\bf 1}$&${\bf 6}$&${\bf adj}$&$~~{\bf 1}$ \\ 
 \hline $D^c$ & ${\bf 1}$&${\bf 3}$&${\bf 1}$&${\bf adj}$&$~~{\bf 1}$ \\
 \hline $M^{\alpha A}$ & ${\bf 2}$&${\bf 1}$&${\bf 4}$&${\bf adj}$&$~~{\bf 1}$ \\
  \hline $\lambda_{\dot\alpha A}$ & ${\bf 1}$&${\bf 2}$&${\bf \ov{4}}$&${\bf adj}$&$~~{\bf 1}$ \\
  \end{tabular}
\caption{\small The neutral sector moduli.}
\end{center}
\end{table}

\subsubsection{The Charged Sector}

 The charged sector fields arise from the zero modes of the open strings
stretching between one of the $N$ D3 branes and one of the $k$ D(-1)-branes.  For each such open string we have two conjugate sectors distinguished by the orientation of the string.
In the bosonic sector we obtain a bosonic moduli field  $\omega_{\dot\alpha}$  with an $SO(4)$ Weyl spinor index which the GSO projection chooses to be of negative
chirality. In the conjugate sector, we get an independent bosonic field $\bar\omega_{\dot\alpha}$ with an index of the same chirality.  As will be discussed below, these charged bosonic moduli are related to the size of the instanton. 

In
the fermionic sector, we obtain two SO(6) Weyl spinors $\mu^A$ and $\bar\mu^A$, one for each conjugate sector, with chirality fixed by the GSO projection such that they both transform in the fundamental representation of $SU(4)\cong SO(6)$. Since these open strings stretch between one of the $N$ D3-branes and  one of the $k$ D($-1$)-instantons, the Chan-Paton factors for these fields are  $N\times k$ (or $k\times N$) matrices, see Table 2.2.

\begin{table} [ht]
\begin{center}
\begin{tabular}{c||c|c|c|clclcl}
&$SU(2)_L$&$SU(2)_R$&$SU(4)$&$U(k)$&$U(N)$  \\ 
\hline   \hline  $\omega_{\dot\alpha}$  & ${\bf 1}$&${\bf 2}$&${\bf 1}$&${\bf \ov{k}}$&$~~{\bf N}$ \\  
\hline $\bar\omega_{\dot\alpha}$ & ${\bf 1}$&${\bf 2}$&${\bf 1}$&${\bf k}$&$~~{\bf \ov{N}}$ \\ 
 \hline $\mu^A$ & ${\bf 1}$&${\bf 1}$&${\bf 4}$&${\bf \ov{k}}$&$~~{\bf N}$ \\
 \hline $\bar\mu^{A}$ & ${\bf 1}$&${\bf 1}$&${\bf 4}$&${\bf k}$&$~~{\bf \ov{N}}$ \\
  \end{tabular}
\caption{\small The charged sector moduli.}
\end{center}
\end{table}

\section{Recovering the ADHM Construction}

In this section we discuss the interactions for the moduli obtained in the previous section.
One way to recover these interactions  is to consider a D9/D5-system and  perform a dimensional reduction of the six-dimensional action down to zero dimensions  \cite{Dorey:2002ik}. Another way is to compute disk scattering amplitudes with open string moduli  inserted along the boundaries \cite{Billo:2002hm}. Either way, the interactions can at low energy be summarized by the following zero dimensional action,
\begin{equation}
S_{\mathrm{moduli}}^k={\mathrm{Tr}}_{k}\,\left[\frac{1}{2g_0^2}\,
S_{g_0} +S_{s}+S_{L.m.}\right]
\label{tot}
\end{equation}
where
\begin{eqnarray}
S_{g_0} & = &  \frac{1}{2} D_c^2 -
\frac{i}{2} \Big( \lambda_{\dot \alpha i} [s^{i},
\lambda^{\dot \alpha}_{4}]- \frac{1}{2}\epsilon^{ijk}\lambda_{\dot \alpha i}
[s^\dagger_{j}, \lambda^{\dot \alpha}_{k}] \Big) 
\nn \\
&&+ [ s^i, s^j] [ s^\dagger_j,s^\dagger_i] + \frac{1}{2}
[ s^i,s^\dagger_i][s^j ,  s^\dagger_j]  
\label{Sg}
\end{eqnarray}
\begin{eqnarray}
S_s & = & -[a_\mu,s^{\dagger}_{i}] [a^\mu,s^i ]-
\frac{i}{2} \Big( M^{\alpha i} [s^{\dagger}_{i}, M^4_\alpha]- \frac{1}{2}\epsilon_{ijk}M^{\alpha i} [s^{j}, M^k_\alpha] \Big) \nn \\
&&+\frac{1}{2}
s^i  \bar\omega_{\dot\alpha } \omega^{\dot\alpha } s^\dagger_i  +\frac{1}{2}  s^\dagger_i \bar\omega_{\dot\alpha }  \omega^{\dot\alpha } s^i  +\frac{i}{2} \bar\mu^i  \mu^4 s^\dagger_i
- \frac{i}{2}  \bar\mu^4    \mu^i s^\dagger_i
+\frac{i}{2} \epsilon_{ijk} \bar\mu^i   \mu^j  s^k \nn \\
\label{s}
\end{eqnarray}
\begin{eqnarray}
S_{L.m.} & = &  i \left(\bar\mu^i \omega_{\dot\alpha} +
\bar\omega_{\dot\alpha} \mu^i + \sigma^\mu_{\beta
\dot\alpha}{[M^{\beta i}, a_\mu]}\right)\! \lambda^{\dot\alpha}_i
\nn \\
&& + i \left(\bar\mu^4 \omega_{\dot\alpha} +
\bar\omega_{\dot\alpha} \mu^4 + \sigma^\mu_{\beta
\dot\alpha}{[M^{\beta 4}, a_\mu]}\right)\! \lambda^{\dot\alpha}_4 \nn \\
&&\phantom{\tr } - i D^c\!\left( \bar\omega^{\dot \alpha}
(\tau^c)^{\dot\beta}_{\dot\alpha} \omega_{\dot\beta} + i
\bar\eta^c_{\mu\nu}  {[a^\mu, a^\nu]}\right) 
\label{Lm}
\end{eqnarray}
where $\tau^c$ are the usual Pauli matrices, $\eta$ (and $\bar\eta$) the 't~Hooft
symbols and $\sigma$ (and $\bar\sigma$) are the chiral (and anti-chiral) off-diagonal blocks in the four-dimensional gamma-matrices, see \cite{Dorey:2002ik,Billo:2002hm} for details. For later convenience we have treated the $A=1,2,3$ components and the $A=4$ component of the fundamental $SU(4)$ indices separately.

In (\ref{tot}) $g_0$ is the zero-dimensional coupling constant, related to $\alpha'$ and $g_s$ according to $1/g^2_0\sim (\alpha')^{2}/g_s$. The dimensionless four-dimensional coupling constant $g_4$ is instead related to $g_s$ according to $1/g_4^2\sim 1/g_s$, implying that both $g_0$ and $g_4$ can not be held fixed simultaneously in the field theory limit $\alpha' \to 0$. Since we are interested in the four-dimensional gauge theory we define the field theory limit as $\alpha' \to 0$ while keeping $g_s$ fixed, implying that $g_0\to\infty$. 

If the moduli fields were to have canonical scaling dimensions, then there would appear an overall factor of $1/g_0^2$ in (\ref{tot}) and all interactions would vanish in the field theory limit. In order to avoid this and to get (\ref{tot}), we have assigned to the moduli fields the following scaling dimensions \cite{Billo:2002hm},  
\begin{eqnarray}
\label{dim}
&&[a^\mu ]  =  [\omega_{\dot{\alpha}} ]=[\overline{\omega}_{\dot{\alpha}} ]=M^{-1}_{s} ~,~ [D^c ]=M^{2}_{s}~,~[s^i]=M_{s} \nn \\
&&\big[ M^{\alpha A} \big]  =  [\mu^A ]=[\overline{\mu}^A]= M^{-1/2}_{s}  ~,~ [\lambda_{ \dot\alpha A} ]  =  M^{3/2}_{s}  ~.
\end{eqnarray}
where $M_s =1/\sqrt{\alpha'}$. For example, note that the neutral fermions $M^{\alpha A} $ and $\lambda_{ \dot\alpha A }$ do not have the same dimension. With this choice of scaling dimensions we get (\ref{tot}) in which  only (\ref{Sg}) vanishes in the field theory limit. Furthermore, in this limit  the neutral fields $D^c$ and $\lambda^{\dot\alpha}_A$ in (\ref{Lm}) become Lagrange multipliers. The algebraic equations $D^c$ and $\lambda^{\dot\alpha}_A$ impose when integrated over are precisely the bosonic and fermionic ADHM constraints. It is from the solutions of these constraints that the YM instantons with arbitrary instanton number can be constructed \cite{Atiyah:1978ri}. In fact, (\ref{tot})  provides the ADHM measure on the moduli space of the $k$ instanton sector of $\mathcal{N}=4$ SYM \cite{Dorey:2002ik}. 

In order to be more precise about the connection to the ADHM construction, we study the (bosonic) classical vacuum structure of (\ref{tot}). For the simple case with a single $k=1$ D($-1$)-instanton all commutators vanish and  we are left with the following equations, 
\begin{equation}
\label{vac}
\omega^{\dot\alpha } s^{i} =0~~~\mathrm{and}~~~\bar\omega^{\dot \alpha}
(\tau^c)^{\dot\beta}_{\dot\alpha} \omega_{\dot\beta}=0~.
\end{equation}
where the second set of equations are the ADHM constraints. From these equations we see that the classical vacuum consists of two distinct branches \cite{Witten:1995gx,Douglas:1995bn}:
 
The Coulomb branch corresponds to allowing for the $s^i$ moduli to have non-vanishing vevs while setting  $\omega^{\dot\alpha }=0 $. Since the $s^i$ moduli correspond to the position of the D($-1$)-instanton in the directions transverse to the D3-branes, this branch corresponds to the situation where the D($-1$)-instanton is located away from the D3-branes in the transverse space. Moreover, since the $\omega^{\dot\alpha } $ moduli are related to the size of the instanton, this branch describes a D($-1$)-instanton with zero size. Thus, a D($-1$)-instanton along the Coulomb branch does not correspond to an ordinary YM instanton.  

However, we can place the D$(-1)$-instanton on top of the D3-branes by setting $s^i=0$ and obtain the Higgs branch. From (\ref{vac}) we see that this choice allows us to give a non-vanishing vev to the $\omega^{\dot\alpha } $ moduli, provided they satisfies the ADHM constraint. The size of the instanton is  given by $\rho^2=\bar\omega_{\dot\alpha }\omega^{\dot\alpha }$, implying that the D($-1$)-instanton is allowed to have a non-vanishing size. This implies that a D$(-1)$-instanton along the  Higgs branch can be identified with an ordinary YM instanton. However, in the zero size limit, where the D$(-1)$-instanton is allowed to move away from the D3-branes, this is no longer true. 

Thus, string theory does not only provide a geometric realization of the ADHM construction but due to its higher dimensional structure it goes beyond, since the D$(-1)$-instanton is a well-defined object regardless of whether it is on top of the D3-branes or away from them. In other words, a D($-1$)-instanton exists even without the presence of a gauge theory, in contrast a YM instanton. This feature will be important in the following chapter.

\section{The Moduli Space Integral}

In analogy with the discussion in the introduction, in the presence of D($-1$)-instantons in the world volume of D3-branes, vacuum disks with boundary along the D($-1$)-instantons are present. Such disks should be taken into account when computing correlation functions with fields from the gauge sector. Even though some of the terms in (\ref{tot}) correspond to mixed disk diagrams where part of the boundary is along the D3-branes, from the point of view of the gauge theory living on the D3-branes, all terms in (\ref{tot}) are vacuum contributions since they do not involve any gauge sector field. It is therefore convenient to consider a more general ``vacuum" disk which represents a sum of disks where the first one is the usual vacuum D($-1$)-disk, given by $2\pi i \tau k$, and the others are the disks in (\ref{tot}). In fact, D($-1$)-instantons are taken into account by multiplying gauge sector correlation functions  by such a generalized vacuum disk. Moreover, the correlation functions can be multiplied by any number of such generalized vacuum disks and by summing up all possible contributions accompanied with a symmetry factor, the generalized disk exponentiates. This means that a correlation function evaluated in the $k$
D($-1$)-instanton sector should be weighted by a factor
\begin{eqnarray}
\label{weight}
e^{2\pi i \tau k-S_{\mathrm{moduli} }^k}
~~.
\end{eqnarray} 
Although all these contributions are disconnected from the worldsheet point of view, they are connected from the point of view of the four-dimensional theory since we will eventually integrate over all the instanton moduli. 

In this generalized vacuum disk, it is natural to include all open string worldsheets, for all genus with (or without) moduli field insertions, since they would also correspond to vacuum contributions from the gauge theory point of view. In the following chapter, since we will be interested in superpotential contributions, due to holomorphy, we only need to consider additional amplitudes of order $\mathcal{O}(g_s^0)$. 
Such additional amplitudes correspond to one-loop vacuum open string amplitudes with at least one boundary on a D($-1$)-instanton \cite{Blumenhagen:2006xt}. Moreover, in the following chapter, we will be interested in local configurations where the only contribution to such one-loop amplitudes arise from massless open string states, since  the contribution from the massive ones vanish \cite{DiVecchia:2005vm,Akerblom:2006hx}. However, since we will integrate out these massless D($-1$)-instanton moduli modes explicitly, to avoid double counting, we should not circulate the massless modes in the loop. Thus, instantonic open string amplitudes other than those we have already included will not be relevant for our purposes.

In QFT, YM instantons can give rise to non-perturbative corrections to the four dimensional action via their saddle point contribution to the euclidean path integral \cite{'tHooft:1976fv,'tHooft:1976up} (see also e.g. \cite{Dorey:2002ik,Vandoren:2008xg}). At weak coupling, the path integral can be expanded in a sum of terms, each of which being an integral over the instanton moduli space of the corresponding instanton sector. Even though we currently lack a second quantized formulation of string theory we can simply translate the  QFT treatment of YM instantons to string theory language. This leads us to define the moduli space integral in the $k$ D($-1$)-instanton sector as the integral of (\ref{weight}),
\begin{eqnarray}
\label{SW}
Z_k &=&M_s^{bk} \int d\{ a,M,\lambda, s,D, \omega,\ov\omega, \mu,\ov\mu  \}_k \,\,e^{2\pi i \tau k-S_{\mathrm{moduli}}^k } \nn \\
&=&M_s^{bk} \,e^{2\pi i \tau k} \int d\mathcal{M}_k \,\, e^{-S_{\mathrm{moduli}}^k }
~~
\end{eqnarray} 
where we have pulled out the contribution from the vacuum D($-1$)-instanton disk since it does not depend on the moduli. The prefactor in (\ref{SW}) is introduced in order to compensate for the dimension of the measure. The factor $b$ in the power to which the prefactor is raised can in general be identified with the one loop beta-function coefficient of the corresponding gauge theory. Note however that in this $\mathcal{N}=4$ conformal case the measure is dimensionless and $b=0$, which can be shown using (\ref{dim}). The complete contribution is obtained by summing up (\ref{SW}) for all $k$. 

The zero modes $x^\mu=\Tr_k \, a^\mu$ and $\theta^{\alpha A} =\Tr_k \, M^{\alpha A}$ correspond to the center of mass motion of the D($-1$)-instantons and arise from the supertranslations broken by their presence.  Since they do not appear in (\ref{tot}) it is convenient to separate the integration over these modes from the others,
\begin{eqnarray}
\label{SW11}
Z_k &=& M_s^{bk} \,e^{2\pi i \tau k} \int d^4 x\, d^8 \theta\, \int \widehat{d\mathcal{M}} \,e^{-S_{\mathrm{moduli}}^k}
~~
\end{eqnarray}
where $\widehat{d\mathcal{M}}$ denotes the integration measure over the centered moduli space, which is the non-trivial part of the integral. From (\ref{SW11}) we see that $x^\mu$ and $\theta^{\alpha A} $  play the role of  superspace coordinates.  
In the next step, we will discuss how to include fields from the gauge sector of the D3-branes.

In QFT, a correlation function evaluated in the $k$ instanton sector involves insertions of fields expressed in terms of the classical profile they acquire in that instanton background. 
As was discussed in the introduction, the classical profile for  gauge sector fields in the world volume theory of the D3-branes can be obtained by considering mixed disks upon which the  fields have tadpoles. In \cite{Billo:2002hm} it is shown that the tadpole for a gauge field corresponds to a mixed disk with a gauge field inserted along the D3-boundary and two moduli fields $\omega$ and $ \bar\omega$ inserted at the two boundary changing points. The tadpole for a scalar field corresponds to a mixed disk with a $\mu$ and a $\bar\mu$ moduli insertion.  
Such tadpoles are one-point functions from the gauge theory point of view but three-point functions from the worldsheet point of view. Hence, a correlation function with gauge sector fields, evaluated in a certain D($-$1)-instanton sector, involves insertions of  tadpoles expressed as mixed disk amplitudes. 

One way to include the gauge sector fields is to simply add these mixed disks to (\ref{tot}). We will only consider mixed disk with insertions of scalar fields since, in the following chapter, we will be interested in superpotentials. In additions to the mixed disks with a single scalar field insertion discussed above, there also exist non-vanishing mixed disk amplitudes with two scalar fields together with $\omega$ and $\bar\omega$ insertions. The total set of terms that describe how the charged moduli couple to the gauge sector scalar fields is given by, 
\begin{eqnarray}
S_\Phi & = &  \frac{1}{2}
 \bar\omega_{\dot\alpha }\Big( \Phi^i  \Phi^\dagger_i  +  \Phi^\dagger_i  \Phi^i \Big) \omega^{\dot\alpha }  +\frac{i}{2} \bar\mu^i \Phi^\dagger_i \mu^4 
- \frac{i}{2} \bar\mu^4   \Phi^\dagger_i \mu^i 
-\frac{i}{2} \epsilon_{ijk} \bar\mu^i  \Phi^j \mu^k ~.\nn \\
\label{SPhi}
\end{eqnarray}    
By  splitting the components $A=1,2,3$ and $A=4$ of the fundamental $SU(4)$ indices on the $\mu$-fields, we see clearly in (\ref{SPhi}) that only the last coupling  depends holomorphically  on the scalars. This will be important when we project the theory to $\mathcal{N}=1$. Note that  the terms  in (\ref{SPhi}) are similar to the terms in the second row in (\ref{s}), which is not surprising since the $s^i$ moduli are the dimensional reduction of the $\Phi^i$ scalars. 

By adding (\ref{SPhi}) to (\ref{tot}) we get the following total moduli action, 
\begin{equation}
S_{\mathrm{D3/D(-1)}}^k=S_{\mathrm{moduli}}^k + {\mathrm{Tr}}_{k}\,\left[S_\Phi\right]\,.
\label{Tot}
\end{equation}
 Moreover, by including the gauge sector  scalars in this way, we generalize the moduli space integral in (\ref{SW11}) to the following one,
\begin{eqnarray}
\label{SW2}
Z_k &=& M_s^{bk} \,e^{2\pi i \tau k} \int d^4 x\, d^8 \theta\, \int \widehat{d\mathcal{M}}_k \,e^{-S_{\mathrm{D3/D(-1)}}^k }
~~
\end{eqnarray}
where the scalars $\Phi$ in $S_{\mathrm{D3/D(-1)}}^k$ can be promoted to superfields, having a $\theta$-expansion arising from additional mixed disks involving insertions of $\theta$-moduli at the D($-1$)-boundary and  superpartners of the scalar fields at the D3-boundary \cite{Green:2000ke}. Hence, the integrand of (\ref{SW2}) depends implicitly on the $x^\mu$ and $\theta^{\alpha A} $ modes.

In order to test (\ref{SW2}) one can realize gauge theories which correspond to QFT setups where ordinary gauge instanton effects are known to occur. In the following chapter we realize $\mathcal{N}=1$ SQCD-like theories in which a non-perturbative superpotential is known to arise from a gauge instanton effect. By reproducing this superpotential we gain some confidence in (\ref{SW2}) and in the remainder of the following chapter we apply it to configurations that are more exotic from the field theory point of view.

\chapter{Non-Perturbative Superpotentials}

In this chapter we will apply the D-instanton calculus discussed in the previous chapter to  $\mathcal{N}=1$ gauge theories. The way we realize the $\mathcal{N}=1$ gauge theories is to place D3-branes at a singular point in the background geometry which we choose to be a non-compact orbifold. This non-compact choice allows us to neglect global issues such as tadpole cancellation and it is motivated by our interest in the gauge sector rather than the gravitational sector. The fact that we choose an orbifold implies that we have simple worldsheet techniques at our disposal. Using this background will allow us to illustrate rather general arguments in an explicit way. 

We start by placing the D3/D($-1$) system described in the previous chapter at the orbifold singularity and discuss how the open string spectrum is affected. We obtain    $\mathcal{N}=1$ quiver gauge theories living on D3-branes as well as the corresponding D($-1$)-instanton sectors. With this input, we can write an $\mathcal{N}=1$ version of the moduli space integral in (\ref{SW2}) which computes D($-1$)-instanton generated superpotentials. By realizing a SQCD-like gauge theory, we are able to test the moduli space integral by computing the non-perturbative Affleck, Dine and Seiberg (ADS) superpotential \cite{Affleck:1983mk} (see also \cite{Taylor:1982bp}) in the case where it is known to be generated by an ordinary gauge instanton effect. We then consider configurations in which the D($-1$)-instanton can not be interpreted as an ordinary gauge instanton and it is therefore referred to as a stringy instanton. 

In \ABFLP , \Petersson, \FP~and \PeterssonN~the precise circumstances under which such a stringy instanton contributes to the superpotential  are analyzed.

\section{$\mathbf{\mathcal{N}=1}$ D-Instanton Calculus}

In this section we describe the D$3/\mathrm{D}(-1)$ system at an orbifold singularity and the $\mathcal{N}=1$ version of the D-instanton calculus discussed in the previous chapter. In order to get   a SQCD-like $\mathcal{N}=1$ gauge theories we choose, as a simple example, the  orbifold to be $\mathbb{C}^3/\mathbb{Z}_2 \times \mathbb{Z}_2$ \cite{Douglas:1996sw,Morrison:1998cs}. 

\subsubsection{The Gauge Sector}

The group $\mathbb{Z}_2 \times \mathbb{Z}_2$ has four elements: the
identity $e$, the generators of the two $\mathbb{Z}_2$ that we denote
with $g_1$ and $g_2$ and their product, denoted by $g_3 =g_1 g_2$. If
we introduce complex coordinates $(z_1, z_2,z_3)\in \mathbb{C}^3$ in the directions transverse to the D3-branes
\beq
z^1 = x^4 + i x^5~~,~~z^2 = x^6 +i x^7~~,~~ z^3 = x^8 + i x^9
\label{zzz67}
\end{equation}
the action of the orbifold group can be defined as in Table 3.1.
\begin{table} [ht]
\begin{center}
\begin{tabular}{c||c|c|c|}
&$z^1$ & $z^2$ & $z^3$ \\ \hline  \hline $e$    &$z^1$ & $z^2$ & $z^3$
\\  \hline $g_1$  &$z^1$ & $-z^2$ & $-z^3$ \\  \hline $g_2$ &$-z^1$ &
$z^2$ & $-z^3$ \\  \hline $g_3$  &$-z^1$ & $-z^2$ & $z^3$ \\ \hline
\end{tabular}
\caption{\small The action of the orbifold generators.}
\end{center}
\end{table}
Since the $\mathbb{Z}_2 \times \mathbb{Z}_2$ generators act non-trivially in all three complex transverse directions, the orbifold acts as a subgroup of $SU(3)$ and therefore, only one quarter of the original supersymmetries are preserved. We thus obtain an $\mathcal{N}=1$ world volume theory on the D3-branes when they are placed at the singular orbifold point. This can also be understood at the level of the spectrum as we will discuss below. 

From Table 3.1, it is easy to realize that a single D3-brane at an arbitrary point in the transverse space is not an invariant configuration. The only invariant configuration for a single D3-brane is when it is completely  stuck at the origin. The fact that the $\mathbb{Z}_2 \times \mathbb{Z}_2$ orbifold has three non-trivial generators implies that a single D3-brane is required to have three image D3-branes in order to be allowed to move away from the origin. An invariant configuration consisting of a single D3-brane and three images is called a regular D3-brane since it is allowed to move freely in the transverse space. Instead, for a configuration in which one or more image D3-branes are missing, there is one or more complex direction in the transverse space along which the brane is not allowed to move. Such a configuration is called a fractional D3-brane since it makes up a fraction of a regular D3-brane.

Moreover, the existence of three non-trivial generators implies the existence of three independent vanishing two-cycles at the singularity. 
This provides us with an alternative description of fractional D3-branes, namely as D5-branes wrapped on two-cycles which have vanishing geometric volume in the orbifold limit. The tension of a fractional D3-brane is however finite in this limit since the NS-NS two form $B$-field in the D5-brane world volume has non-vanishing flux through the vanishing two-cycle (see e.g. \cite{Bertolini:2003iv}). This can be seen by considering (\ref{Dp}) for $p=5$ with world volume $\mathcal{M}_{p+1}=\mathbb{R}^4\times \Sigma_2$ where $\Sigma_2$ is one of the vanishing two-cycles through which the $B$-field has a non-vanishing flux,
\begin{equation}
\label{B2}
\int_{\Sigma_2}B=\frac{1}{4}\,4\pi\alpha'\,.
\end{equation}  
By using (\ref{B2}) in (\ref{Dp}) in the low energy limit, we get that the complexified gauge coupling constant is given by $\tau/4$, i.e. one quarter of the complexified gauge coupling on a regular D3-brane. This implies that the tension and R-R charge for a fractional D3-brane is one quarter of that of a regular D3-brane. Moreover, the sum of the world volume actions for the four different types of fractional D3-branes yields the world volume action for a regular D3-brane.

The gauge sector of the orbifold theory can be obtained as an orbifold projection of the $\Ncal= 4$ SYM \cite{Douglas:1996sw,Morrison:1998cs}. A generic open string state consists of an oscillator part and a Chan-Paton (CP) part. The orbifold action on the oscillators is given according to Table 3.1. The orbifold action on the CP factors corresponds to choosing matrix representations for the generators and acting on the CP matrices. If we start from $N$ D3-branes in flat space the orbifold action on the $N\times N$ CP factors can be obtained by using the following matrix representations $\gamma(g)$ for the first two non-trivial orbifold generators in Table 3.1,
\beq
\gamma(g_1)  =\begin{pmatrix}1 & 0 & 0   & 0   \cr  0 & 1 & 0   & 0 \cr 0 & 0
& -1   &  0 \cr  0   & 0   & 0 & -1 \cr\end{pmatrix}~~~,~~~
\gamma(g_2) =\begin{pmatrix}1 & 0 & 0   & 0   \cr   0 & -1 & 0   & 0 \cr 0 & 0
& 1   &  0 \cr  0   & 0   & 0 & -1 \cr\end{pmatrix}~
\label{chanpatonz2z2}
\eeq
where the 1's denote $N_\ell \times N_\ell$ unit matrices ($\ell=1,...,4$) and $\sum_{\ell=1}^{4}N_\ell =N$ \cite{Bertolini:2001gg}. Since we should keep only open string states that are invariant under the combined action of the orbifold, we enforce the following conditions,
\beq
A_\mu = \gamma(g)A_\mu \gamma(g)^{-1}~~~,~~~
\Phi^i = \pm \gamma(g)\Phi^i\gamma(g)^{-1}
\label{orbaction}
\eeq
where the sign $\pm$ is given by the action of the orbifold generators $g$
in Table 3.1. Note that there is no extra sign arising from the oscillators corresponding to the gauge fields since their indices are not pointing in the orbifold directions.  
With the choice (\ref{chanpatonz2z2}), the gauge fields
are block diagonal matrices of different size $(N_1,N_2,N_3,N_4)$, giving rise to a $\prod_{\ell=1}^{4}U(N_\ell)$ gauge group. 

The three complex scalars $\Phi^i$
have the following form,
\beq
\Phi^1 = \begin{pmatrix}0 & \times &
0   & 0   \cr  \times & 0 & 0   & 0 \cr 0 & 0   & 0   &  \times \cr  0
& 0   & \times & 0 \cr\end{pmatrix},~~ \Phi^2 = \begin{pmatrix}0 & 0
& \times &  0  \cr     0   & 0   & 0   & \times \cr \times & 0   & 0
& 0  \cr  0   & \times & 0 & 0 \cr\end{pmatrix}, ~~ \Phi^3 =
\begin{pmatrix}0 & 0   & 0   & \times \cr  0   & 0 & \times & 0   \cr
0 & \times & 0   & 0   \cr  \times & 0   & 0   & 0 \cr\end{pmatrix}
\label{structure}
\eeq
where the crosses  represent the non-zero entries $\Phi_{\ell m}$, corresponding to scalars transforming in the fundamental representation of $U(N_\ell)$ and in the anti-fundamental representation of $U(N_m)$. For the gauginos of the gauge sector, $\Lambda^{\alpha A}$ and $\bar\Lambda_{\dot\alpha A}$, we impose conditions corresponding to the fermionic versions of (\ref{orbaction}). In these conditions the orbifold action on the CP part is the same while the orbifold action on the oscillator part corresponds to rotating the (anti-)fundamental $SU(4)$ spinor indicies. These (spinorial) rotation matrices can be chosen such that the gaugino with component $A=4$ is block diagonal, and the gauginos with $A=1,2,3$ have the same structure as in (\ref{structure}). Hence, the gaugino with $A=4$ component joins the gauge field $A^\mu$ and form $\mathcal{N}=1$ vector multiplet  while the other gauginos join the scalars $\Phi^i$ and form $\mathcal{N}=1$ chiral superfields, which we also denote by $\Phi^i$.

\begin{figure}[ht]
\begin{center}
{\includegraphics[width=55mm]{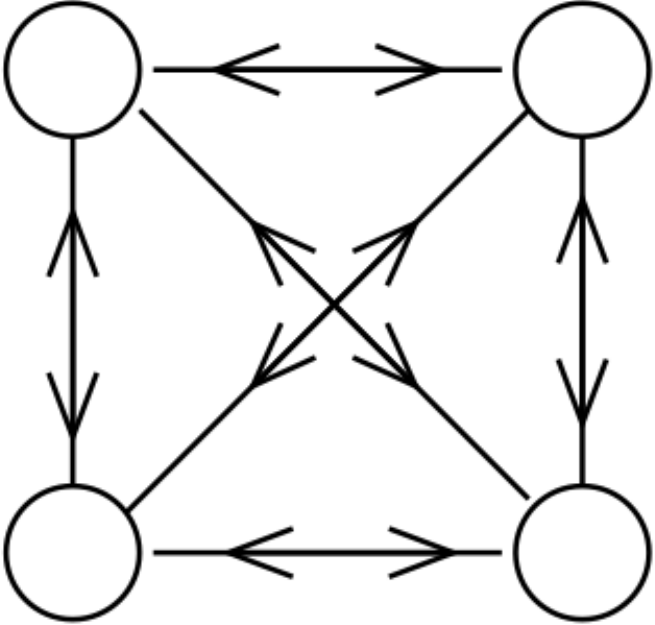}}
\caption{\small  Quiver diagram for the
$\mathbb{Z}_2 \times \mathbb{Z}_2$ orbifold theory. Round circles
correspond to $U(N_\ell)$ gauge factors while the lines connecting
quiver nodes represent the bi-fundamental chiral superfields
$\Phi_{\ell m}$.}
\label{quiverD3}
\end{center}
\end{figure}
The field content of the resulting theories can be conveniently summarized in a quiver diagram,
see Figure~\ref{quiverD3}, which, together with the cubic superpotential
\beqs  W &=&
\Phi_{12}\Phi_{23}\Phi_{31}-\Phi_{13}\Phi_{32}\Phi_{21}+\Phi_{13}\Phi_{34}\Phi_{41}-
\Phi_{14}\Phi_{43}\Phi_{31} \nonumber \\ &&
+\Phi_{14}\Phi_{42}\Phi_{21}-\Phi_{12}\Phi_{24}\Phi_{41}+\Phi_{24}\Phi_{43}\Phi_{32}-
\Phi_{23}\Phi_{34}\Phi_{42} \nn \\
         \label{wtree}
\eeqs 
uniquely specifies the theory. This superpotential can be directly obtained from the $\Ncal
= 4$ SYM theory, written in $\Ncal
= 1$ language,
\beq
W_{{\cal N}=4} = \tr~ \Phi^1 [ \Phi^2 , \Phi^3]
 \label{wN4}\,.
\eeq
These theories are non-chiral $\Ncal=1$
four-node quiver gauge theory in which non-chirality implies that 
the four gauge group ranks can be chosen independently \cite{Bertolini:2001gg}.

A stack of $N$ regular D3-branes amounts to having the same
rank assignment $(N,N,N,N)$ on every node of the quiver. Such a stack is allowed to move away from the singularity in any direction. The gauge group in this case is $U(N)^4$ and the world volume theory is an $\Ncal =1$ superconformal gauge theory. For any other rank assignment, it instead corresponds to a stack of fractional D3-branes and the world volume theory is no longer conformal. The lack of conformal invariance can be seen from the one-loop beta-function coefficient of e.g. the first gauge node in Figure ~\ref{quiverD3}, $b_1=3N_c-N_f=3N_1-N_2-N_3-N_4$.

\subsubsection{Instanton sector}

The orbifold projection of the neutral sector is very similar to the gauge sector since it can be viewed as a dimensional reduction of the gauge sector.
This implies that the zero modes structure of the D($-1$)-instantons can be directly obtained from the gauge sector structure.  In particular, there are four nodes and hence four different types of fractional D($-1$)-instantons. Regular D($-1$)-instantons have the
same rank (instanton number) at every node $(k,k,k,k)$ while all other situations can be
thought of as fractional D($-1$)-instantons. Generically, we can
 characterize an instanton configuration in our orbifold by $(k_1, k_2,
k_3, k_4)$.

The bosonic
modes comprise a $4\times 4$ block diagonal matrix $a^\mu$, while the three complex modes $s^i$  have the same structure as (\ref{structure}), but
now where each block entry is a $k_\ell\times k_m$ matrix. For the fermionic
zero-modes  $M^{\alpha A}$ and $\lambda_{\dot\alpha A}$ 
we again get that for $A=4$ they are block diagonal while for $A=1,2,3$ they have
the structure of (\ref{structure}).

In the charged sector, the bosonic zero-modes $\omega_{\dot\alpha}$, $\bar\omega_{\dot\alpha}$ are diagonal in the gauge factors since their $\dot\alpha$-index is not affected by the orbifold action in the transverse space. These modes are block diagonal matrices with entries $N_\ell \times k_\ell$ and $k_\ell \times N_\ell$ respectively. The charged fermions $\mu^{A}$, $\bar\mu^{A}$ are matrices with block entries $N_\ell \times k_m$ and $k_m \times N_\ell$, respectively, and they display the same
structure as (\ref{structure}) for $A=1,2,3$ and are 
diagonal for $A=4$.

\subsubsection{The Moduli Space Integral}

Consider now the moduli space integral (\ref{SW2}) in this orbifold configuration with arbitrary fractional D3-brane rank assignment $(N_1.N_2,N_3,N_4)$.  For simplicity, let us only consider a single fractional D($-1$)-instanton at node 1, i.e. with rank assignment $(k_1,k_2,k_3,k_4)=(1,0,0,0)$, denoted by D($-1)_1$. 

For this simple choice, the only massless modes present in the neutral
sector are four $x^\mu$ modes, from the upper-left component of
$a^\mu$, three auxiliary modes $D^c$, two fermions $\theta^\alpha$ from the upper-left
component of $M^{\alpha 4}$ and two more fermions
$\lambda_{\dot\alpha}$ from the upper-left component of
$\lambda_{\dot\alpha 4}$.  

In the charged sector, there are 4$N_1$ bosonic moduli $\omega_{\dot\alpha}$, $\bar\omega_{\dot\alpha}$ from the strings stretching between the D(-1$)_1$-instanton and the $N_1$ D3${}_1$-branes. Furthermore, there are $2N_\ell$ charged  fermionic modes $\mu$, $\bar\mu$ from the open strings stretching between the D(-1$)_1$-instanton and the $N_\ell$ D3${}_\ell$-branes at node $\ell$, where $\ell=1,2,3,4$.

From the scaling dimension of the moduli fields (\ref{dim}) we obtain the dimension of the measure for the moduli space integral corresponding to this instanton configuration,
\begin{eqnarray}
\label{dimmeas2}
\Big[ d \{ x, \theta,  \lambda,D ,\omega , \ov\omega , \mu ,\ov\mu \Big] &=& M_s^{-(n_{x } -\frac{1}{2}n_{\theta} +\frac{3}{2}n_{\lambda} -2 n_{D}+ n_{\omega,\ov\omega} - \frac{1}{2}n_{\mu , \ov\mu } )} 
\nn \\
&=& 
M_{s}^{-(3N_1-N_2-N_3-N_4)}=M_{s}^{-b_1}~.
\end{eqnarray}
In order to compensate for the dimension of the measure, we thus need a prefactor with dimension $M_{s}^{b_1}$ where we identify  the power  $b_1$ with the one-loop beta-function coefficient  for the coupling constant of the gauge group at node 1.

In analogy with the interpretation of a fractional D3-brane as a wrapped D5-brane, we can also interpret a D($-1)_1$-instanton as a ED1-brane wrapped on a vanishing two-cycle.  
By using similar arguments as before we find that the fractional D$(-1)_1$-instanton action is given by
 \begin{eqnarray}
\label{ED1}
S_{D(-1)_1} & = & -\frac{2\pi i \tau}{4}\,
\end{eqnarray}
which is one quarter of the action for a regular D($-1$)-instanton.

For a single D$(-1)_1$-instanton, the $\mathcal{N}=1$ version of the moduli space integral (\ref{SW2}) has the following structure,
\begin{eqnarray}
\label{4dSW}
Z_{k_1=1} &=&  \int d^4 x \, d^2 \theta \,W
~~
\end{eqnarray} 
where the non-perturbative superpotential is given by the centered moduli space integral of the orbifold projected moduli, 
\begin{equation}
\label{W}
W=M_{s}^{b_1 }\,e^{\frac{2\pi i \tau}{4}}\int d\widehat{\mathcal{M}}_{k_1=1}\,\,e^{-S_{\mathrm{D3/D(-1)}}^{k_1=1} }\,.
\end{equation}
The action $S_{\mathrm{D3/D(-1)}}^{k_1=1}$ is the orbifold projected version of (\ref{Tot}). In order to see if a non-perturbative superpotential is generated by a D$(-1)_1$-instanton in a particular $\mathcal{N}=1$ gauge theory we need to choose a rank assignment for the fractional D3-branes and evaluate (\ref{W}).

\section{Gauge Instantons}

\begin{figure}[ht]
\begin{center}
\includegraphics[width=70mm]{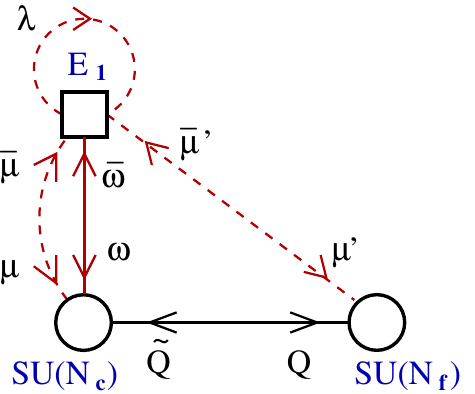}
\caption{\small  The quiver diagram describing a gauge D($-1)_1$-instanton located on the first node, together with $N_c$ fractional D3-branes. Gauge theory nodes are represented by round circles, instanton nodes by squares.}
\label{quiverADS}
\end{center}
\end{figure}

In order to test (\ref{W}) we choose a rank assignment that corresponds to an $\mathcal{N}=1$ SQCD configuration where it is known from field theory that a non-perturbative superpotential is generated from a gauge instanton effect. In order to realize such a setting we occupy two of the four nodes, $(N_1,N_2,N_3,N_4)=(N_c,N_f,0,0)$, where $N_c$ and $N_f$ are arbitrary. At low energies, the $U(1)$ factors of the $U(N_c)\times U(N_f)$ gauge group are decoupled and the resulting effective gauge group is $SU(N_c)\times SU(N_f)$. The only chiral fields present are the two components of $\Phi^1$
connecting the first and second node
\begin{equation}
\Phi^1 =
\begin{pmatrix}0 & Q & 0   & 0   \cr  \tilde Q & 0 & 0   & 0 \cr 0 & 0
& 0   &  0 \cr  0   & 0   & 0 & 0 \cr\end{pmatrix}~.
\label{Phi1}
\end{equation}
By placing a single D($-1)_1$-instanton at the first node, i.e.  $(k_1,k_2,k_3,k_4)=(1,0,0,0)$, the action (\ref{Tot}) greatly simplifies and the only non-vanishing terms are
\begin{equation}
S_{L.m.} = i
\left(\bar\mu \omega_{\dot\alpha} +  \bar\omega_{\dot\alpha }
\mu \right) \lambda^{\dot\alpha}  - i D^c \bar\omega^{\dot
\alpha}(\tau^c)_{\dot\alpha}^{\dot\beta} \omega_{\dot\beta}~,
\label{S1ADS}
\end{equation}
and
\begin{equation}
S_{\Phi} =
\frac12\,\bar\omega_{{\dot\alpha}} \big(Q {Q^\dagger}
+\tilde Q^\dagger \tilde Q\big) \omega^{{\dot\alpha}} -
\frac{i}2\,\bar \mu \tilde Q^\dagger{} \mu'{}+\frac{i}2\,
\bar \mu' Q^\dagger \mu~
          \label{mixed}
\end{equation}
where $\mu$ and $\bar\mu$ denote the massless charged fermion modes of the open strings between the D($-1)_1$-instanton and the $N_c$ D3-branes at node 1 while $\mu'$ and $\bar\mu'$ arise from the strings between the D($-1)_1$-instanton and the $N_f$ D3-branes at node 2. 
Note that the couplings in (\ref{mixed}) depend non-holomorphically on the chiral superfields. 
Moreover, since we are interested in a field theory result, we have taken the field theory limit, in which $S_{g_0}=(1/2)D_c^2$ in (\ref{tot}) vanishes.  The zero mode structure can be conveniently summarized in a generalized quiver diagram as represented in Figure~\ref{quiverADS}, which accounts for both the D3-brane configuration and the D($-1)_1$-instanton zero modes.

For this configuration $S_{\mathrm{D3/D(-1)}}^{k_1=1}$ in (\ref{W}) is given by the sum of (\ref{S1ADS}) and (\ref{mixed})
and the corresponding moduli space integral is
\begin{equation}
\label{nn}
W =\Lambda^{3N_c-N_f} \int
d\{\omega,\bar\omega,\mu,\bar\mu\}\, \delta(\bar\mu_u
\omega^u_{\dot\alpha} +  \bar\omega_{\dot\alpha u}
\mu^u)\,\delta(\bar\omega_{u}^{\dot
\alpha}(\tau^c)_{\dot\alpha}^{\dot\beta}  \omega_{\dot\beta}^{u})
\,e^{-S_\Phi}~
\end{equation}
where we have integrated over the Lagrange multipliers $D$ and $\lambda$ in (\ref{S1ADS}) and obtained the bosonic
and fermionic ADHM constraints in the form of $\delta$-functions. In (\ref{nn})
we have combined the dimensionful $M_{s}^{b_1}$ and the D($-1)_1$-instanton action into a factor $\Lambda^{b_{1}}$ which can be identified with the dynamical scale of the SQCD theory. 

Due to the presence of
extra $\mu$ modes in the integrand from the fermionic delta function,
 only when  $N_f=N_c-1$ we obtain a non-vanishing result. By evaluating (\ref{nn}) we obtain (see e.g. \cite{Dorey:2002ik,Akerblom:2006hx} for details),
\begin{equation}
W_{ADS} =  \frac{\Lambda^{2N_c +1}}{\det
\left[ \tilde Q Q \right] }~
\label{ADS}
\end{equation}
which is just the expected ADS superpotential for $N_f =N_c-1$ \cite{Taylor:1982bp,Affleck:1983mk}, the only case where such non-perturbative contribution is generated by a
genuine one-instanton effect and not by gaugino condensation.

\section{Stringy Instantons}

\begin{figure}[ht]
\begin{center}
\includegraphics[width=70mm]{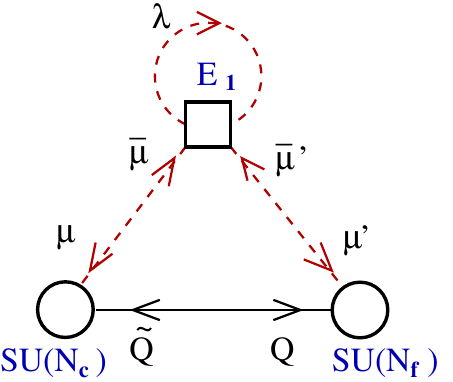}
\caption{\small  The quiver diagram describing a stringy D($-1)_3$-instanton, located on the third node, and fractional D3-branes, located at the first and second nodes. }
\label{quiverBB}
\end{center}
\end{figure}

Let us now consider a system with the same fractional D3-brane rank assignment
$(N_c,N_f,0,0)$ but instead with a single fractional D($-1$)-instanton at the third node $(0,0,1,0)$. Such a D($-1)_3$-instanton sits on a node which does not have any gauge theory associated to it and hence it can not be interpreted as an ordinary gauge instanton. The quiver diagram, with the relevant zero-modes structure, is given in Figure~\ref{quiverBB}.
It is important to note that the D$(-1)_3$-instanton action, which is given by (\ref{ED1}), is well defined even in the absence of fractional D3-branes at the same node, i.e. $N_1=0$. Moreover, the dimension of the measure (\ref{dimmeas2}), which in this case requires a prefactor $M_{s}^{-N_c-N_f}$, is also well defined even though the power to which the prefactor is raised can no longer be identified with the one-loop beta-function coefficient of any gauge theory. 

In this case, the charged modes $\omega_{\dot\alpha }$ and $\bar\omega_{\dot\alpha }$ vanish due to their diagonal structure. Since these charged modes correspond to the size of the instanton, such a D$(-1)_3$-instanton has zero size, in analogy with a D($-1$)-instanton along the Coulomb branch discussed in the previous chapter. The absence of the bosonic charged modes can also be understood by interpreting the D$(-1)_3$-instanton as an  ED1-brane wrapped on the two-cycle corresponding to node 3. In the case where all the spacefilling D5-branes are wrapping other two-cycles, i.e. when the fractional D3-branes are located at other nodes, the open strings between the ED1-brane and the D5-branes have eight Neumann-Dirichlet directions. This implies that the  NS sector is massive and hence no charged bosons arise. This is in contrast to the case when the ED1-brane and the D5-branes wrap the same cycle. In this case, there are four Neumann-Dirichlet directions and charged bosons appear.  The neutral zero-modes  are however the same as
before.

Because of the absence of bosonic charged modes, there is no longer any terms like (\ref{S1ADS}) and the zero-dimensional action is simply given by,
\begin{eqnarray}
S_\Phi &=& \frac{i}{2}\,\bar \mu \,Q \,\mu' - \frac{i}{2}\,
\bar\mu'\,\tilde Q \,\mu~.
\label{stringy}
\end{eqnarray}
where $\mu$ and $\bar\mu$ denote charged fermions from the strings between the D($-1)_3$-instanton and the $N_c$ D3-branes at node 1 while $\mu'$ and $\bar\mu'$ arise from the strings between the D($-1)_3$-instanton and the $N_f$ D3-branes at node 2. In (\ref{stringy}) we see that the
 only remaining charged fermion zero modes are those that couple holomorphically to the chiral superfields, corresponding to the last term in (\ref{SPhi}). The diagonal charged fermions that originate from the $\mu^4$ and $\bar\mu^4$ modes and that couple anti-holomorphically to the chiral superfields are projected out. 

Since there are only charged fermions that couple to the matter fields one expects, by integrating out these fermion modes (for $N_c=N_f$ in this case), that the resulting instanton generated superpotential has a polynomial structure. This is the main reason why superpotentials arising from these types of stringy instantons are of phenomenological interest. For example, in configurations where the surrounding gauge theory is engineered in an appropriate way, such non-perturbative superpotentials can correspond a linear Polonyi term relevant for supersymmetry breaking, a Majorana mass term relevant for right handed neutrinos or certain Yukawa couplings relevant for $SU(5)$ GUT models \cite{Blumenhagen:2006xt}~--\cite{Cvetic:2007sj}.

However, when inserting (\ref{stringy}) in the moduli space integral (\ref{W}),
\begin{eqnarray}
\label{exotic}
W &=& M_s^{-N_c-N_f} e^{\frac{2\pi i \tau}{4}}\int d\{\lambda,D,\mu,\bar\mu\}\, e^{-S_\Phi}\nn \\
&=&M_s^{-2N_c} e^{\frac{2\pi i \tau}{4}}\int d\{\lambda,D\}\,\det \left[ Q \tilde Q \right]
\end{eqnarray}
we see an immediate obstacle. Since the $\lambda$ mode does not appear\footnote{The auxiliary field $D$ also does not appear in the integrand. It does however not raise any concern, since before taking the field theory limit, $D$ appeared
quadratically  in the action (\ref{tot}) and, by integrating it out, we obtain an
overall normalization constant.} in the  integrand,
the integral (\ref{exotic}) is trivially zero. The $\lambda$ mode does not appear due to the absence of charged bosonic modes $\omega$ and $\bar\omega$. 

Moreover, the absence of the $\lambda$ mode is expected for the following reason: A spacefilling D5-brane preserves half of the eight supercharges of the background. If an ED1-instanton wraps the same cycle as the D5-brane it breaks half of the four supercharges preserved by the spacefilling D5-brane and hence, it carries two goldstino neutral fermionic zero modes $\theta^\alpha$. In contrast, if the ED1-instanton wraps a cycle not occupied by any spacefilling D5-brane, it breaks half of eight supercharges preserved by the background and therefore, it carries four goldstino neutral fermionic zero modes $\theta^\alpha$ and $\lambda^{\dot{\alpha}}$.  To contribute to the superpotential, an ED-instanton is only allowed to carry the two $\theta^\alpha$ goldstino modes that make up the chiral superspace measure \cite{Witten:1996bn}. Hence, if we want the stringy instanton to generate a non-perturbative superpotential we must modify our configuration in some way.  

\ABFLP , \Petersson~and \FP~discuss three different ways to modify the configuration such that a superpotential is generated. 

In \ABFLP~we show that, by introducing an orientifold plane on which the the D($-1)_3$-instanton is placed,  the $\lambda$ mode is projected out (see also \cite{Argurio:2007qk,Bianchi:2007wy}). The reason is because the D-instanton in this case breaks half of the four supercharges preserved by the orientifold. 

Another mechanism is provided in \Petersson~where the orientifold plane is replaced by a single fractional D3-brane on which the D($-1)_3$-instanton is placed (see also \cite{Aganagic:2007py,GarciaEtxebarria:2007zv}). Again, the D($-1)_3$-instanton breaks half of the four supercharges preserved by the fractional D3-brane. In this case, charged bosons appear and hence, the $\lambda$ mode is lifted by interactions like (\ref{S1ADS}). Even though the D($-1)_3$-instanton action in this case is given in terms of the gauge coupling on the single fractional D3-brane, this instanton does not admit a direct interpretation as an ordinary gauge instanton since the world volume theory on the fractional D3-brane is an abelian (commutative) gauge theory.\footnote{See \cite{Argurio:2012iw} for a my more recent work in which these stringy instantons are interpreted in terms of standard field theory.}  

In \FP~we discuss the possibility of lifting the $\lambda$ mode by turning on background fluxes \cite{Billo':2008pg,Billo':2008sp} (see also \cite{Tripathy:2005hv,Martucci:2005rb,Bergshoeff:2005yp,Blumenhagen:2007bn,Camara:2003ku,Grana:2003ek}). In this case, supersymmetry breaking three-form fluxes induce a mass term for the $\lambda$ modes in the zero dimensional instanton action.

\chapter{Multi-Instantons in Toric Gauge Theories}

In the previous chapters we have seen how D-instantons are treated in gauge theories realized as world volume theories of spacefilling D-branes probing an orbifold singularity. 
It is however important to try to go beyond the orbifold limit, particularly
having in mind applications to phenomenology and the gauge/gravity correspondence, where
orbifold gauge theories provide too restrictive a class of models. 

In this
context it is more interesting to consider gauge theories arising from
D-branes probing a generic toric singularity
\cite{Morrison:1998cs,Park:1999ep,Beasley:2001zp,Feng:2001bn,Feng:2002zw,Feng:2001xr,Martelli:2004wu,Benvenuti:2004dy,Bertolini:2004xf,Franco:2005rj}. However, in the non-orbifold case, worldsheet techniques are not straightforwardly available and therefore, we make use of other methods to treat D-instantons in such backgrounds. These methods involve  performing partial resolutions starting from an orbifold, which, from the gauge theory point of view, imply that  Fayet-Iliopoulos (FI) terms are turned on. As a consequence, 
some of the chiral superfields acquire vevs in order to satisfy the D-flatness conditions. Moreover, some other matter fields then acquire a mass
and are integrated out, typically yielding new terms in the superpotential
which are of order higher than cubic, which is the generic case for orbifolds since the superpotential is obtained by projecting the cubic $\mathcal{N}=4$ superpotential. In this way, we obtain non-orbifold gauge theories. 

The higgsing procedure can be
applied (with some care) to the instanton sector as well. This method works quite generally and it applies to rigid
instantons as well as instantons with internal neutral modes. In fact, the
role played by these extra neutral modes in the multi-instanton case is
crucial for the higgsing procedure to work. Indeed, as it will become clear,
single instantons in a toric geometry will generically descend
from multi-instantons in the unhiggsed parent theory.

In this chapter we give a general set of rules for how to construct the moduli action for D-instantons in gauge theories arising from branes probing any toric singularities.  In \AFP~ we apply these rules and provide many explicit examples such as the non-chiral Suspended Pinch Point (SPP) and the conifold as well as the chiral first three del Pezzo's ($dP_1$, $dP_2$ and $dP_3$). This construction allows us to compute novel D-instanton effects in these backgrounds.

\section{The Gauge Sector Point of View}

\begin{figure}
\begin{center}
\includegraphics[width=60mm]{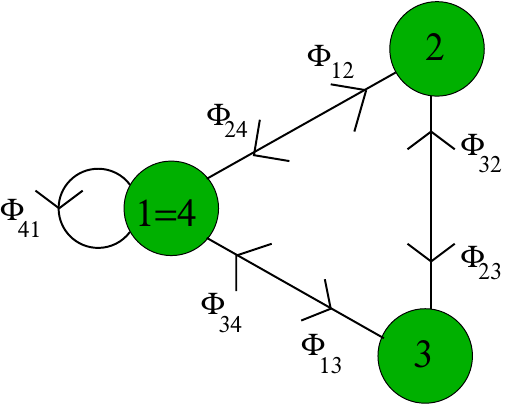}
\caption{\small{The SPP theory higgsed down from the $\mathbb{Z}_2\times\mathbb{Z}_2$ theory.}}
\label{SPPD}
\end{center}
\end{figure}

The basic idea behind our construction is the well known fact
(see e.g.~\cite{Morrison:1998cs,Park:1999ep,Beasley:2001zp,Feng:2001bn,Feng:2002zw,Feng:2001xr}) that any quiver
gauge theory describing D-branes at a toric singularity can be obtained by
higgsing a sufficiently large orbifold.  Since we have in this thesis focused on the 
$\mathbb{C}^3/\mathbb{Z}_2\times\mathbb{Z}_2$ quiver gauge theory, let us use this theory as a simple example of the general procedure. We start by turning on an FI parameter such that the following chiral superfield aquires a vev in order to satisfy the D-flatness condition,
\beq
\Phi_{14} = m.
\eeq
This requires the condition $N_1 =N_4 =N$ and breaks the two gauge groups corresponding to nodes 1 and 4 to the diagonal subgroup,
\begin{equation}
U(N)_1 \times U(N)_4 \to U(N)_{(14)}~.
\end{equation}
The chiral superfield $\Phi_{41}$ will therefore transform in the adjoint representation of $SU(N)_{(14)}$.
From the superpotential in (\ref{wtree}) we immediately see that the fields $\Phi_{31}$, $\Phi_{43}$, $\Phi_{42}$
and $\Phi_{21}$ become massive. One should integrate them out through
their F-flatness equations, which read:
\beqs
\Phi_{31}=\frac{1}{m}\Phi_{32}\Phi_{24}, &&
\Phi_{43}=\frac{1}{m}\Phi_{12}\Phi_{23}, \nn \\
\Phi_{42}=\frac{1}{m}\Phi_{13}\Phi_{32}, &&
\Phi_{21}=\frac{1}{m}\Phi_{23}\Phi_{34} ~.\label{intout}
\eeqs
Inserting these values back into (\ref{wtree}) gives us the superpotential for the gauge theory arising from D-branes probing the SPP  singularity,
\beq  W_{SPP} =
\frac{1}{m}\Phi_{24}\Phi_{12}\Phi_{23}\Phi_{32}-\frac{1}{m}\Phi_{13}\Phi_{32}\Phi_{23}\Phi_{34}
+\Phi_{13}\Phi_{34}\Phi_{41}-\Phi_{24}\Phi_{41}\Phi_{12}
         \label{wSPP}
\eeq
where all the remaining fields (except for $\Phi_{41}$) transform in bifundamental representations of two of the factors in the gauge group $U(N)_{(14)}\times U(N_2) \times U(N_3 )$, see Figure \ref{SPPD}.

\begin{figure}
\begin{center}
\includegraphics[width=100mm]{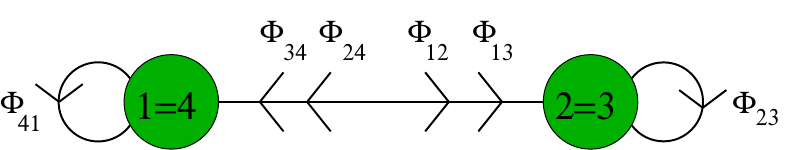}
\caption{\small{The $\mathbb{C}^2/\mathbb{Z}_2$ theory higgsed down from the SPP theory.}}
\label{Z2hD}
\end{center}
\end{figure}

We can continue this procedure and obtain the quiver gauge theory for the $\mathbb{C}^2/\mathbb{Z}_2$ orbifold if we start from the SPP theory and give an additional vev to the chiral superfield $\Phi_{32}$,
\begin{equation}
\Phi_{32} = m ~.
\label{vev32}
\end{equation}
This means that we have the condition $N_3 =N_2 =M$ and that we ``pinch'' the two gauge groups corresponding to nodes 2 and 3 together,
\begin{equation}
U(M)_2 \times U(M)_3 \to U(M)_{(23)}~.
\end{equation}
As before, the chiral superfield $\Phi_{23}$ will now transform in the adjoint representation of $U(M)_{(23)}$. We see from (\ref{wSPP}) that (\ref{vev32}) does not induce any new mass terms, but gives us the $\mathbb{C}^2/\mathbb{Z}_2$ superpotential, 
\beq
W_{\mathbb{Z}_2} =
\Phi_{24}\Phi_{12}\Phi_{23}-\Phi_{13}\Phi_{23}\Phi_{34}
+\Phi_{13}\Phi_{34}\Phi_{41}-\Phi_{24}\Phi_{41}\Phi_{12}
\eeq
where $\Phi_{12}$, $\Phi_{13}$ are in the ($\square$,$\overline{\square}$) of the gauge group $U(N)_{(14)}\times U(M)_{(23)}$, while $\Phi_{24}$, $\Phi_{34}$ are in the ($\overline{\square}$,$\square$) and $\Phi_{41}$, $\Phi_{23}$ are in the adjoint of the respective gauge groups, see Figure \ref{Z2hD}.

\begin{figure}
\begin{center}
\includegraphics[width=60mm]{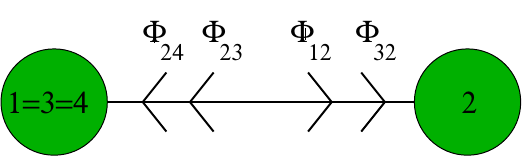}
\caption{\small{The conifold theory higgsed down from the SPP theory.}}
\label{ConiD}
\end{center}
\end{figure}

\begin{figure}
\begin{center}
\includegraphics[width=20mm]{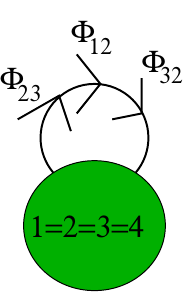}
\caption{\small{The $\mathcal{N}=4$ theory higgsed down from the conifold theory.}}
\label{N4fZ2Z2D}
\end{center}
\end{figure}

Alternatively, we could obtain the conifold gauge theory by starting again from the SPP theory but  giving a vev instead to the chiral superfield $\Phi_{34}$,
\begin{equation}
\Phi_{34} = m .
\label{vev34}
\end{equation}
This implies the condition $N_1 =N_4 =N_3 =N$, such that the three gauge groups corresponding to nodes 1, 3 and 4 now become
\begin{equation}
U(N)_1 \times U(N)_3 \times U(N)_4 \to U(N)_{(134)}~.
\end{equation}
 We see from (\ref{wSPP}) that (\ref{vev34}) induces a mass term for the bifundamental chiral superfield $\Phi_{13}$ and the adjoint field $\Phi_{41}$. Hence, we solve for these fields and get the following expressions,
\begin{equation}
\Phi_{13}=\frac{1}{m}\Phi_{12}\Phi_{24}, \qquad
\label{int34}
\Phi_{41}=\frac{1}{m}\Phi_{32}\Phi_{23}~.
\end{equation}
Inserting (\ref{vev34}) and (\ref{int34}) into (\ref{wSPP}) yields the superpotential for the conifold,
\beqs  W_{con} &=&
\frac{1}{m}\Phi_{12}\Phi_{23}\Phi_{32}\Phi_{24}-\frac{1}{m}\Phi_{12}\Phi_{24}\Phi_{32}\Phi_{23}
\label{wConifold}
\eeqs
where $\Phi_{12}$, $\Phi_{32}$ are in the ($\square$,$\overline{\square}$) of
the gauge group $U(N)_{(134)}\times U(N_2)$, $\Phi_{24}$, $\Phi_{23}$ are in
the ($\overline{\square}$,$\square$) and there are no more adjoint fields, see Figure \ref{ConiD}.
From here, giving a vev to, say, $\Phi_{24}=m$, leads straightforwardly
to the $\Ncal=4$ theory and its cubic superpotential, see Figure \ref{N4fZ2Z2D}.

These simple examples illustrate the general fact that non-orbifold gauge theories can be obtained by higgsing a simple orbifold gauge theory. Moreover, by further higgsing, it is possible to obtain a smaller orbifold theory. In the next section we apply the higgsing procedure in the instanton sector.

\section{Constructing D-instanton Moduli Spaces}

Let us now study what consequences  the above higgsing procedure have on the instanton sector. From the couplings of the charged modes to the chiral superfields in the gauge theory, seen in (\ref{SPhi}),
it is clear that the matter field vevs will give masses to some of the charged zero
modes. Moreover, a corresponding neutral zero mode
must also obtain a vev and correspondingly some neutral zero modes will also become massive. The reason is that the neutral sector, which is just the reduction to zero dimensions of the quiver gauge theory, couples to the same background closed string mode which
generates the FI term in the matter sector and therefore also generates a similar
term in the instanton sector. As a consequence, the structure of the surviving neutral zero modes
mirrors exactly the structure of quiver gauge and matter fields. This means that if we give a vev to the chiral superfield $\Phi_{ab}$ in the bi-fundamental representation of the gauge groups associated to node $a$ and $b$ we will also give a vev to the corresponding neutral scalar zero mode $s_{ab}$. 
Moreover, one finds that the couplings between the bosonic charged zero modes
and the superfields, as well as the anti-holomorphic couplings between the charged
fermionic zero modes and the superfields is exactly
as in the orbifold case.
Namely, for every pair of nodes $a$ and $b$ for which the relevant fields exist there will be the following couplings:
\beq
      \bar\omega_{aa}\Phi_{ab}\Phi^\dagger_{ba}\omega_{aa}, \quad
      \bar\omega_{aa}\Phi^\dagger_{ab}\Phi_{ba}\omega_{aa}, \quad
      \bar\mu_{aa}\Phi^\dagger_{ab}\mu_{ba}, \quad
      \bar\mu_{ab}\Phi^\dagger_{ba}\mu_{aa}.
\eeq
In the case of multiple instantons there are similar couplings between the charged moduli and the neutral moduli $s$.  However, the extra neutral moduli will not be present in the case of a single (fractional) instanton and this is the configuration that is mostly studied in practical applications.
  
\begin{figure}
\begin{center}
\includegraphics[width=130mm]{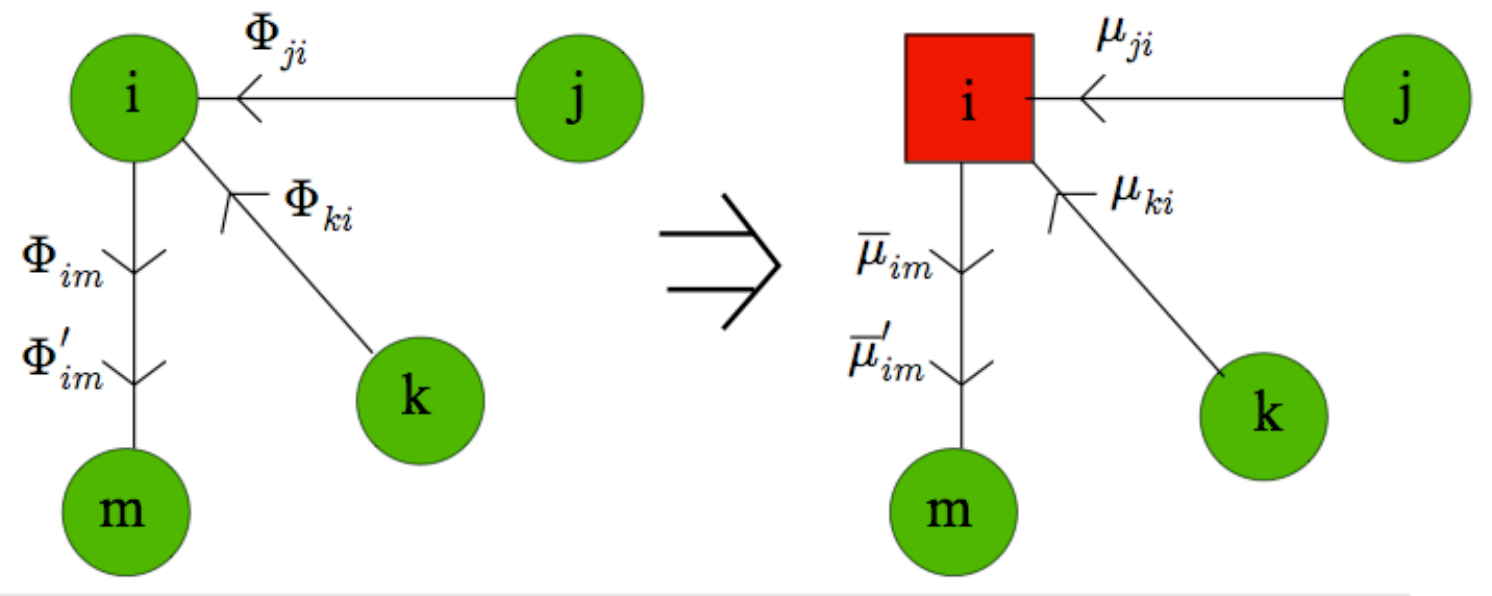}
\caption{\small{The structure of the charged fermion zero modes follows from the structure of the quiver.}}
\label{nnn}
\end{center}
\end{figure}

 For a D-instanton located at a particular node in the gauge theory the charged fermionic zero mode structure is given by the arrows that connect  that node. Diagrammatically, in Figure \ref{nnn} it is shown that by replacing a gauge node for an instanton node along with replacing the chiral superfields for charged fermionic zero modes we obtain the complete charged sector. 
  
  The holomorphic coupling between the charged fermionic
zero modes and the superfields is
obtained by taking each term in the superpotential and, while keeping
the same quiver index structure, substituting two fermionic charged zero
modes and all combinations of matter fields $\Phi$ and neutral bosonic zero modes $s$ allowed by the symmetries.
Again ignoring the bosonic modes $s$ for the time being, this rule means that, if one encounters, say, the term $\tr\, \Phi_{12} \Phi_{23} \Phi_{34} \Phi_{41}$ in the superpotential, one should expect the four terms:
\beq
   \tr \big( \bar\mu_{12} \Phi_{23} \Phi_{34} \mu_{41} + \bar\mu_{23} \Phi_{34} \Phi_{41} \mu_{12} +
   \bar\mu_{34} \Phi_{41} \Phi_{12} \mu_{23} +   \bar\mu_{41} \Phi_{12} \Phi_{23} \mu_{34} \big)
\eeq
in the instanton action, see also \cite{Kachru:2008wt}. It is easy to convince oneself that the bosonic neutral modes $s$ can be accommodated in an analogous way.

In \AFP, these rules are applied to several chiral and non-chiral gauge theories and we construct the corresponding general D-instanton moduli actions. The possibility to continue the Higgsing procedure down to a known orbifold configurations allows us to check the validity of these rules.

\chapter{Particle/Instanton Dualities and Gauge Couplings}

In this chapter we discuss  multi-instanton contributions to certain string theory couplings and how they can be computed in dual  pictures. In order to understand the underlying symmetry and duality properties of the theory one is often required to take into account exact  quantum corrections. Since many couplings receive instanton corrections from all instanton numbers, it is necessary to have methods to compute multi-instanton contributions.   
In this chapter we will use the idea of resumming multi-instanton effects by computing one-loop diagrams in a T-dual theory, in which the instantons turn into D-particles running in a loop \cite{Ooguri:1996me}.  
In particular, we propose a general method for computing one-loop amplitudes with BPS particles in configurations which involve spacefilling D-branes and orientifold planes. In order to test these methods we apply them to configurations which have dual descriptions in which exact results are known.  Examples of such configurations concern eight dimensional gauge theories with sixteen supercharges which, on the heterotic side, arise from compactifications on a two-torus with Wilson lines. In the heterotic theory, the quantum corrections to the quartic gauge field strength couplings ($F^4$) are completely captured by a perturbative one-loop computation. 

On the dual type IIB side, where the eight dimensional gauge theory arises as the world volume theory on D7-branes,
these couplings receive both perturbative and non-perturbative corrections. The non-perturbative corrections are generated by D$(-1)$-instantons on top of the D7-branes, which can be viewed as a ``exotic" or ``stringy'' instantons. The reason is that the open strings between the D$(-1)$-instantons and the D7-branes have eight Neumann-Dirichlet directions. Therefore, the NS sector is massive and the only charged massless modes are the fermionic ones, in analogy with the discussion in chapter 3 concerning the case when an ED1-instanton and the spacefilling D5-branes were wrapping different two-cycles.

Such D$(-1)$-instanton corrections were explicitly computed in \cite{Billo:2009di}, up to instanton number five, using localization techniques \cite{Moore:1998et}. At instanton numbers higher than five the computations become technically very challenging and hence, it would be desirable to have alternative methods. We show  that the complete quantum corrections to this coupling, for all instanton numbers, are captured by a simple one-loop amplitude in the T-dual theory. This dual theory corresponds to a nine dimensional gauge theory, arising as the world volume theory of D8-branes in type IIA. By computing a one-loop amplitude with D0-particles in  this theory and performing a T-duality,  the D0-particles circulating the loop turn into multi-instanton corrections \cite{Gutperle:1999dx}. Moreover, it is possible to lift this configuration to the Horava-Witten picture in M-theory, in which the corresponding one-loop computation involves $E_8$ gauge bosons. 

The type IIA computation allows for a direct comparison with the dual heterotic worldsheet instanton computation, since both can be regarded as one-loop diagrams of BPS particles \cite{Harvey:1995fq}. This comparison allows us to revisit the proposal in \cite{Blumenhagen:2008ji} of the existence and nature of certain polyinstanton effects, which also exist in these 8d models. 
Our analysis suggests that these effects do not conflict with heterotic-type I duality. 
This chapter sets the stage for  \PSU\, where we apply these ideas and compute exact perturbative and non-perturbative quantum corrections to gauge $F^4$ couplings, gravitational $R^4$ couplings and mixed $R^2F^2$ couplings. We also discuss the case with general Wilson lines and moreover, how K3 compactifications turn the eight-dimensional conclusions drawn for the quartic gauge corrections into similar four-dimensional  conclusions for gauge kinetic functions in the resulting  $\cN=2$ models.

\section{Dualities in Eight Dimensions}

In this section, we will discuss quartic gauge couplings in eight dimensions from various dual perspectives. Let us start with the  heterotic SO(32) string theory compactified on a two-torus with Wilson lines. For later purposes, it is instructive to view this compactification as a two step procedure, where the ten-dimensional theory is first compactified on a circle with Wilson lines, breaking the gauge group to SO(16$)^2$. This nine-dimensional theory is then further compactified on another circle with Wilson lines, breaking the SO(16$)^2$ gauge group to SO(8$)^4$. In the eight dimensional gauge theory we find quartic invariants with the structure $\tr F^4$, $\tr (F^2)^2$ and also a Pfaffian quartic invariant Pf$\, F $. The the heterotic one-loop computation for these couplings  and is given by the following expression (up to an overall coefficient) \cite{Bachas:1997mc,Kiritsis:1997hf,Lerche:1998nx,Lerche:1998gz,Lerche:1998pf,Foerger:1998kw,Kiritsis:1999ss,Gutperle:1999xu,Kiritsis:2000zi,Gava:1999ky},
\begin{equation}
 \label{het}
 \begin{aligned}
& -6\,t_8\tr F^4\,
\log\left|\frac{\eta(4\Th)}{\eta(2\Th)}\right|^4 -\frac{3}{2}\,  t_8
(\tr F^2)^2 \log \left(\im\Th\, \im\Uh
\frac{|\eta(2\Th)|^8 |\eta(\Uh)|^4}{|\eta(4\Th)|^4}  \right)\\
& \hspace{60pt} -48\,t_8\Pf F\,\log \left|\frac{\eta(\Th +
1/2)}{\eta(\Th)}\right|^4
 \end{aligned}
\end{equation}
where $\Th$ and $\Uh$ are the K\"ahler modulus
and the complex structure of the 2-torus, respectively, $\eta$ is the Dedekind function and $t_8$ is the usual rank eight tensor (arising in various string amplitudes) contracting the spacetime indices, see e.g. \cite{Billo:2009di} for details. Since the $t_8$-tensor will always appear in the couplings we will in the following suppress it. By expanding (\ref{het}) we see that it has the following structure,  
\begin{eqnarray}
 \label{hetexp}
  &&\,-\,\tr F^4\,\Big\{2\pi i\Th  
    -12 \sum_{k=1}^\infty \sum_{\ell|k} \frac{1}{\ell} \big( \qh^{4k}- \qh^{2k}\big) +
    \mathrm{c.c.} \Big\}   \label{expheta}\nn \\ 
 && -\, (\tr F^2)^2\, \Big\{ \frac{3}{2}
    \log\left(\im\Th\, \im\Uh\, |\eta(\Uh)|^4\right) \nn \\
   && \phantom{-\, (\tr F^2)^2\, \Big\{}+ 3 \,\Big( \sum_{k=1}^\infty \sum_{\ell|k} \frac{1}{\ell}\big(\qh^{4k}-2  \qh^{2k}\big)+
    \mathrm{c.c.}  \Big)    \Big\}\label{exphetb}\nn \\
 && -192\,\Pf F \, \Big(\sum_{k=1}^\infty \sum_{\ell|2k-1} \frac{1}{\ell}\,
    \qh^{2k-1} + \mathrm{c.c}  \Big)\label{exphetc}
\end{eqnarray}
where  $ \qh = e^{2\pi i\, \Th}$. 

This eight dimensional SO$(8)^4$ heterotic theory with sixteen supercharges admits a dual description in terms of the type IIB theory compactified on a two-torus with an orientifold projection that flips the world sheet parity and inverts the two torus directions\footnote{There is also a factor $(-)^{F_L}$, where $F_L$ is the left moving fermion number, involved in the orientifold projection which is however not relevant for our purposes.}. 
The orientifold projection breaks half of the 32 background supercharges and defines four fixed points on the torus, corresponding to the location of  four O7-planes. For global tadpole cancellation we are, as usual, required to introduce 32 spacefilling D7-branes and moreover, for local tadpole cancelation we are required to place eight D7-branes on top of each O7-plane. This implies that the eight dimensional world volume theory is an SO$(8)^4$ gauge theory with sixteen supercharges. 

The duality map between the heterotic picture and the type IIB picture relates the
complexified K\"ahler modulus $\Th$  of the torus on the heterotic side and the
axio-dilaton field $\tau$ on the type IIB side, while the complex
structure remains the same. This implies that the $\qh$-terms for the three quartic invariants in (\ref{hetexp}) have the interpretation of D$(-1)$-instanton corrections  since the D($-1$)-instanton action appears in the exponent on the type IIB side. In addition, by looking at the structure of (\ref{hetexp}), one expects to find a tree level term for tr$\,F^4$ and a one-loop term for ($\tr F^2)^2$. These perturbative terms as well as the non-perturbative terms, up to D$(-1)$-instanton number five, was computed in \cite{Billo:2009di} and found to be in agreement with the heterotic results. 

In this chapter, we are interested in additional dual descriptions in which other methods can be used to compute the  quantum correction to the quartic gauge coupling. For example, by T-dualizing along one of the two-torus directions in the type IIB picture we obtain the IIA theory compactified on a circle with an orientifold projection that flips the world sheet parity and inverts the circle direction \cite{Polchinski:1995df}. Such a projection breaks half of the 32 background supercharges and defines two fixed points on the circle, corresponding to the location of two O8-planes. For local tadpole cancellation, in this case we are required to place 16 D8-branes on each of the two fixed points. In this way we obtain a nine dimensional $SO(16)^2$ gauge theory with sixteen supercharges. In the reverse sense, the type IIB picture is recovered by compactifying one of the world volume directions along the O8/D8-branes and choose Wilson lines such that the D7-branes are symmetrically distributed in the transverse two-torus in the T-dual theory. Moreover, since D($-1$)-instantons are related to D0-particles via T-duality one expects that the non-perturbative corrections arising from D($-1$)-instantons on the type IIB side should be captured by a one-loop computation with D0-particles in the world volume theory of the O8/D8-branes compactified on a circle \cite{Gutperle:1999dx}. In addition to the BPS D0-particles, we also expect perturbative open string BPS states stretching between the D8-branes. If the endpoints of the strings are attached to the same stack of D8-branes, the corresponding states have integer winding charge in the $({\bf 120,1})+({\bf 1,120})$ representation of the $SO(16)^2$ gauge group. Instead, for strings that stretch between the two stacks, the corresponding states have half-integer winding charge in the $({\bf 16},{\bf 16})$ representation. 

In order to understand the charges and representations of the D0-particles it is instructive to lift the system to M-theory. One way to obtain the M-theory picture is to start from the fact that the original heterotic SO(32) theory admits a T-dual description in terms of the heterotic $E_8\times E_8$ theory. The strong coupling limit of the heterotic $E_8\times E_8$ theory is described in terms of M-theory compactified on an $S^1_{10}/\mathbb{Z}_2$-interval (where we have chosen the interval to lie in the tenth direction). This Horava-Witten background \cite{Horava:1995qa} can be seen as an 11-dimensional theory with two ten dimensional boundaries (along $x^1,x^2,\cdots,x^9$ and $x^{11}$), corresponding to the endpoints of the $S^1_{10}/\mathbb{Z}_2$-interval, at each of which an $E_8$ gauge multiplet lives and where half of the 32 bulk supersymmetries are preserved. 

In order to relate this picture to the type IIA picture we focus on one of these ten dimensional boundary $E_8$ gauge theories. We compactify the eleventh direction on a circle $S^1_{11}$ and obtain a nine-dimensional theory with an infinite KK tower of $E_8$ bosons. By turning on Wilson lines along $S^1_{11}$ that break $E_8\to SO(16)$, the KK tower splits into two towers: One for states in the {\bf 120} with even ($2n$) KK momenta and one for the states in the {\bf 128} with odd ($2n-1$) KK momenta.  In the $2n$-tower, the lowest (massless) states are identified with the gauge bosons of the SO(16) gauge group, while the higher levels are identified with D0-particles with R-R charge $2n$. In the $(2n-1)$-tower all levels are identified with D0-particles with R-R charge $(2n-1)$. 

In order to understand this connection to the type IIA picture, let us study how one of the two SO(16) gauge groups in type IIA picture is enhanced to $E_8$.  Consider a single D0-particle on top of the stack of O8/D8-branes. Since a single (fractional) D0-particle does not have an orientifold image, it is stuck at this point. Such a particle has fermionic zero modes in the {\bf 16} of the SO(16) gauge group, arising from the massless modes of the open strings between the D0-particle and the D8-branes. Quantization of these zero modes implies that the D0-particle transform in the {\bf 128} of the gauge group (the second spinor representation is projected out by the residual $\mathbb{Z}_2$ gauge symmetry of the D0-particle) \cite{Danielsson:1996es,Kachru:1996nd}. We therefore expect that the $(2n-1)$ KK tower of gauge bosons in the M-theory picture corresponds to D0-particles with R-R charge $(2n-1)$ in the IIA picture. In the strong coupling limit, or equivalently, the large $S^1_{11}$-radius limit,  the D0-particles in the {\bf 128} become light and combine with the gauge particles and the light D0-particles in the {\bf 120}  to form the {\bf 248} carried by the $E_8$ gauge bosons.

Hence, since we know the charges and representations for all BPS states in the nine dimensional $SO(16)^2$ theory on the O8/D8-branes we have all the input we need to compute the one-loop amplitudes.

\section{One-Loop Diagrams}
\label{generalities}

We are interested in computing the one-loop diagram of BPS particles in the $\mathbb{R}^8 \times S^1_9$ world volume of a stack of O8/D8-branes in the type IIA picture, with four external insertions of gauge fields strengths. We begin by considering the general structure of a one-loop contribution from a BPS particle with mass $\mu$ in $\mathbb{R}^d \times S^1_9$ with $k$ insertions (where we have for later convenience chosen the circle to be in the 9th direction and have radius $R_9$). The one-loop diagram is given by the following expression,
\begin{equation}
\label{amp}
 \int_{0}^{\infty} \frac{dt}{t}\,t^k \,\sum_{\ell_9} \int d^{d}\mathbf{p} ~ e^{-\pi t \left( \mathbf{p}^2+\frac{(\ell_9-\tilde{a})^2}{R^{2}_{9}} +\mu^2 \right)} \Tr_{\mathbf{ R}} F^k
\end{equation}
where the sum is over the KK-momenta $\ell_9$ along $S^1_9$ and $\mu$ is the 9d mass of the BPS particle. In (\ref{amp}), $\tilde{a}=a+A$, where $A$ is given by ${\bf w \cdot }{\bf A}_{9}$, where ${\bf w}$ is the weight vector of the corresponding $SO(16)$ representation of the particle and ${\bf A}_{9}$ is the Wilson line for the $SO(16)$ gauge group along $S^1_9$. For a D0-particle state, $a=c$, where $c$ is the   R-R one-form along the circle while for an open string winding state, $a=b$, where $b$ corresponds to the NS-NS two-form.  Moreover, for a bound state of $n$ D0-particles the mass is given by $\mu=n/g_s$ while for a winding state the mass is $\mu=w_{10}R_{10}$, where $w_{10}$ is the winding charge and $R_{10}$ is the radius of the tenth direction transverse to the O8/D8-branes. For BPS protected amplitudes, the vertex operator part can simply be replaced by a factor $t^4$ and insertions encoding the coupling to the external gauge bosons. In this case, the insertion is $\tr_{\mathbf{R}} F^4$ for a BPS particle in a representation ${\mathbf{R}}$ of the gauge group.

By integrating over the continuous momenta and performing a Poisson resummation on $\ell_9$ the prefactor of (\ref{amp}) becomes
\begin{eqnarray}
\label{generalamp}
\mathcal{A}&=&R_9 \sum_{w_9}  \int_{0}^{\infty} \frac{dt}{t} ~t^{k-(d+1)/2} ~e^{-\frac{\pi R_9^2 w_9^2}{t}-\pi t \mu^2 }e^{-2\pi i w_9 \tilde{c}}~\nn \\
&=& \mathcal{A}_{w_9=0}+  \mathcal{A}_{w_9\neq 0}
\end{eqnarray}
 The zero winding part is given by
\begin{equation}
\label{gral2}
\mathcal{A}_{w_9=0} = R_9   \int \frac{dt}{t} ~t^{k-(d+1)/2} ~e^{-\pi t \mu^2 } 
\end{equation}
and a non-zero winding part by
\begin{equation}
\label{ }
\mathcal{A}_{w_9\neq 0}= R_9 \sum_{w_9 \neq 0}  \int_{0}^{\infty} \frac{dt}{t} ~t^{k-(d+1)/2} ~e^{-\frac{\pi R_9^2 w_9^2}{t}-\pi t \mu^2 }e^{-2\pi i w_9 \tilde{c}}.
\end{equation}
where winding in this case refers to the winding of the worldline of the BPS-particle  along $S^1_9$.

We will now focus on the case when the BPS particle is a D0-particles with mass $\mu=n/g_s$ ($n\neq0$) and RR charge $c=nc_0$.  For such a particle, the non-zero winding contribution becomes
\begin{eqnarray}
\label{BesselGen}
 \mathcal{A}_{w_9\neq 0}^{\mathrm{D0}}&=&   2R_9 \sum_{w_9 \neq 0}  \left| \frac{n}{ R_9 g_s w_9} \right|^{(d+1)/2-k} \nn \\
 &&\times \,K_{(d+1)/2-k} \left( \left| \frac{2\pi R w_9 n}{g_s} \right| \right) e^{-2\pi i w_9 n c_0}e^{-2\pi i w_9 A}
\end{eqnarray}
 where we have used the following relation for modified Bessel functions,
 \begin{equation}
\label{ }
\int_{0}^{\infty} \frac{dt}{t}t^{-s} e^{-\frac{A}{t}-tB}=2\left| \frac{B}{A} \right|^{s/2} K_s \left( 2\sqrt{|AB|} \right).
\end{equation}
If we now specify to the case $d=8$, $k=4$ we see that this non-zero winding contribution takes the form of an instanton expansion,
\begin{eqnarray}
\label{InstGen}
\mathcal{A}_{w_9\neq 0 }^{\mathrm{D0}}&=&2R_9 \sum_{w_9\neq0}  \left| \frac{n}{ R_9 g_s w_9} \right|^{1/2} K_{1/2} \left(  \frac{2\pi R |w_9 n |}{g_s}  \right) e^{-2\pi i w_9 n c}e^{-2\pi i w_9 A}\nn \\
&=& \sum_{w_9 > 0} \frac{1}{w_9} \, q^{w_9\,n}\, e^{2\pi i \, w_9A}  + \mathrm{c.c.} 
\end{eqnarray}
where we have used $K_{1/2}(x)=\sqrt{\frac{\pi}{2x}}e^{-x}$ and where $\tau= i \frac{R_{9}}{g_s}+c_0$. We now see that by integrating out the non-zero winding contribution of a D0-particle with charge $n$ we obtain an instanton expansion in the eight dimensional theory. In order to take into account D0-particles with all charges in all representations of $SO(16)$, we simply sum over all $n$ and over all weight vectors ${\bf w}$. 

It is interesting to note that 
for our case (when $(d+1)/2-k=1/2$ in (\ref{BesselGen})) there are no perturbative corrections around the instanton since the first term in the series is exact. In contrast, if we were to calculate for example the contribution from a D0-particle to the $R^4$-term in $\mathbb{R}^9 \times S^1_9$, in order to reproduce the $R^4$ term in ten dimensional type IIB \cite{Green:1997tv}, we should instead insert $k=4$ Riemann tensors and set $A=0$. For this case, since $(d+1)/2-k=(9+1)/2-4=1$, we would instead get the Bessel function $K_1(x)=\sqrt{\frac{\pi}{2x}}e^{-x}\left(1+\mathcal{O}(1/x)+\cdots \right)$ which has an infinite series of perturbative corrections.

 \section{The Type IIA Picture}

\subsubsection{The $SO(16)$ model}
\label{sosixteen}

Consider first the simple situation where there are no Wilson lines on the circle $S^1_9$, so that the eigth-dimensional gauge group is $SO(16)^2$. The non-perturbative  corrections are easily computed from the D0-brane one-loop diagrams, using the BPS bound state spectrum information: For each stack of O8/D8-branes, there exists a D0-brane BPS bound state of mass $|2n|/g_s$, in the representation ${\bf{120}}$ of the corresponding $SO(16)$, and one state of mass $|2n-1|/g_s$ in the ${\bf{128}}$, for each non-zero integer $n$. Focusing on a single $SO(16)$, the contribution is
\begin{eqnarray}
\label{so16nonpert}
\Delta^{D0}_{SO(16)}&= &2 \sum_{w_9,n>0}  \frac{1}{w_9}\, q^{w_9(2n)} \tr_{\mathbf{120}} F^{4} +2  \sum_{w_9,n>0}  \frac{1}{w_9}\, q^{w_9(2n-1)} \tr_{\mathbf{ 128}} F^{4}  + \mathrm{c.c.} \nn \\
& = & 
8 \, \tr F^{4} \,\sum_{k>0}\sum_{\ell \large| k} \frac{1}{\ell}\, \left[ 2 q^{2k} - q^k \right]\, 
\nn \\
&&+\,  3\,( \tr F^{2})^2 \, \sum_{k>0}\sum_{\ell \large| k} \frac{1}{\ell} \,\left[ - q^{2k} + 2 q^k \right] \, +\mathrm{c.c.} 
\end{eqnarray}
where we have rewritten the expression (\ref{so16nonpert}) in terms of traces in the vector representation by using the following trace identities \cite{Erler:1993zy}, 
\begin{eqnarray}
\label{traceso16}
&& \quad \tr_{\mathbf{ 120}} F^{4}_{SO(16)}= 8 \tr F^{4}_{SO(16)}  +3  ( \tr F^{2}_{SO(16)})^2 \nn  \\ 
&& \quad \tr_{\mathbf{ 128}} F^{4}_{SO(16)}= - 8 \tr F^{4}_{SO(16)}  +6  ( \tr F^{2}_{SO(16)})^2 \,.
\end{eqnarray}
 The result (\ref{so16nonpert}) agrees with the result in \cite{Gutperle:1999dx} for the $SO(16)^2$ case on the heterotic side. 

\subsubsection{The $SO(8)^2$ Model}
\label{soeight}

Let us now consider turning on $\IZ_2$ valued Wilson lines which break each nine-dimensional $SO(16)$ factor to $SO(8)^2$.  The 9d BPS states  in $SO(16)$ representations pick up different phases according to their behaviour under the decomposition
\begin{eqnarray}
&&  \mathbf{ 120}  \to  (\mathbf{ 28},\mathbf{ 1})_{+}+(\mathbf{ 1},\mathbf{ 28})_{+}+(\mathbf{ 8_v} ,\mathbf{ 8_v})_{-} \nn\\
&& \mathbf{ 128} \to (\mathbf{ 8_s},\mathbf{ 8_s})_{+}+(\mathbf{ 8_c},\mathbf{ 8_c})_{-} 
\label{SO(8)reps}
\end{eqnarray}
where the subindex $\pm$ corresponds to having $e^{2\pi i\,A}=\pm 1$ in (\ref{InstGen}). 

We focus on $F^4$ corrections associated to only one of the $SO(8)$ factors, in which case only states charged under this factor contribute. The result is 
\beqa
\label{aha}
\Delta_{SO(8)}^{D0} &=& 2\,\Bigl[\sum_{n,w_9>0}\, \frac{1}{w_9}\, q^{2n\, w_9}\, \tr_{\mathbf{28}}\,F^4\, +\, 8\, \sum_{n,w_9>0}\, \frac{1}{w_9}\, q^{2n\, w_9}\, (-1)^{w_9}\, \tr_{\mathbf{8_v}}\,F^4\, \nonumber \\
&&+\,8\, \sum_{n,w_9>0}\, \frac{1}{w_9}\, q^{(2n-1)\, w_9}\, \tr_{\mathbf{8_s}}\,F^4\, \nn \\
&&+\, 
8\, \sum_{n,w_9>0}\, \frac{1}{w_9}\, q^{(2n-1)\, w_9}\,(-1)^{w_9}\,  \tr_{\mathbf{8_c}}\,F^4\Bigr]+\mathrm{c.c.} 
\eeqa
Using the trace identities, 
\begin{eqnarray}
\label{traceso8}
&& \quad \tr_{\mathbf{8_s}} F_{SO(8)}^2\, =\,  \tr_{\mathbf{8_c}} F_{SO(8)}^2\, =\, \tr F^2 \nn \\
&& \quad \tr_{\mathbf{28}} F^{4}_{SO(8)}\, = \, 3\, (\tr F_{SO(8)}^2)^2 \nn \\
&& \quad \tr_{\mathbf{8_s}} F^{4}_{SO(8)}\, = \, -\frac 12\, \tr F_{SO(8)}^4\, +\,\frac 38\, (\tr F_{SO(8)}^2)^2\, -\, 12 \Pf F \nn \\
&& \quad \tr_{\mathbf{8_c}} F^{4}_{SO(8)}\, = \, -\frac 12\, \tr F_{SO(8)}^4\, +\,\frac 38\, (\tr F_{SO(8)}^2)^2\, +\, 12 \Pf F \nn \\
\end{eqnarray}
we obtain
\beqa
\Delta_{SO(8)}^{D0} 
&=& 12 \, \tr F^4\,  \sum_{k>0}\sum_{\ell \large| k} \, \frac{1}{\ell}\, \left[ \, q^{4k}\, -\, q^{2k}\, \right]\, \nn \\
&&-\, 3\, (\tr F^2)^2\, \sum_{k>0}\sum_{\ell \large| k} \, \frac{1}{\ell} \, \left[ \, q^{4k}\, -\, 2q^{2k}\, \right]\,  \nn\\
&&- 192\, \Pf F\, \sum_{k>0}\sum_{\ell \large| 2k-1}\, \frac{1}{\ell}\, q^{2k-1}+\mathrm{c.c.}
\eeqa
in complete agreement with (\ref{hetexp}).

\subsubsection{Perturbative Contributions}

Perturbative corrections to the gauge couplings  arise in two different ways. First of all, bound states of D0 branes with zero winding in the circle $S_9^1$ yield a tree-level contribution which can be easily obtained from \eqref{gral2}, 
\begin{eqnarray}
\label{tree-level}
\Delta^{\mathrm{Pert}}_{w_9=0}&=&R_9\sum_{n\in \mathbb{Z}}  \int \frac{dt}{t} ~t^{-\frac{1}{2}} \left[e^{-\pi t \frac{(2n)^2}{g_s^2}}\,\tr_{\mathbf{120}} F^4 +
e^{-\pi t \frac{(2n-1)^2}{g_s^2}}\,\tr_{\mathbf{128}} F^4 \right] \nn \\
&=& \frac{\pi\, \tau_2}{3} \left[ 2\,\tr_{\mathbf{120}}F^4 - \tr_{\mathbf{128}}F^4\right]~
\end{eqnarray}
where we have Poisson resummed and performed the integral.
Using the trace identities in (\ref{traceso8}), we obtain the following contribution for the $SO(8)^4$ model:
\begin{eqnarray}
&\Delta_{SO(8)}^{\mathrm{Pert}}&=
-2\pi i \tau \,\tr \,F^4+ \mathrm{c.c.}
\end{eqnarray}
in agreement with (\ref{hetexp}).

The spectrum of BPS particles in the type IIA picture also contains perturbative states, corresponding to open strings winding in the interval $S^1_{10}/\IZ_2$. Since we know their winding charges and gauge group representations, they can be integrated out in the same way as as the D0-particle states.   
Their contribution is calculated  in \PSU~ and the result corresponds to perturbative one-loop correction for the $(\tr F^2)^2$, in agreement with (\ref{hetexp}). Hence, we have reproduced the complete heterotic result in (\ref{hetexp}) by integrating out the entire BPS spectrum in the type IIA picture.

\section{The M-Theory Picture}
\label{mtheory}

The non-perturbative contributions in the type IIA picture are due to D0-particles. These states admit an interpretation in the Horava-Witten M-theory lift in terms of momentum modes of the $E_8$ vector multiplets on the boundaries. The non-perturbative contribution should therefore admit a simple description as a one-loop diagram of massless $E_8$ gauge particles in the ten-dimensional boundary compactified on two-torus down to eight dimensions. This description makes the modular properties of the result manifest. It also allows for generalizations of the computation in models with general Wilson lines, not necessarily related to the perturbative type IIA picture.

\subsubsection{The One-Loop Diagram with $E_8$-Bosons}
\label{mtheoryoneloop}

Consider Horava-Witten theory on a two-torus in the $x^9$ and $x^{11}$ directions, i.e. M-theory on $\mathbb{R}^8 \times S^{1}_{(9)}\times (S^{1}_{(10)} / \mathbb{Z}_2) \times S^{1}_{(11)}$. We are interested in one-loop amplitudes  with four external insertions of gauge field strengths.
We begin by studying the massless $E_8$ gauge bosons which live at  ten-dimensional boundaries of the $S^{1}_{(10)} / \mathbb{Z}_2$ interval. By summing over the KK-momenta they carry in the $ T^{2}_{(9,11)} = S^{1}_{9} \times  S^{1}_{11} $ directions and taking into account the Wilson lines along the $ T^{2}_{(9,11)}$ we arrive at the following expression for the one-loop amplitude,
\begin{eqnarray}
\label{Mgauge}
\mathcal{A}^{HW}  &=&
\int_{0}^{\infty} \frac{dt}{t} t^4 \sum_{\ell_I} \int d^{8}\mathbf{p} ~ e^{-\pi t \left( \mathbf{p}^2+G^{IJ} \widetilde{\ell}_{I} \widetilde{\ell}_{J}\right)} \nn\\ 
&=&
 \int_{0}^{\infty} \frac{dt}{t} \sum_{\ell_9 , \ell_{11}}~ e^{-\pi t  \frac{1}{V_{(2) }\tau_2} | \widetilde{\ell}_{9} -\tau\widetilde{\ell}_{11} |^2}
\end{eqnarray}
where we have integrated over the continuous momenta and written out the metric on the $ T^{2}_{(9,11)}$ in terms of its compex structure $\tau$ and volume $V_{(2)}$. Moreover, we have denoted 
\begin{equation}
\label{ }
\widetilde{\ell}_{I} =
\begin{pmatrix}
  \widetilde{\ell}_{9}     \\
  \widetilde{\ell}_{11}   
\end{pmatrix}
=
\begin{pmatrix}
 \ell_9 - \mathbf{\Lambda}\cdot \mathbf{A}_9     \\
 \ell_{11}   - \mathbf{\Lambda} \cdot \mathbf{A}_{11}
 \end{pmatrix}
 \end{equation}
where $ \mathbf{\Lambda}$ is one of the weight vectors of the adjoint ($\mathbf{248}$) representation of $E_8$ and $\mathbf{A}_{I}$ denotes the Wilson lines along the $I=9,11$ directions of the $T^{2}_{(9,11)}$. In order to get the total contribution from all $E_8$ gauge bosons we simply sum over all 248 weight vectors. The action of the modular group is manifest in this expression, so the invariance group of the result is the subgroup of $SL(2,\IZ)$ preserving the Wilson line structure. 

In order to recover the type IIA picture we should choose Wilson lines along the eleventh direction such that the $E_8$ gauge group is broken to $SO(16)$. With this choice of Wilson lines, the ${\mathbf{248}}\to{\bf {120}}+{\bf {128}}$, corresponding to $\mathbf{\Lambda\cdot A}_{11}$ being in $\IZ$ or $\IZ+\frac 12$. Since the KK momenta $\ell_{11}$ corresponds to the charge of the D0-particle in the type IIA picture, the shift generated by $\mathbf{\Lambda\cdot A}_{11}$ corresponds to the split into two KK-towers with either momenta $2n$ in the {\bf 120} or with momenta $(2n-1)$ in the {\bf 128}. Hence, upon Poisson resummation of the KK momenta $\ell_9$, we obtain (\ref{so16nonpert}) for non-zero winding $w_9$. 

Furthermore, in order to recover the $SO(8)$ case we also turn on Wilson lines along the ninth direction such that  $SO(16)\to SO(8)^2$. With such additional Wilson lines the {\bf {120}} and {\bf {128}} decomposes as shown in (\ref{SO(8)reps}) where the $\pm$ corresponds to $\mathbf{\Lambda\cdot A}_{9}$ being either $\IZ$ or $\IZ+\frac 12$. Hence, the factor $A$ in (\ref{InstGen}) is  given by $\mathbf{\Lambda\cdot A}_{9}$ and the sum over the representations, for example in (\ref{aha}), corresponds to the sum over the weight vectors that are relevant for the particular $SO(8)$ factor under consideration. 

From the one-loop diagram with $E_8$ gauge bosons in the M-theory picture with this choice of Wilson lines, we recover the complete contributions from D0-particles in the type IIA picture. 
As a generalization one could choose more general Wilson lines $\mathbf{A}_9$ in order to end up with a type IIB picture where the D7-branes are not located in the most symmetric way. Also, by allowing for general Wilson lines $\mathbf{A}_{11}$ one ends up with D8-branes at general positions in the transverse interval. Thus, the position of D7-branes and the D8-branes are naturally geometrized in terms of the Wilson lines $\mathbf{A}_9$ and $\mathbf{A}_{11}$ in the M-theory picture.

\section{Polyinstantons}
\label{polyinstanton}

In the previous sections we saw that the contribution from a single  BPS D0-brane can correspond to a multi-instanton contribution on the type IIB side. This is particularly manifest for BPS D0-brane bound states of $k$ elementary D0-branes. 
In a four-dimensional setup \cite{Blumenhagen:2008ji}, the authors considered a different kind of multiple instanton effect, dubbed polyinstanton, which also has an eigth-dimensional analog in our setup.
The polyinstantons in \cite{Blumenhagen:2008ji} were claimed to violate heterotic-type I duality. 
In this section we address this puzzle for eigth-dimensional polyinstantons, shedding light from a new perspective, valid also in the four-dimensional setup. The bottom line is that polyinstanton processes can be interpreted as reducible Feynman diagrams which do not contribute to the microscopic one-particle-irreducible (1PI) action.

\begin{figure}[!htp]
\centering
\includegraphics[scale=0.43]{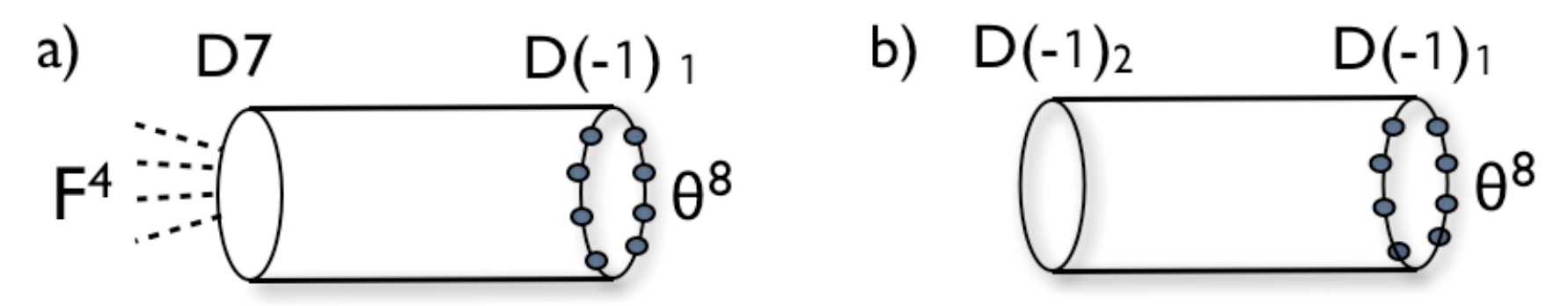}
\caption{\small a) D$(-1)$-instanton correction to the $F^4$ coupling on a D7-brane. b) A related diagram describes a D$(-1)$-instanton correction to the action of a second D$(-1)$-instanton.}
\label{poly}
\end{figure}

Let us start by introducing the eight-dimensional polyinstanton corrections to e.g. $F^4$, in complete analogy with the $F^2$ corrections \cite{Blumenhagen:2008ji}. The microscopic diagram leading to a D7-brane $F^4$ correction from a D$(-1)$-instanton  includes a cylinder diagram with a boundary on the D7-brane (with four fields strength insertions) and a boundary on the D$(-1)$ (with eight fermion zero mode insertions saturating the instanton Goldstinos), see figure \ref{poly}a. As shown in figure \ref{poly}b, there is a similar diagram, with the D7-brane replaced by a second D$(-1)$-brane instanton, and with no insertions on the correspoding boundary. Labeling the two instantons $1,2$ to avoid confusion,
this diagram represents the correction from D$(-1)_1$ to the action of D$(-1)_2$. Considering now an $F^4$ term induced by D$(-1)_2$, the inclusion of this correction would naively lead to a contribution to the eight-dimensional effective action schematically of the form
\beqa
 \int d^8x\, \tr F^4\, e^{-(S_2+e^{-S_1})}\, =\, \int d^8x\, \tr F^4\,  \sum_{n=0}^\infty \, \frac{1}{n!}\, e^{-S_2}\, (e^{-S_1})^n
\label{naive}
\eeqa
where we interpret the contribution from the D$(-1)_1$-instanton as a non-perturbative correction to the D$(-1)_2$-instanton action. Microscopically, the $n^{th}$ term corresponds to a polyinstanton process with one D$(-1)_2$-instanton and $n$ independent D$(-1)_1$-instantons. The zero modes of the latter are saturated through D$(-1)_1$-D$(-1)_2$ cyclinders as in Figure \ref{poly}b.
The result involves an integration over the relative positions of the instantons in the eigth-dimensional space, just like in the four-dimensional case \cite{Blumenhagen:2008ji}. The contribution is therefore in principle not localized on coincident instantons, as opposed to the multi-instantons studied in \cite{GarciaEtxebarria:2007zv}. In particular, since the polyinstantons in general sit at different locations in the internal space, the saturation of fermion zero modes can take place independently of the distances between the instantons.

\subsubsection{Polyinstantons and heterotic-type II orientifold duality}
\label{polyduality}

The eigth-dimensional corrections arising from  D$(-1)$-instantons correspond under T-duality to one-loop diagrams of nine-dimensional BPS D0-particles. These are directly translated to one-loop diagrams of nine-dimensional BPS states in the heterotic dual, reproducing the genus one worldsheet instanton contributions. This contribution in principle includes certain D-brane multi-instantons, namely those T-dual to nine-dimensional particles which form BPS bound states at threshold, and whose hallmark is that their contribution is localized on configurations of coincident instantons. 

Polyinstanton processes instead involve instantons whose T-dual particles do not combine into nine-dimensional  one-particle BPS bound states. This is manifest as in general the individual instantons sit at different points in the internal space, and this separation can persist in the dual type IIA picture, e.g. when they map to D0-branes on different $SO(16)$ boundaries. Therefore they are manifestly not included in the one-loop diagram of one-particle BPS states, and hence in the heterotic genus one worldsheet contribution.

There is a clear way out of this potential clash with duality. The heterotic genus one worldsheet diagram (and so the type IIA one-loop diagram) computes the one-loop correction to the 1PI action. Namely it includes the effects of massless states (and is hence non-holomorphic) but does not include {\rm reducible} contributions. These can be later generated by computing tree level diagrams using the effective vertices of the 1PI action. We will now argue that  polyinstanton effects actually correspond to such reducible diagrams, and hence do not contribute to the 1PI action, restoring agreement between type II orientifolds and their heterotic duals.

\subsubsection{Polyinstantons in spacetime as reducible diagrams}
\label{poly1pi}

\begin{figure}[!htp]
\centering
\includegraphics[scale=0.55]{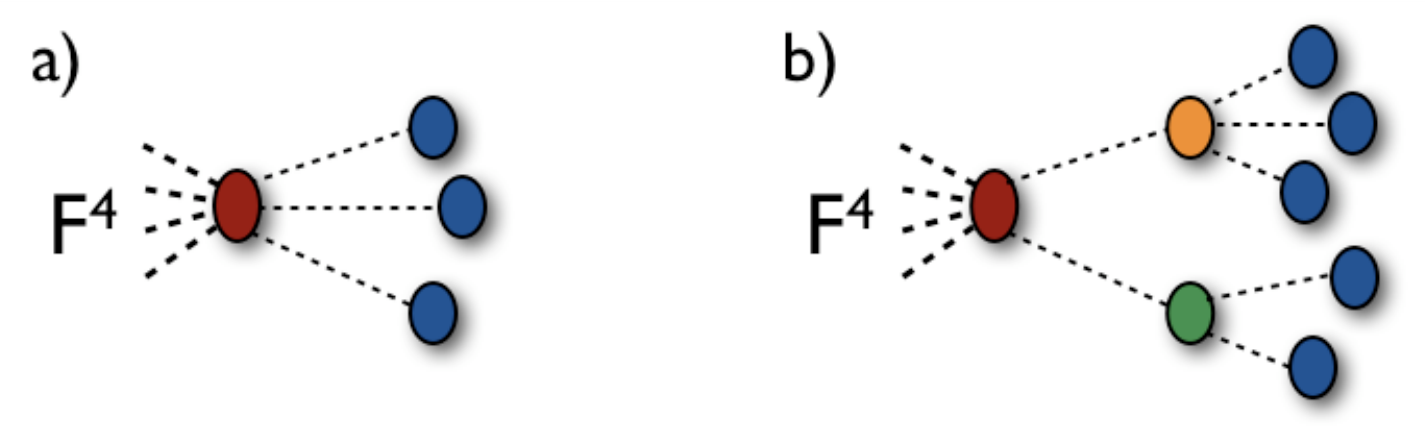}
\caption{\small Polyinstanton processes as reducible spacetime Feynman diagrams. Colored blobs denote elementary instanton interactions in the 1PI action, joined by propagating closed string modes. Summing over polyinstanton processes like a), with arbitrary numbers of blue blobs, reproduces an effective exponential correction to the action of the red blob instanton. Figure b) shows richer polyinstanton processes, involving the elementary interactions described by (\ref{interaction1}) and (\ref{interaction2}).}
\label{reducible}
\end{figure}

In type IIB, the polyinstantons correspond to individual D$(-1)$-instantons joined by cylinder diagrams. As the instantons are in general placed at  different locations, it is natural to interpret the cylinders as a tree level closed string exchange, and the corresponding processes as reducible. Thus, polyinstantons do not induce new terms in the microscopic 1PI action, but are rather generated by other elementary effective vertices in the 1PI action. The picture is particularly clear in the type IIA picture, where the polyinstanton is given by a Feynman diagram with a loop of BPS particles with four field strength insertions, joined by  closed string propagators to other loops of BPS particles (which can be subsequently joined to other propagators and loops), see Figure \ref{reducible}. 

The exponential combinatorics of polyinstantons in (\ref{naive})  is simply the combinatorics of spacetime Feynman diagrams with two basic kinds of interaction diagrams, see Figure \ref{reducible}a. For massless closed string states, these can be explicitly obtained in the factorization limit. One basic interaction vertex corresponds to an instanton coupling to $F^4$ with emission of $n$ massless closed string states. It simply follows from expanding the $F^4$ instanton corrections in the fluctuations of the  dynamical modulus controling the instanton action, in our case $\tau$. Expanding it into a vev plus fluctuation, $\tau_0+\tau$,  each $F^4$ instanton correction produces terms of the following form,
\beqa
q^N \, \tr F^4\; \to \; \sum_{n=0}^\infty \frac{(2\pi i N)^n}{n!}\, q_0^N \,\tr F^4\, \tau^n~.
\label{interaction1}
\eeqa
The second kind of vertex is the emission of a massless closed string from an instanton. 
Although unfamiliar, this contribution indeed exists, as follows. The fluctuation $\tau$ is a complex scalar belonging to a multiplet whose on-shell structure has the form
\beqa
\Phi_\tau\, =\, \tau \, \, +\, \ldots\, + \theta^8\, \partial^4 {\ov \tau}~.
\eeqa
This follows from the orientifold truncation of the chiral on-shell superfield $\Phi$, satisfying ${\bar D}\Phi=0$, $D^4\Phi={\bar D}^4{\bar\Phi}=0$, in \cite{Green:1998by}. This gives the supersymmetric completion of the instanton action, and the last term
corresponds to an interaction saturating all the instanton goldstino zero modes with one insertion of ${\ov\tau}$. Therefore one generates the couplings
\beqa
\int \, d^8x\,d^8\theta\, e^{2\pi iN(\tau_0+\Phi_\tau)}\, \to\, 
\int \, d^8x\,\, q_0^N\, \, \partial^4{\ov\tau}\, +\, \cdots~.
\label{interaction2}
\eeqa
The term we required has been separated out explicitly. The other terms can be used to construct more involved diagrams, as in Figure \ref{reducible}b.
In addition to the instanton generated couplings with a massless closed string field in (\ref{interaction1}) and (\ref{interaction2}), the 1PI action also contains couplings to massive closed string fields. Polyinstanton effects correspond to reducible diagrams with massive particle exchange. However, since such diagrams are reducible they do not contribute to the microscopic 1PI action and hence, polyinstanton effects do not appear in the effective action we map to the heterotic side.  

In \PSU\,, by computing one-loop amplitudes with (bulk) D0-particles that do not carry any representation under the gauge groups, we obtain exact quantum corrections also to the gravitational $R^4$ coupling and mixed $R^2F^2$ coupling. Moreover, we compactify our eight-dimensional type IIA picture on a K3 manifold and discuss four-dimensional $\cN=2$  gauge kinetic functions. We also consider general Wilson lines and discuss the modular properties of the one-loop amplitudes in the M-theory picture.

\clearpage
\pagestyle{plain}
\def\href#1#2{#2}

\begin{thebibliography}{99}

\bibitem{Argurio:2007vqa}
  R.~Argurio, M.~Bertolini, G.~Ferretti, A.~Lerda and C.~Petersson,
  ``Stringy instantons at orbifold singularities,''
  JHEP {\bf 0706} (2007) 067
  [arXiv:0704.0262 [hep-th]].
  
\bibitem{Petersson:2007sc}
  C.~Petersson,
  ``Superpotentials From Stringy Instantons Without Orientifolds,''
  JHEP {\bf 0805} (2008) 078
  [arXiv:0711.1837 [hep-th]].
  
\bibitem{Argurio:2008jm}
  R.~Argurio, G.~Ferretti and C.~Petersson,
  ``Instantons and Toric Quiver Gauge Theories,''
  JHEP {\bf 0807} (2008) 123
  [arXiv:0803.2041 [hep-th]].
  
\bibitem{Ferretti:2009tz}
  G.~Ferretti and C.~Petersson,
  ``Non-Perturbative Effects on a Fractional D3-Brane,''
  JHEP {\bf 0903} (2009) 040
  [arXiv:0901.1182 [hep-th]].
  
\bibitem{Petersson:2009zz}
  C.~Petersson,
  Nucl.\ Phys.\ Proc.\ Suppl.\  {\bf 192-193} (2009) 169.
  
\bibitem{Petersson:2010qu}
  C.~Petersson, P.~Soler and A.~M.~Uranga,
  ``D-instanton and polyinstanton effects from type I' D0-brane loops,''
  JHEP {\bf 1006} (2010) 089
  [arXiv:1001.3390 [hep-th]].
  
\bibitem{Argurio:2006my}
  R.~Argurio, G.~Ferretti and C.~Petersson,
  ``Massless fermionic bound states and the gauge/gravity correspondence,''
  JHEP {\bf 0603} (2006) 043
  [hep-th/0601180].
  
\bibitem{Gran:2008vi}
  U.~Gran, B.~E.~W.~Nilsson and C.~Petersson,
  ``On relating multiple M2 and D2-branes,''
  JHEP {\bf 0810} (2008) 067
  [arXiv:0804.1784 [hep-th]].

\bibitem{Glashow:1961tr}
  S.~L.~Glashow,
  ``Partial Symmetries Of Weak Interactions,''
  Nucl.\ Phys.\  {\bf 22} (1961) 579.
  
\bibitem{Weinberg:1967tq}
  S.~Weinberg,
  ``A Model Of Leptons,''
  Phys.\ Rev.\ Lett.\  {\bf 19} (1967) 1264.
  
\bibitem{Salam}
  A.~Salam,
  Originally printed in ``Svartholm: Elementary Particle Theory, Proceedings of the Nobel Symposium held 1968 at Lerum, Sweden'', 367-377.
  
\bibitem{Weinberg:1995mt}
  S.~Weinberg,
  ``The Quantum theory of fields. Vol. 1: Foundations,''
{\it  Cambridge, UK: Univ. Pr. (1995) 609 p}

\bibitem{Weinberg:1996kr}
  S.~Weinberg,
  ``The quantum theory of fields. Vol. 2: Modern applications,''
{\it  Cambridge, UK: Univ. Pr. (1996) 489 p}

\bibitem{Peskin:1995ev}
  M.~E.~Peskin and D.~V.~Schroeder,
  ``An Introduction To Quantum Field Theory,''
{\it  Reading, USA: Addison-Wesley (1995) 842 p}

\bibitem{Amsler:2008zzb}
  C.~Amsler {\it et al.} [ Particle Data Group Collaboration ],
  ``Review of Particle Physics,''
  Phys.\ Lett.\  {\bf B667 } (2008)  1.


\bibitem{Englert:1964et}
  F.~Englert and R.~Brout,
  ``Broken Symmetry and the Mass of Gauge Vector Mesons,''
  Phys.\ Rev.\ Lett.\  {\bf 13} (1964) 321.

\bibitem{Higgs:1964pj}
  P.~W.~Higgs,
  ``Broken Symmetries and the Masses of Gauge Bosons,''
  Phys.\ Rev.\ Lett.\  {\bf 13} (1964) 508.
  
\bibitem{Guralnik:1964eu}
  G.~S.~Guralnik, C.~R.~Hagen and T.~W.~B.~Kibble,
  ``Global Conservation Laws and Massless Particles,''
  Phys.\ Rev.\ Lett.\  {\bf 13} (1964) 585.
  

\bibitem{Wess:1992cp}
  J.~Wess and J.~Bagger,
  ``Supersymmetry and supergravity,''
{\it  Princeton, USA: Univ. Pr. (1992) 259 p}

\bibitem{Martin:1997ns}
  S.~P.~Martin,
  ``A Supersymmetry Primer,''
  arXiv:hep-ph/9709356.

\bibitem{Weinberg:2000cr}
  S.~Weinberg,
  ``The quantum theory of fields.  Vol. 3: Supersymmetry,''
{\it  Cambridge, UK: Univ. Pr. (2000) 419 p}

\bibitem{Komatsu:2008hk}
  E.~Komatsu {\it et al.} [ WMAP Collaboration ],
  ``Five-Year Wilkinson Microwave Anisotropy Probe (WMAP) Observations: Cosmological Interpretation,''
  Astrophys.\ J.\ Suppl.\  {\bf 180 } (2009)  330-376.
  [arXiv:0803.0547 [astro-ph]].
  
\bibitem{Petersson:2011in}
  C.~Petersson and A.~Romagnoni,
  ``The MSSM Higgs Sector with a Dynamical Goldstino Supermultiplet,''
  JHEP {\bf 1202} (2012) 142
  [arXiv:1111.3368 [hep-ph]].

\bibitem{Petersson:2012dp}
  C.~Petersson, A.~Romagnoni and R.~Torre,
  ``Higgs Decay with Monophoton + MET Signature from Low Scale Supersymmetry Breaking,''
  JHEP {\bf 1210} (2012) 016
  [arXiv:1203.4563 [hep-ph]].
  
\bibitem{Bellazzini:2012mh}
  B.~Bellazzini, C.~Petersson and R.~Torre,
  ``Photophilic Higgs from sgoldstino mixing,''
  Phys.\ Rev.\ D {\bf 86} (2012) 033016
  [arXiv:1207.0803 [hep-ph]].

\bibitem{Petersson:2012nv}
  C.~Petersson, A.~Romagnoni and R.~Torre,
  ``Liberating Higgs couplings in supersymmetry,''
  Phys.\ Rev.\ D {\bf 87} (2013) 1,  013008
  [arXiv:1211.2114 [hep-ph]].
    
\bibitem{Dudas:2012fa}
  E.~Dudas, C.~Petersson and P.~Tziveloglou,
  ``Low Scale Supersymmetry Breaking and its LHC Signatures,''
  Nucl.\ Phys.\ B {\bf 870} (2013) 353
  [arXiv:1211.5609 [hep-ph]].
  
\bibitem{Berg:2012cg}
  M.~Berg, I.~Buchberger, D.~M.~Ghilencea and C.~Petersson,
  ``Higgs diphoton rate enhancement from supersymmetric physics beyond the MSSM,''
  Phys.\ Rev.\ D {\bf 88} (2013) 2,  025017
  [arXiv:1212.5009 [hep-ph]].
  
  
\bibitem{Dudas:2013mia}
  E.~Dudas, C.~Petersson and R.~Torre,
  ``Collider signatures of low scale supersymmetry breaking: A Snowmass 2013 White Paper,''
  arXiv:1309.1179 [hep-ph].
  
\bibitem{D'Hondt:2013fql}
  J.~D'Hondt, K.~De Causmaecker, B.~Fuks, A.~Mariotti, K.~Mawatari, C.~Petersson and D.~Redigolo,
  ``Multilepton signals of gauge mediated supersymmetry breaking at the LHC,''
  Phys.\ Lett.\ B {\bf 731} (2014) 7
  [arXiv:1310.0018 [hep-ph]].
  
\bibitem{Ferretti:2013wya}
  G.~Ferretti, A.~Mariotti, K.~Mawatari and C.~Petersson,
  ``Multiphoton signatures of goldstini at the LHC,''
  JHEP {\bf 1404} (2014) 126
  [arXiv:1312.1698 [hep-ph]].
  
\bibitem{Calibbi:2014pza}
  L.~Calibbi, A.~Mariotti, C.~Petersson and D.~Redigolo,
  ``Selectron NLSP in Gauge Mediation,''
  JHEP {\bf 1409} (2014) 133
  [arXiv:1405.4859 [hep-ph]].
  
\bibitem{Petersson:2014faa}
  C.~Petersson,
  ``Multilepton and multiphoton signatures of supersymmetry at the LHC,''
  arXiv:1405.5882 [hep-ph].
  
\bibitem{Petersson:2014oga}
  C.~Petersson,
  ``Disappearing charged tracks in association with displaced leptons from supersymmetry,''
  arXiv:1411.0632 [hep-ph].
  
\bibitem{Ferretti:2015dea}
  G.~Ferretti, R.~Franceschini, C.~Petersson and R.~Torre,
  ``Spot the stop with a b-tag,''
  Phys.\ Rev.\ Lett.\  {\bf 114} (2015) 201801
  [arXiv:1502.01721 [hep-ph]].
  
\bibitem{Ferretti:2015ala}
  G.~Ferretti, R.~Franceschini, C.~Petersson and R.~Torre,
  ``Light stop squarks and b-tagging,''
  PoS CORFU {\bf 2014} (2015) 076
  [arXiv:1506.00604 [hep-ph]].
  
\bibitem{Petersson:2015rza}
  C.~Petersson and R.~Torre,
  ``ATLAS diboson excess from low scale supersymmetry breaking,''
  JHEP {\bf 1601} (2016) 099
  [arXiv:1508.05632 [hep-ph]].
  
\bibitem{Petersson:2015mkr}
  C.~Petersson and R.~Torre,
  ``The 750 GeV diphoton excess from the goldstino superpartner,''
  arXiv:1512.05333 [hep-ph].
    

\bibitem{WeinbergCos}
S.~Weinberg,
 ``Gravitation and Cosmology: Principles and Applications of the General Theory of Relativity,''
{\it  John Wiley and Sons (1972) 657 p}

\bibitem{Misner:1974qy}
  C.~W.~Misner, K.~S.~Thorne and J.~A.~Wheeler,
  ``Gravitation,''
{\it  San Francisco 1973, 1279p}

\bibitem{Wald:1984rg}
  R.~M.~Wald,
  ``General Relativity,''
{\it  Chicago, Usa: Univ. Pr. ( 1984) 491p}

\bibitem{Green:1987sp}
  M.~B.~Green, J.~H.~Schwarz and E.~Witten,
  ``Superstring Theory. Vol. 1: Introduction,''
{\it  Cambridge, Uk: Univ. Pr. ( 1987) 469 P. ( Cambridge Monographs On Mathematical Physics)}

\bibitem{Green:1987mn}
  M.~B.~Green, J.~H.~Schwarz and E.~Witten,
  ``Superstring Theory. Vol. 2: Loop Amplitudes, Anomalies And Phenomenology,''
{\it  Cambridge, Uk: Univ. Pr. ( 1987) 596 P. ( Cambridge Monographs On Mathematical Physics)}

\bibitem{Polchinski:1998rq}
  J.~Polchinski,
  ``String theory. Vol. 1: An introduction to the bosonic string,''
{\it  Cambridge, UK: Univ. Pr. (1998) 402 p}

\bibitem{Polchinski:1998rr}
  J.~Polchinski,
  ``String theory. Vol. 2: Superstring theory and beyond,''
{\it  Cambridge, UK: Univ. Pr. (1998) 531 p}

\bibitem{Witten:1995ex}
  E.~Witten,
  ``String theory dynamics in various dimensions,''
  Nucl.\ Phys.\  B {\bf 443} (1995) 85
  [arXiv:hep-th/9503124].

\bibitem{Hull:1994ys}
  C.~M.~Hull and P.~K.~Townsend,
  ``Unity of superstring dualities,''
  Nucl.\ Phys.\  B {\bf 438} (1995) 109
  [arXiv:hep-th/9410167].

\bibitem{Horava:1995qa}
  P.~Horava and E.~Witten,
  ``Heterotic and type I string dynamics from eleven dimensions,''
  Nucl.\ Phys.\  B {\bf 460} (1996) 506
  [arXiv:hep-th/9510209].
  
  
\bibitem{Polchinski:1995df}
  J.~Polchinski and E.~Witten,
  ``Evidence for Heterotic - Type I String Duality,''
  Nucl.\ Phys.\  B {\bf 460} (1996) 525
  [arXiv:hep-th/9510169].

\bibitem{Polchinski:1995mt}
  J.~Polchinski,
  ``Dirichlet-Branes and Ramond-Ramond Charges,''
  Phys.\ Rev.\ Lett.\  {\bf 75} (1995) 4724
  [arXiv:hep-th/9510017].
  
\bibitem{Berkooz:1996km}
  M.~Berkooz, M.~R.~Douglas and R.~G.~Leigh,
  ``Branes intersecting at angles,''
  Nucl.\ Phys.\  B {\bf 480} (1996) 265
  [arXiv:hep-th/9606139].
  
\bibitem{Blumenhagen:2000wh}
  R.~Blumenhagen, L.~Goerlich, B.~Kors and D.~Lust,
  ``Noncommutative compactifications of type I strings on tori with  magnetic
  background flux,''
  JHEP {\bf 0010} (2000) 006
  [arXiv:hep-th/0007024].
  
\bibitem{Aldazabal:2000dg}
  G.~Aldazabal, S.~Franco, L.~E.~Ibanez, R.~Rabadan and A.~M.~Uranga,
  ``D = 4 chiral string compactifications from intersecting branes,''
  J.\ Math.\ Phys.\  {\bf 42} (2001) 3103
  [arXiv:hep-th/0011073].
 
\bibitem{Sagnotti:1987tw}
  A.~Sagnotti,
  ``Open strings and their symmetry groups,''
  Presented at the Cargese Summer Institute on Non-Perturbative Methods in Field Theory, Cargese, France, Jul 16-30, 1987. 
  [arXiv:hep-th/0208020].
  
\bibitem{Gimon:1996rq}
  E.~G.~Gimon and J.~Polchinski,
  ``Consistency Conditions for Orientifolds and D-Manifolds,''
  Phys.\ Rev.\  D {\bf 54} (1996) 1667
  [arXiv:hep-th/9601038].
  

\bibitem{Cvetic:2001nr}
  M.~Cvetic, G.~Shiu and A.~M.~Uranga,
  ``Chiral four-dimensional N = 1 supersymmetric type IIA orientifolds from
  intersecting D6-branes,''
  Nucl.\ Phys.\  B {\bf 615} (2001) 3
  [arXiv:hep-th/0107166].

\bibitem{Maldacena:1997re}
  J.~M.~Maldacena,
  ``The large N limit of superconformal field theories and supergravity,''
  Adv.\ Theor.\ Math.\ Phys.\  {\bf 2} (1998) 231
  [Int.\ J.\ Theor.\ Phys.\  {\bf 38} (1999) 1113]
  [arXiv:hep-th/9711200].
  
\bibitem{Klebanov:1998hh}
  I.~R.~Klebanov and E.~Witten,
  ``Superconformal field theory on threebranes at a Calabi-Yau  singularity,''
  Nucl.\ Phys.\  B {\bf 536} (1998) 199
  [arXiv:hep-th/9807080].
  
  
\bibitem{Klebanov:2000nc}
  I.~R.~Klebanov and A.~A.~Tseytlin,
  ``Gravity Duals of Supersymmetric SU(N) x SU(N+M) Gauge Theories,''
  Nucl.\ Phys.\  B {\bf 578} (2000) 123
  [arXiv:hep-th/0002159].
  
  \bibitem{Klebanov:2000hb}
  I.~R.~Klebanov and M.~J.~Strassler,
  ``Supergravity and a confining gauge theory: Duality cascades and
  chiSB-resolution of naked singularities,''
  JHEP {\bf 0008} (2000) 052
  [arXiv:hep-th/0007191].
  
  \bibitem{Berenstein:2005xa}
  D.~Berenstein, C.~P.~Herzog, P.~Ouyang and S.~Pinansky,
  ``Supersymmetry breaking from a Calabi-Yau singularity,''
  JHEP {\bf 0509} (2005) 084
  [arXiv:hep-th/0505029].

\bibitem{Franco:2005zu}
  S.~Franco, A.~Hanany, F.~Saad and A.~M.~Uranga,
  ``Fractional branes and dynamical supersymmetry breaking,''
  arXiv:hep-th/0505040.

\bibitem{Bertolini:2005di}
  M.~Bertolini, F.~Bigazzi and A.~L.~Cotrone,
  ``Supersymmetry breaking at the end of a cascade of Seiberg dualities,''
  Phys.\ Rev.\ D {\bf 72} (2005) 061902
  [arXiv:hep-th/0505055].
  
  \bibitem{Bagger:2007jr}
  J.~Bagger and N.~Lambert,
  ``Gauge Symmetry and Supersymmetry of Multiple M2-Branes,''
  Phys.\ Rev.\  D {\bf 77} (2008) 065008
  [arXiv:0711.0955 [hep-th]].
  
  \bibitem{Gustavsson:2007vu}
  A.~Gustavsson,
  ``Algebraic structures on parallel M2-branes,''
  arXiv:0709.1260 [hep-th].
  
\bibitem{Gomis:2008uv}
  J.~Gomis, G.~Milanesi and J.~G.~Russo,
  ``Bagger-Lambert Theory for General Lie Algebras,''
  JHEP {\bf 0806} (2008) 075
  [arXiv:0805.1012 [hep-th]].
 
\bibitem{Benvenuti:2008bt}
  S.~Benvenuti, D.~Rodriguez-Gomez, E.~Tonni and H.~Verlinde,
  ``N=8 superconformal gauge theories and M2 branes,''
  JHEP {\bf 0901} (2009) 078
  [arXiv:0805.1087 [hep-th]].
  
\bibitem{Aharony:2008ug}
  O.~Aharony, O.~Bergman, D.~L.~Jafferis and J.~Maldacena,
  ``N=6 superconformal Chern-Simons-matter theories, M2-branes and their
  gravity duals,''
  JHEP {\bf 0810} (2008) 091
  [arXiv:0806.1218 [hep-th]].

\bibitem{Shenker:1990uf}
  S.~H.~Shenker,
  ``The Strength of nonperturbative effects in string theory,''
  Presented at the Cargese Workshop on Random Surfaces, Quantum Gravity and Strings, Cargese, France, May 28 - Jun 1, 1990.

\bibitem{Dine:1986zy}
  M.~Dine, N.~Seiberg, X.~G.~Wen and E.~Witten,
  ``Nonperturbative Effects on the String World Sheet,''
  Nucl.\ Phys.\  B {\bf 278} (1986) 769.

\bibitem{Dine:1987bq}
  M.~Dine, N.~Seiberg, X.~G.~Wen and E.~Witten,
  ``Nonperturbative Effects on the String World Sheet. 2,''
  Nucl.\ Phys.\  B {\bf 289} (1987) 319.

\bibitem{Polchinski:1994fq}
  J.~Polchinski,
  ``Combinatorics of boundaries in string theory,''
  Phys.\ Rev.\  D {\bf 50} (1994) 6041
  [arXiv:hep-th/9407031].
  
\bibitem{Gibbons:1995vg}
  G.~W.~Gibbons, M.~B.~Green and M.~J.~Perry,
  ``Instantons and Seven-Branes in Type IIB Superstring Theory,''
  Phys.\ Lett.\  B {\bf 370} (1996) 37
  [arXiv:hep-th/9511080].
  
\bibitem{Green:1997tv}
  M.~B.~Green and M.~Gutperle,
  ``Effects of D-instantons,''
  Nucl.\ Phys.\  B {\bf 498} (1997) 195
  [arXiv:hep-th/9701093].
  
\bibitem{Green:1981ya}
  M.~B.~Green and J.~H.~Schwarz,
  ``Supersymmetrical Dual String Theory. 3. Loops And Renormalization,''
  Nucl.\ Phys.\  B {\bf 198} (1982) 441.
  
\bibitem{Gross:1986iv}
  D.~J.~Gross and E.~Witten,
  ``Superstring Modifications Of Einstein's Equations,''
  Nucl.\ Phys.\  B {\bf 277} (1986) 1.
  
\bibitem{Becker:1995kb}
  K.~Becker, M.~Becker and A.~Strominger,
  ``Five-Branes, Membranes And Nonperturbative String Theory,''
  Nucl.\ Phys.\  B {\bf 456} (1995) 130
  [arXiv:hep-th/9507158].
  
\bibitem{Harvey:1996ir}
  J.~A.~Harvey and G.~W.~Moore,
  ``Fivebrane instantons and R**2 couplings in N = 4 string theory,''
  Phys.\ Rev.\  D {\bf 57} (1998) 2323
  [arXiv:hep-th/9610237].
  
   \bibitem{Belavin:1975fg}
  A.~A.~Belavin, A.~M.~Polyakov, A.~S.~Shvarts and Yu.~S.~Tyupkin,
  ``Pseudoparticle solutions of the Yang-Mills equations,''
  Phys.\ Lett.\  B {\bf 59} (1975) 85.

  
 \bibitem{Witten:1995gx}
  E.~Witten,
  ``Small Instantons in String Theory,''
  Nucl.\ Phys.\  B {\bf 460} (1996) 541
  [arXiv:hep-th/9511030].

\bibitem{Douglas:1995bn}
  M.~R.~Douglas,
  ``Branes within branes,''
  arXiv:hep-th/9512077.
  
  
  
\bibitem{Billo':2008pg}
  M.~Billo', L.~Ferro, M.~Frau, F.~Fucito, A.~Lerda and J.~F.~Morales,
  ``Non-perturbative effective interactions from fluxes,''
  JHEP {\bf 0812} (2008) 102
  [arXiv:0807.4098 [hep-th]].
  
\bibitem{DiVecchia:1997pr}
  P.~Di Vecchia, M.~Frau, I.~Pesando, S.~Sciuto, A.~Lerda and R.~Russo,
  ``Classical p-branes from boundary state,''
  Nucl.\ Phys.\  B {\bf 507} (1997) 259
  [arXiv:hep-th/9707068].
  
   
\bibitem{Billo:2002hm}
  M.~Billo, M.~Frau, I.~Pesando, F.~Fucito, A.~Lerda and A.~Liccardo,
  ``Classical gauge instantons from open strings,''
  JHEP {\bf 0302} (2003) 045
  [arXiv:hep-th/0211250].
  
\bibitem{Atiyah:1978ri}
  M.~F.~Atiyah, N.~J.~Hitchin, V.~G.~Drinfeld and Yu.~I.~Manin,
  ``Construction of instantons,''
  Phys.\ Lett.\  A  {\bf 65}, 185 (1978).
  
  \bibitem{Witten:1996bn} E.~Witten, ``Non-Perturbative Superpotentials  In String Theory,''
Nucl.\ Phys.\  B {\bf 474}, 343 (1996)
  [arXiv:hep-th/9604030].  
  
  
  
  \bibitem{Blumenhagen:2006xt}
  R.~Blumenhagen, M.~Cvetic and T.~Weigand,
  ``Spacetime instanton corrections in 4D string vacua - the seesaw mechanism
  for D-brane models,''
  Nucl.\ Phys.\  B {\bf 771} (2007) 113
  [arXiv:hep-th/0609191].

\bibitem{Ibanez:2006da}
  L.~E.~Ibanez and A.~M.~Uranga,
  ``Neutrino Majorana masses from string theory instanton effects,''
  JHEP {\bf 0703} (2007) 052
  [arXiv:hep-th/0609213].

\bibitem{Florea:2006si}
  B.~Florea, S.~Kachru, J.~McGreevy and N.~Saulina,
  ``Stringy instantons and quiver gauge theories,''
  JHEP {\bf 0705} (2007) 024
  [arXiv:hep-th/0610003].



\bibitem{Abel:2006yk}
  S.~A.~Abel and M.~D.~Goodsell,
  ``Realistic Yukawa couplings through instantons in intersecting brane
  worlds,''
  JHEP {\bf 0710} (2007) 034
  [arXiv:hep-th/0612110].

\bibitem{Akerblom:2006hx}
  N.~Akerblom, R.~Blumenhagen, D.~Lust, E.~Plauschinn and M.~Schmidt-Sommerfeld,
  ``Non-perturbative SQCD Superpotentials from String Instantons,''
  JHEP {\bf 0704} (2007) 076
  [arXiv:hep-th/0612132].

\bibitem{Bianchi:2007fx}
  M.~Bianchi and E.~Kiritsis,
  ``Non-perturbative and Flux superpotentials for Type I strings on the $Z_3$
  orbifold,''
  Nucl.\ Phys.\  B {\bf 782} (2007) 26
  [arXiv:hep-th/0702015].

\bibitem{Cvetic:2007ku}
  M.~Cvetic, R.~Richter and T.~Weigand,
  ``Computation of D-brane instanton induced superpotential couplings -
  Majorana masses from string theory,''
  Phys.\ Rev.\  D {\bf 76} (2007) 086002
  [arXiv:hep-th/0703028].

\bibitem{Argurio:2007qk}
  R.~Argurio, M.~Bertolini, S.~Franco and S.~Kachru,
  ``Metastable vacua and D-branes at the conifold,''
  JHEP {\bf 0706} (2007) 017
  [arXiv:hep-th/0703236].



\bibitem{Bianchi:2007wy}
  M.~Bianchi, F.~Fucito and J.~F.~Morales,
  ``D-brane Instantons on the $T^6/Z_3$ orientifold,''
  JHEP {\bf 0707} (2007) 038
  [arXiv:0704.0784 [hep-th]].

\bibitem{Ibanez:2007rs}
  L.~E.~Ibanez, A.~N.~Schellekens and A.~M.~Uranga,
  ``Instanton Induced Neutrino Majorana Masses in CFT Orientifolds with
  MSSM-like spectra,''
  JHEP {\bf 0706} (2007) 011
  [arXiv:0704.1079 [hep-th]].

\bibitem{Akerblom:2007uc}
  N.~Akerblom, R.~Blumenhagen, D.~Lust and M.~Schmidt-Sommerfeld,
  ``Instantons and Holomorphic Couplings in Intersecting D-brane Models,''
  JHEP {\bf 0708} (2007) 044
  [arXiv:0705.2366 [hep-th]].

\bibitem{Antusch:2007jd}
  S.~Antusch, L.~E.~Ibanez and T.~Macri,
  ``Neutrino Masses and Mixings from String Theory Instantons,''
  JHEP {\bf 0709} (2007) 087
  [arXiv:0706.2132 [hep-ph]].


\bibitem{Blumenhagen:2007zk}
  R.~Blumenhagen, M.~Cvetic, D.~Lust, R.~Richter and T.~Weigand,
  ``Non-perturbative Yukawa Couplings from String Instantons,''
  arXiv:0707.1871 [hep-th].

 \bibitem{Aharony:2007db}
  O.~Aharony, S.~Kachru and E.~Silverstein,
  ``Simple Stringy Dynamical SUSY Breaking,''
  Phys.\ Rev.\  D {\bf 76} (2007) 126009
  [arXiv:0708.0493 [hep-th]].


\bibitem{Billo:2007sw}
  M.~Billo, M.~Frau, I.~Pesando, P.~Di Vecchia, A.~Lerda and R.~Marotta,
  ``Instantons in N=2 magnetized D-brane worlds,''
  JHEP {\bf 0710} (2007) 091
  [arXiv:0708.3806 [hep-th]].
  
 
\bibitem{Billo:2007py}
  M.~Billo, M.~Frau, I.~Pesando, P.~Di Vecchia, A.~Lerda and R.~Marotta,
  ``Instanton effects in N=1 brane models and the Kahler metric of twisted
  matter,''
  JHEP {\bf 0712} (2007) 051
  [arXiv:0709.0245 [hep-th]].


\bibitem{Camara:2007dy}
  P.~G.~Camara, E.~Dudas, T.~Maillard and G.~Pradisi,
  ``String instantons, fluxes and moduli stabilization,''
  Nucl.\ Phys.\  B {\bf 795} (2008) 453
  [arXiv:0710.3080 [hep-th]].

\bibitem{Cvetic:2007qj}
  M.~Cvetic and T.~Weigand,
  ``Hierarchies from D-brane instantons in globally defined Calabi-Yau
  Orientifolds,''
  arXiv:0711.0209 [hep-th].

\bibitem{Ibanez:2007tu}
  L.~E.~Ibanez and A.~M.~Uranga,
  ``Instanton Induced Open String Superpotentials and Branes at
  Singularities,''
  arXiv:0711.1316 [hep-th].



\bibitem{Blumenhagen:2007sm}
  R.~Blumenhagen, S.~Moster and E.~Plauschinn,
  ``Moduli Stabilisation versus Chirality for MSSM like Type IIB
  Orientifolds,''
  JHEP {\bf 0801} (2008) 058
  [arXiv:0711.3389 [hep-th]].


\bibitem{Cvetic:2007sj}
  M.~Cvetic, R.~Richter and T.~Weigand,
  ``D-brane instanton effects in Type II orientifolds: local and global
  issues,''
  arXiv:0712.2845 [hep-th].
  
    
\bibitem{Gukov:1999ya}
  S.~Gukov, C.~Vafa and E.~Witten,
  ``CFT's from Calabi-Yau four-folds,''
  Nucl.\ Phys.\  B {\bf 584} (2000) 69
  [Erratum-ibid.\  B {\bf 608} (2001) 477]
  [arXiv:hep-th/9906070].
  
\bibitem{Dasgupta:1999ss}
  K.~Dasgupta, G.~Rajesh and S.~Sethi,
  ``M theory, orientifolds and G - flux,''
  JHEP {\bf 9908} (1999) 023
  [hep-th/9908088].
  
\bibitem{Giddings:2001yu}
  S.~B.~Giddings, S.~Kachru and J.~Polchinski,
  ``Hierarchies from fluxes in string compactifications,''
  Phys.\ Rev.\  D {\bf 66} (2002) 106006
  [arXiv:hep-th/0105097].

\bibitem{Kachru:2003aw}
  S.~Kachru, R.~Kallosh, A.~D.~Linde and S.~P.~Trivedi,
  ``De Sitter vacua in string theory,''
  Phys.\ Rev.\  D {\bf 68} (2003) 046005
  [arXiv:hep-th/0301240].
  
\bibitem{Polyakov:1976fu}
  A.~M.~Polyakov,
 ``Quark Confinement And Topology Of Gauge Groups,''
 Nucl.\ Phys.\  B {\bf 120} (1977) 429.
  
\bibitem{Ooguri:1996me}
  H.~Ooguri and C.~Vafa,
  ``Summing up D-instantons,''
  Phys.\ Rev.\ Lett.\  {\bf 77} (1996) 3296
  [arXiv:hep-th/9608079].
  
\bibitem{Green:1997as}
  M.~B.~Green, M.~Gutperle and P.~Vanhove,
  ``One loop in eleven dimensions,''
  Phys.\ Lett.\  B {\bf 409} (1997) 177
  [arXiv:hep-th/9706175].

\bibitem{Green:2000ke}
  M.~B.~Green and M.~Gutperle,
  ``D-instanton induced interactions on a D3-brane,''
  JHEP {\bf 0002} (2000) 014 [arXiv:hep-th/0002011].


\bibitem{Dorey:2002ik} N.~Dorey, T.~J.~Hollowood, V.~V.~Khoze and
  M.~P.~Mattis, ``The calculus of many instantons,''
Phys.\ Rept.\ {\bf 371} (2002) 231 [arXiv:hep-th/0206063].


\bibitem{Witten:1995im} E.~Witten,
``Bound states of strings and  p-branes,''
Nucl.\ Phys.\  B {\bf 460} (1996) 335
  [arXiv:hep-th/9510135].  


 \bibitem{Brink:1976bc}
  L.~Brink, J.~H.~Schwarz and J.~Scherk,
  ``Supersymmetric Yang-Mills Theories,''
  Nucl.\ Phys.\  B {\bf 121}, 77 (1977).
  
\bibitem{DiVecchia:2005vm}
  P.~Di Vecchia, A.~Liccardo, R.~Marotta and F.~Pezzella,
  ``On the gauge / gravity correspondence and the open/closed string
  duality,''
  Int.\ J.\ Mod.\ Phys.\  A {\bf 20} (2005) 4699
  [arXiv:hep-th/0503156].
  
    \bibitem{'tHooft:1976fv}
  G.~'t Hooft,
  ``Computation of the quantum effects due to a four-dimensional
  pseudoparticle,''
  Phys.\ Rev.\  D {\bf 14}, 3432 (1976)
  [Erratum-ibid.\  D {\bf 18}, 2199 (1978)].
  
\bibitem{'tHooft:1976up}
  G.~'t Hooft,
  ``Symmetry breaking through Bell-Jackiw anomalies,''
  Phys.\ Rev.\ Lett.\  {\bf 37} (1976) 8.

\bibitem{Vandoren:2008xg}
  S.~Vandoren and P.~van Nieuwenhuizen,
  ``Lectures on instantons,''
  arXiv:0802.1862 [hep-th].


\bibitem{Affleck:1983mk} I.~Affleck, M.~Dine and N.~Seiberg,
  ``Dynamical Supersymmetry Breaking In Supersymmetric QCD,'' Nucl.\
  Phys.\  B {\bf 241} (1984) 493.  

   \bibitem{Taylor:1982bp} T.~R.~Taylor, G.~Veneziano and
  S.~Yankielowicz, ``Supersymmetric QCD And Its Massless Limit: An
   Effective Lagrangian Analysis,''
  Nucl.\ Phys.\  B {\bf 218} (1983)
  493.  
  
    \bibitem{Douglas:1996sw} M.~R.~Douglas and G.~W.~Moore,
``D-branes,  Quivers, and ALE Instantons,''
[arXiv:hep-th/9603167].
  
\bibitem{Morrison:1998cs}
  D.~R.~Morrison and M.~R.~Plesser,
  ``Non-spherical horizons. I,''
  Adv.\ Theor.\ Math.\ Phys.\  {\bf 3} (1999) 1
  [arXiv:hep-th/9810201].

\bibitem{Bertolini:2003iv}
  M.~Bertolini,
  ``Four Lectures On The Gauge/Gravity Correspondence,''
  Int.\ J.\ Mod.\ Phys.\  A {\bf 18} (2003) 5647
  [arXiv:hep-th/0303160].


    \bibitem{Bertolini:2001gg} M.~Bertolini, P.~Di Vecchia, G.~Ferretti
  and R.~Marotta, ``Fractional branes and N = 1 gauge theories,''
  Nucl.\ Phys.\ B {\bf 630} (2002) 222 [arXiv:hep-th/0112187].
  
  
  \bibitem{Aganagic:2007py}
  M.~Aganagic, C.~Beem and S.~Kachru,
  ``Geometric Transitions and Dynamical SUSY Breaking,''
 [arXiv:0709.4277 [hep-th]].
  
\bibitem{GarciaEtxebarria:2007zv}
  I.~Garcia-Etxebarria and A.~M.~Uranga,
  ``Non-perturbative superpotentials across lines of marginal stability,''
  [arXiv:0711.1430 [hep-th]].
  
\bibitem{Argurio:2012iw}
  R.~Argurio, D.~Forcella, A.~Mariotti, D.~Musso and C.~Petersson,
  ``Field Theory Interpretation of N=2 Stringy Instantons,''
  arXiv:1211.1884 [hep-th].
    
\bibitem{Billo':2008sp}
  M.~Billo', L.~Ferro, M.~Frau, F.~Fucito, A.~Lerda and J.~F.~Morales,
  ``Flux interactions on D-branes and instantons,''
  JHEP {\bf 0810} (2008) 112
  [arXiv:0807.1666 [hep-th]].
  
\bibitem{Tripathy:2005hv}
  P.~K.~Tripathy and S.~P.~Trivedi,
  ``D3 Brane Action and Fermion Zero Modes in Presence of Background Flux,''
  JHEP {\bf 0506} (2005) 066
  [arXiv:hep-th/0503072].
  
\bibitem{Martucci:2005rb}
  L.~Martucci, J.~Rosseel, D.~Van den Bleeken and A.~Van Proeyen,
  ``Dirac actions for D-branes on backgrounds with fluxes,''
  Class.\ Quant.\ Grav.\  {\bf 22} (2005) 2745
  [arXiv:hep-th/0504041].
      
\bibitem{Bergshoeff:2005yp}
  E.~Bergshoeff, R.~Kallosh, A.~K.~Kashani-Poor, D.~Sorokin and A.~Tomasiello,
  ``An index for the Dirac operator on D3 branes with background fluxes,''
  JHEP {\bf 0510} (2005) 102
  [arXiv:hep-th/0507069].
  
\bibitem{Blumenhagen:2007bn}
  R.~Blumenhagen, M.~Cvetic, R.~Richter and T.~Weigand,
  ``Lifting D-Instanton Zero Modes by Recombination and Background Fluxes,''
  JHEP {\bf 0710} (2007) 098
  [arXiv:0708.0403 [hep-th]].
    
\bibitem{Camara:2003ku}
  P.~G.~Camara, L.~E.~Ibanez and A.~M.~Uranga,
  ``Flux-induced SUSY-breaking soft terms,''
  Nucl.\ Phys.\  B {\bf 689} (2004) 195
  [arXiv:hep-th/0311241].
  
\bibitem{Grana:2003ek}
  M.~Grana, T.~W.~Grimm, H.~Jockers and J.~Louis,
  ``Soft Supersymmetry Breaking in Calabi-Yau Orientifolds with D-branes and
  Fluxes,''
  Nucl.\ Phys.\  B {\bf 690} (2004) 21
  [arXiv:hep-th/0312232].
  
  
  
\bibitem{Park:1999ep}
  J.~Park, R.~Rabadan and A.~M.~Uranga,
  ``Orientifolding the conifold,''
  Nucl.\ Phys.\  B {\bf 570} (2000) 38
  [arXiv:hep-th/9907086].

\bibitem{Beasley:2001zp}
  C.~E.~Beasley and M.~R.~Plesser,
  ``Toric duality is Seiberg duality,''
  JHEP {\bf 0112} (2001) 001
  [arXiv:hep-th/0109053].

\bibitem{Feng:2001bn}
  B.~Feng, A.~Hanany, Y.~H.~He and A.~M.~Uranga,
  ``Toric duality as Seiberg duality and brane diamonds,''
  JHEP {\bf 0112} (2001) 035
  [arXiv:hep-th/0109063].

\bibitem{Feng:2002zw}
  B.~Feng, S.~Franco, A.~Hanany and Y.~H.~He,
  ``Symmetries of toric duality,''
  JHEP {\bf 0212}, 076 (2002)
  [arXiv:hep-th/0205144].



\bibitem{Feng:2001xr}
  B.~Feng, A.~Hanany and Y.~H.~He,
  ``Phase structure of D-brane gauge theories and toric duality,''
  JHEP {\bf 0108}, 040 (2001)
  [arXiv:hep-th/0104259].



\bibitem{Martelli:2004wu}
  D.~Martelli and J.~Sparks,
  ``Toric geometry, Sasaki-Einstein manifolds and a new infinite class of
  AdS/CFT duals,''
  Commun.\ Math.\ Phys.\  {\bf 262} (2006) 51
  [arXiv:hep-th/0411238].

\bibitem{Benvenuti:2004dy}
  S.~Benvenuti, S.~Franco, A.~Hanany, D.~Martelli and J.~Sparks,
  ``An infinite family of superconformal quiver gauge theories with
  Sasaki-Einstein duals,''
  JHEP {\bf 0506} (2005) 064
  [arXiv:hep-th/0411264].

\bibitem{Bertolini:2004xf}
  M.~Bertolini, F.~Bigazzi and A.~L.~Cotrone,
  ``New checks and subtleties for AdS/CFT and a-maximization,''
  JHEP {\bf 0412}, 024 (2004)
  [arXiv:hep-th/0411249].

\bibitem{Franco:2005rj}
  S.~Franco, A.~Hanany, K.~D.~Kennaway, D.~Vegh and B.~Wecht,
  ``Brane dimers and quiver gauge theories,''
  JHEP {\bf 0601}, 096 (2006)
  [arXiv:hep-th/0504110].
  
\bibitem{Kachru:2008wt}
  S.~Kachru and D.~Simic,
  ``Stringy Instantons in IIB Brane Systems,''
  arXiv:0803.2514 [hep-th].
  
\bibitem{Billo:2009di}
  M.~Billo, L.~Ferro, M.~Frau, L.~Gallot, A.~Lerda and I.~Pesando,
  ``Exotic instanton counting and heterotic/type I' duality,''
  JHEP {\bf 0907} (2009) 092
  [arXiv:0905.4586 [hep-th]].
  
\bibitem{Moore:1998et}
  G.~W.~Moore, N.~Nekrasov and S.~Shatashvili,
  ``D-particle bound states and generalized instantons,''
  Commun.\ Math.\ Phys.\  {\bf 209} (2000) 77
  [arXiv:hep-th/9803265].
  
\bibitem{Gutperle:1999dx}
  M.~Gutperle,
  ``A note on heterotic/type I' duality and D0 brane quantum mechanics,''
  JHEP {\bf 9905} (1999) 007
  [arXiv:hep-th/9903010].

   
\bibitem{Harvey:1995fq}
  J.~A.~Harvey and G.~W.~Moore,
  ``Algebras, BPS States, and Strings,''
  Nucl.\ Phys.\  B {\bf 463} (1996) 315
  [arXiv:hep-th/9510182].

  
\bibitem{Blumenhagen:2008ji}
  R.~Blumenhagen and M.~Schmidt-Sommerfeld,
  ``Power Towers of String Instantons for N=1 Vacua,''
  JHEP {\bf 0807} (2008) 027
  [arXiv:0803.1562 [hep-th]].
   
\bibitem{Bachas:1997mc}
  C.~Bachas, C.~Fabre, E.~Kiritsis, N.~A.~Obers and P.~Vanhove,
  ``Heterotic/type-I duality and D-brane instantons,''
  Nucl.\ Phys.\  B {\bf 509} (1998) 33
  [arXiv:hep-th/9707126].

\bibitem{Kiritsis:1997hf}
  E.~Kiritsis and N.~A.~Obers,
  ``Heterotic/type-I duality in D < 10 dimensions, threshold corrections  and
  D-instantons,''
  JHEP {\bf 9710} (1997) 004
  [arXiv:hep-th/9709058].
   
\bibitem{Lerche:1998nx}
  W.~Lerche and S.~Stieberger,
  ``Prepotential, mirror map and F-theory on K3,''
  Adv.\ Theor.\ Math.\ Phys.\  {\bf 2} (1998) 1105
  [Erratum-ibid.\  {\bf 3} (1999) 1199]
  [arXiv:hep-th/9804176].

\bibitem{Lerche:1998gz}
  W.~Lerche, S.~Stieberger and N.~P.~Warner,
  ``Quartic gauge couplings from K3 geometry,''
  Adv.\ Theor.\ Math.\ Phys.\  {\bf 3} (1999) 1575
  [arXiv:hep-th/9811228].

\bibitem{Foerger:1998kw}
  K.~Foerger and S.~Stieberger,
  ``Higher derivative couplings and heterotic-type I duality in eight
  dimensions,''
  Nucl.\ Phys.\  B {\bf 559} (1999) 277
  [arXiv:hep-th/9901020].



\bibitem{Lerche:1998pf}
  W.~Lerche and S.~Stieberger,
  ``On the anomalous and global interactions of Kodaira 7-planes,''
  Fortsch.\ Phys.\  {\bf 48} (2000) 155
  [arXiv:hep-th/9903232].
  
\bibitem{Gutperle:1999xu}
  M.~Gutperle,
  ``Heterotic/type I duality, D-instantons and a N = 2 AdS/CFT
  correspondence,''
  Phys.\ Rev.\  D {\bf 60} (1999) 126001
  [arXiv:hep-th/9905173].
  
\bibitem{Kiritsis:1999ss}
  E.~Kiritsis,
  ``Duality and instantons in string theory,''
  arXiv:hep-th/9906018.
  
\bibitem{Gava:1999ky}
  E.~Gava, K.~S.~Narain and M.~H.~Sarmadi,
  ``Instantons in N = 2 Sp(N) superconformal gauge theories and the AdS/CFT
  correspondence,''
  Nucl.\ Phys.\  B {\bf 569} (2000) 183
  [arXiv:hep-th/9908125].
  
  
  
\bibitem{Kiritsis:2000zi}
  E.~Kiritsis, N.~A.~Obers and B.~Pioline,
  ``Heterotic/type II triality and instantons on K3,''
  JHEP {\bf 0001} (2000) 029
  [arXiv:hep-th/0001083].

 

\bibitem{Danielsson:1996es}
  U.~H.~Danielsson and G.~Ferretti,
  ``The heterotic life of the D-particle,''
  Int.\ J.\ Mod.\ Phys.\  A {\bf 12} (1997) 4581
  [arXiv:hep-th/9610082].

\bibitem{Kachru:1996nd}
  S.~Kachru and E.~Silverstein,
  ``On gauge bosons in the matrix model approach to M theory,''
  Phys.\ Lett.\  B {\bf 396} (1997) 70
  [arXiv:hep-th/9612162].
 
\bibitem{Erler:1993zy}
  J.~Erler,
  ``Anomaly Cancellation In Six-Dimensions,''
  J.\ Math.\ Phys.\  {\bf 35} (1994) 1819
  [arXiv:hep-th/9304104].
 
\bibitem{Green:1998by}
  M.~B.~Green and S.~Sethi,
  ``Supersymmetry constraints on type IIB supergravity,''
  Phys.\ Rev.\  D {\bf 59} (1999) 046006
  [arXiv:hep-th/9808061].

\end{thebibliography}


\end{document}